%% file: DSS-stable_and_unstable_PO.tex
\documentclass[12pt]{iopart}
\pdfoutput=1
\expandafter\let\csname equation*\endcsname\relax
\expandafter\let\csname endequation*\endcsname\relax
\usepackage{amsmath,amssymb}
\usepackage{mathrsfs}
\usepackage{cite}
\usepackage{chemarrow}
\usepackage{graphicx,color,colortbl}
\usepackage{braket}
\usepackage{arydshln}
\usepackage{ntheorem}
\usepackage[all]{xy}
\usepackage{dcolumn}

\newcommand{\sign}{\mathrm{sign}}
\newcommand{\F}{\mathcal{F}}
\renewcommand{\Finv}{\mathcal{F}^{-1}}

\newtheorem*{theorem-non}{}

\newcommand{\ra}{\rightarrow}
\usepackage{etoolbox}

\makeatletter
\def\@mkboth#1#2{}
\newlength\appendixwidth
\preto\appendix{\addtocontents{toc}{\protect\patchl@section}}
\newcommand{\patchl@section}{%
	\settowidth{\appendixwidth}{\textbf{Appendix }}%
	\addtolength{\appendixwidth}{1.5em}%
	\patchcmd{\l@section}{1.5em}{\appendixwidth}{}{\ddt}%
}
\makeatother

\makeatletter
\newcommand{\mainmatter}{%
	\setcounter{footnote}{0}%
	\patchcmd{\@makefntext}{\fnsymbol}{\arabic}{}{}%
	\patchcmd{\@thefnmark}{\fnsymbol}{\arabic}{}{}%
	\def\@makefnmark{\textsuperscript{\arabic{footnote}}}%
}
\makeatother

\begin{document}
	\title[
	Devil's Staircase---Unstable and Stable PO in AKP]
	{Creation Mechanism of Devil's Staircase Surface and Unstable and Stable Periodic Orbit in the Anisotropic Kepler Problem}
	\author{Tokuzo Shimada$^1$,Keita Sumiya$^2$, Kazuhiro Kubo$^3$}
	\address{$^1$Department of physics, School of Science and Technology,
		Meiji University, \\Higashimita 1-1-1, Tama, Kawasaki, Kanagawa, 214-8571, Japan}
	\address{$^2$Faculty of Global Media Studies, Komazawa University, Komazawa 1-23-1, Setagaya,Tokyo,154-0012, Japan}
	\address{$^3$Department of Physics, Bar-Ilan University, Ramat-Gan 52900, Israel}
	\ead{\mailto{tshimada@gravity.mind.meiji.ac.jp},\mailto{ksumiya@komazawa-u.ac.jp},\mailto{kkubo3@gmail.com}}

	\mainmatter

	\section{Introduction}
	\label{sec:1-introduction}
	We study two dimensional AKP (\ref{eq:2d-hamiltonian}) which admits a binary coding
	of the orbit.
	We focus our attention on  the coexistence of the unstable and stable POs of
	the same code.  We clarify the mechanism of a salient bifurcation process where,
	under decreasing system anisotropy,
	an unstable PO ($U$) bifurcates into a stable PO ($S$) and
	a new type of unstable PO ($U'$). Distinguishing self-retracing and non-self-retracing type of
	PO by `$R$' and `$NR$', the bifurcation is $U(R) \ra S(R) + U'(NR)$.
	
	Let us first have a quick look at AKP. It is a vital testing ground of quantum chaos ever since
	Gutzwiller cultivated it
	\cite{gutzwiller71, gutzwiller77,gutzwiller80,gutzwiller81, gutzwiller88a, gutzwiller89, yellowbook}
	to test his periodic orbit theory (POT) in its high anisotropy regime
	($\gamma \equiv \nu/\mu \le 0.2$).
	Because of the Coulomb interaction, it is not a KAM system and
	reveals an abrupt transition from the integrable limit if
	$\gamma$ is reduced from one \cite{wintgen88a}.
	The region ($0.7 \le \gamma \le 0.85$) interestingly exhibits
	critical statistics, which is reminiscent of the Anderson localization
	in condensed matter physics \cite{wintgen88a,verbaarschot03,garcia08, intech12}.
	Recently, we have successfully extracted information of 2-dim low-rank unstable POs
	(sequence, period and even the Lyapunov exponent) from quantized 3-dim AKP levels
	via inverse use of POT and suitable symmetrization \cite{ptep14}.
	We have also shown that the quantum scar in AKP is robust and survives under
	successive avoiding level crossings during large variation of the anisotropy \cite{intech13}.
	It is worth to note that AKP can be realized by a semiconductor with donor impurity and
	experimental manufacturing has been advanced\cite{zhou09,zhou10}.
	
	Now, there is a long standing question regarding the PO in AKP.
	It was shown by Gutzwiller \cite{gutzwiller77}
	and rigorously probed by Devaney \cite{devaney78} that, for an arbitrary given binary sequence
    (or code), 	there exists at least one initial point for a PO which evolves realizing the sequence,
    provided that $\gamma < 8/9$
	\footnote{
		In \cite{gutzwiller77}, a candidate `trajectory' for the code is constructed
		by combining arcs (each satisfies the equation of motion). Joining them into a smooth
        PO requires  the existence of the maximum of joint virial function, which is shown to be
        probable. In \cite{devaney78}, stable and unstable manifolds of distinct hyperbolic
		singularities of AKP Hamilton flow is shown to cross transversely along bi-collision orbits.
	}.
	Here the PO is isolated and unstable.
	
	The {\it uniqueness} of a PO for a given code, on the other hand, was conjectured
	by Gutzwiller from extensive numerical investigation, while
	Broucke found two stable POs as a counter example \cite{broucke85} in the low anisotropy
    region\footnote{%
		{\it Devaney instigated to find a counter example in the form of a stable PO}\cite{yellowbook2}.
		One is $(+--)^2$ (rank 3 and exists for $\gamma >0.5723511979186147$,
		and the other is $(+-++-)^2$ (rank $5$) and $\gamma > 0.7216809344182386$.
		The threshold values here are from our high precision measurement.
	}.
	An overview by 1990 is given in the Gutzwiller's book\cite{yellowbook2}.
	
	Recently, Contopoulos and Harsoula have found that
	there are many stable POs not necessarily in the low anisotropy regime.
	This implies  that AKP is not an Anosov system \cite{contopoulos05}.
 	Unfortunately, their PO search is limited to the case of perpendicular emission from
    the heavy axis, and hence is not capable of correctly detecting the $U'(NR)$
    in the case of the bifurcation $U(R) \ra S(R) + U'(NR)$, since $U'(NR)$ does not have any
    perpendicular crossing of the heavy axis.

	This paper is organized as follows. In \sref{sec:2-integrating}, three basics points are established.\\
	(1) One-time map. A suitably compactified 2-dim rectangle $D$
	is chosen as an initial value domain of the AKP flow and we work out the
	one-time map ${\cal F}: D_0 \ra D_1$. 	In fact,
    the map ${\cal F}|_I$ restricted to the collision manifold $\mathrm{I} (\subset D)$ is investigated
    by Gutzwiller\cite{gutzwiller77} and we integrate his wisdom into a convenient form.
	It is extended to the interior combining numerical analysis (\fref{fig:boundary-map} and \ref{fig:one_time_map_scheme3d}).
	\\
	(2) A level $N$ devil's staircase surface (DSS) and associated
    tiling of $D_0$ by the base {\it ribbons} of the steps.
	 \\
	(3) PO trapping. An initial point of a PO of a given code must lie in the cross-sectional
	area of a future (F) and a past (P)  ribbon of the respective F and P code\footnote{%
		This is noted by Gutzwiller (p.169 in \cite{yellowbook}).
		At the same place, the difficulty to use it to locate an unstable PO due to high Lyapunov exponent is mentioned
		(even for rank 5). Our approach using ribbons (rather than constructing unstable and stable
		manifold) can easily overcome it.}.
	\\
	Then, in \sref{sec:3-DSS-creation}, how the level $N+1$ DSS is created from level $N$ one
    is clarified. The mechanism gives, on one hand, a simple proof of the properness of the tiling (monotonicity of the surface)---traversing
    ribbons from left to right, the height of step always increases.
    On the other hand, it points out the possibility of a non-shrinking ribbon at large $N$.
	\\
	In \sref{sec:4:stable-and-unstable-po}, it is shown explicitly  that there are two fates
   of ribbons at large $N$  depending on the code and the anisotropy.\\
	(A) In most cases, both the F and P ribbons shrink at large $N$.
    Then the trapped PO,  in the cross-sectional area,  is singled out at the crossing of F and P asymptotic curves. \\
	(B) Below certain anisotropy, there emerges an exceptional ribbon, which
    stops shrinkage at finite $N$.
	\footnote{%
		This is also noticed by Gutzwiller with respect to Broucke's island as
		{\it a short but non-vanishing interval} (of the 1-dim devil's staircase).
		(p.409 in \cite{yellowbook}. Also see \cite{gutzwiller88a}).
		In terms of unstable manifold analysis, the island opens after
		{\it some of its {\rm(binary tree's)} branches have been bundled together.}
	}.
	We show this occurs just when the future and past ribbons
	become tangent each other.
	Then, the initial points of unstable and stable PO are both contained in
	the cross-sectional area of non-shrinking future and past ribbons.
	The emergence of such non-shrinking ribbons in turn induces
	a salient code-preserving bifurcation: $U(R) \ra S(R)+U'(NR)$ within $Y$ symmetric PO
    (\fref{fig:broucke_evolution_PO3-6}--\ref{fig:broucke_location}).
	We explain this pattern by symmetry and topology consideration of PO 	
    (\fref{fig:y-sym-classification})
	and conjecture that it should occur for $Y$-symmetric, rank odd PO.
	
	\Sref{sec:5-applications} is devoted to the application of the
	ribbon tiling to the PO search in AKP.
	In order to search PO of all possible symmetry types, brute force shooting would require
	the search in the vast two-dimensional free parameter space, which is not feasible for
	high rank PO.
	However, two-step search---firstly search the relevant ribbon and
	then the PO within it---clearly reduces the task to quasi-one-dimensional.
	Based on this idea, we devise a concrete algorithm of exhaustive search, which is
    sensitive to the coexistence of stable and unstable POs.
	We firstly take a high rank PO ($n=15$) as an example
	and show it does bifurcate as $U(R) \ra S(R)+U'(NR)$.
	We pursue the final POs to the very low anisotropy region ($\gamma>8/9$), and find
	a second bifurcation $S(R) \ra S'(R)+S''(NR)$. (See \fref{fig:PO15-1_bifurcation}).
    We find these successive bifurcations are in accordance with the topology and symmetry
	classification.
	Then, we turn to an extensive PO search, rank extending to $n=10$, and
	at high anisotropy $\gamma=0.2$. This is important since a question is
	cast on the possible existence of a stable PO even at high anisotropy regime \cite{contopoulos05}.
    We find all the 19284 isolated unstable POs just predicted from symmetry analysis
	(with a correction for the new symmetric type ($O$ type) PO at rank seven and nine).
    and nothing else. This confirms (within the set-up resolution) the uniqueness of a PO at a given
    code (modulo symmetry equivalence) at $\gamma=0.2$
	The above unstable POs are used to verify Gutzwiller's approximate action formula.
	Actions of 13648 POs at rank $n=10$ are predicted (with its only two parameters fixed
	by 44 POs at $n=5$) with amazing accuracy (MSD=0.0536). (See \fref{fig:action_test}
    and \tref{tab:msd}).

\section{Devil's Staircase Surface and the One-time Map}
\label{sec:2-integrating}

\subsection{Gutzwiller's rectangle}
\label{subsec:2-1:Gutzwiller-rectangle}

The Hamiltonian of two-dimensional AKP is given by
\begin{equation}
	\label{eq:2d-hamiltonian}
	H = \frac{u^2}{2\mu} + \frac{v^2}{2\nu} -\frac{1}{r},
	~~(r\equiv\sqrt{x^2+y^2}, ~\mu > \nu)
\end{equation}
where $u\equiv p_x, v\equiv p_y$ and the mass anisotropy
is proportional to $1-\gamma$ with $\gamma \equiv \nu/\mu< 1$.
Because $\ddot{y}/\ddot{x}=(\mu/\nu){y}/{x}>{y}/{x}$,
the orbit tends to cross the heavy $x$-axis more frequently. Thus,
the Poincar\'e surface of section (PSS) is specified
by the condition $y=0$ in
the phase space,
and we can encode the orbit by the code of $a_i=\pm 1$,  the sign of $x_i$ at
the $i$-th crossing of the orbit with the $x$-axis.
The future code is $(a_0,a_1,\cdots)$ and the past is $(a_0,a_{-1},\cdots)$
\footnote{%
	To treat the future and past DSS on an equal footing
	we include $a_0$ in both sequences.
	The choice of the present ($i=0$) is immaterial due to the time-translation invariance.
}.
~The constant energy condition determines the kinematically allowed region
on the PSS (the physical region for short).  As the potential is a homogeneous in coordinates, the system
has a scaling property. Any value for the energy is equivalent and we take $H=-1/2$
by convention. Under $y=0$ and $H=-1/2$,
the domain of initial coordinates in the PSS is
\begin{equation}
	\mathfrak{D}=\Set{ {(x,u)}|{|x|} \leq \frac{2}{1+u^{2}/\mu} ~\mathrm{and}~ -\infty \le u \le +\infty },
	\label{eq:lips}
\end{equation}
which is a `lips'-like region in \fref{fig:gutzwiller-lips}(a)\footnote{%
	In fact the physical region is a double-cover of \fref{fig:gutzwiller-lips}(a)
	because $v$ can be either
	positive or negative. Hereafter we limit our consideration
	to $\mathfrak{D}$ with $v\ge 0$, since the case $v<$ is obtained by a simple reflection
	with respect to the $y-$axis.}.
\input gutzwiller-lips.tex

The part of $\mathfrak{D}$ where $r=0$ (the location of the collision) is the core line of $\mathfrak{D}$
(the $u$-axis)
where $v=\infty$ and $-\infty \leq u \leq \infty$.
Following Gutzwiller we compactify $\mathfrak{D}$ by an area preserving map
\begin{equation}
	X=x\left(1+u^{2}/\mu\right),~~U=\sqrt{\mu}\arctan \left(u/\sqrt{\mu}\right)
	\label{eq:xu-transformation}
\end{equation}
into a rectangle
\begin{equation}
	D =\Set{ (X,U)|  {|X|}\leq 2~\mathrm{and}~{|U|} \leq B \equiv\frac{\sqrt{\mu} \pi}{2}}
	\label{eq:rectangle}
\end{equation}
depicted in \fref{fig:gutzwiller-lips}(b).
The collision occurs on the I-shaped backbone of $D$ :
\begin{equation}
	\label{eq:backbone}
	\mathrm{I}=\Set{(X,U)|{|X|}\leq 2 ~\mathrm{and}~ U=\pm B } \cup \Set{(X,U)|X=0 ~\mathrm{and}~ {|U|} \leq  B}.
\end{equation}
We call $\mathrm{I}$ collision manifold.

\subsection{Symbolic coding of orbits and Devil's staircase surface}
\label{subsec:2-2:symbolic-coding-and-devil's-staircase-surface}

The future and past surfaces over $ D $
at level $N$ are respectively described by the height functions
\begin{equation}
	\zeta^{F}_N(X_0,U_0)  \equiv \sum_{j=0}^{N} \frac{1}{2^{j+1}} a_j (X_0,U_0), ~
	\zeta^{P}_N(X_0,U_0) \equiv \sum_{j=0}^{N} \frac{1}{2^{j+1}} a_{-j}(X_0,U_0).
	\label{eq:zeta}
\end{equation}
\fref{fig:coarse-grained-dss} shows the first three future DSS as examples.
They are calculated from $a_j(X_0,U_0)$
of each orbit starting from a site $(X_0,U_0)$ of a fine lattice set on $D$.
\input coarse-grained-dss.tex
\noindent
By definition (\ref{eq:zeta}), the level $N$ surface has $2^{N+1}$ heights.
Now, the examples reveal rather simple structure.
Firstly, there is only one {\it step} for each height in a surface. We call the base of each
step a {\it ribbon} (see \fref{fig:step-ribbon}).
\input step-ribbon.tex
\noindent
Then, each ribbon fully extends from $U_0=-B$ to $U_0=B$
and there is no isolated bubbles in the base domain.
Therefore, the $2^{N+1}$ ribbons altogether constitute a {\it tiling} of base domain $D$.
Furthermore, the height of the step is monotonously increasing,
if we traverse ribbons from left to right.
Let us describe this as ribbons are {\it properly tiling} $D$.

The reason why the surface is organized in this way at every $N$ is by no means
trivial and we give a proof in \sref{sec:3-DSS-creation}.
A remark is in order about our approach based on based on finite $N$
(or coarse-grained) DSS.
The edges of ribbons constitute the unstable (stable) manifolds of the collision
for the future (past) DSS, because the change of the code is invoked by the collision.
(See \fref{fig:c-curve}).
They can be calculated by Gutzwiller's technique of the collision
parameter \cite{gutzwiller77}.
In a way, our approach  is dual to Gutzwiller's.
However, the reason of monotonic ordering of the manifolds
is very hard to see in the latter. More importantly,
the proof on proper tiling (by mathematical induction regarding $N$)
in turn clarifies how the higher level surfaces are
generated by the repeated application of one-time map.
And, through this generating mechanism, we can
clarify the coexistence of PO having the same code.

Now, let us consider how a PO is related to ribbons.
The code of a PO is cyclic, that is,
\begin{eqnarray}
	a^F_{\mathrm{PO}}
	=(a_0,a_1,\cdots, a_{2n-1}; a_{2n}=a_{0},  a_{2n+1}=a_{1}, \cdots).
	\label{eq:po-sequence}
\end{eqnarray}
Here the first $2n$ bits are the primary part of the PO and
the integer $n$ is the rank of the PO\footnote{%
	\ The length of PO must be even ($2n$),  because it must
	cross the $x$-axis even times to complete its period.
}. 
If the initial point of the PO is $(X^*_0,U^*_0)\in$D,
its height is naturally defined by
\begin{eqnarray}
	\zeta^{F}_{\mathrm{PO}}(X^*_0,U^*_0) =
	\frac{1}{1-\frac{1}{2^{2n}}}
	\left\{
	\sum_{j=0}^{2n-1} \frac{1}{2^{j+1}} a_j (X^*_0,U^*_0)
	\right\},
	\label{eq:zetaPO}
\end{eqnarray}
where the pre-factor takes care of the repetition of the primary cycle.
Then we can prove the following:
\begin{theorem-non}{}
	A step of the level $N$ DSS whose height is given by the
	first $N+1$ bits of the string (\ref{eq:po-sequence});
	\begin{equation}
		\zeta^{F}_N(a)
		= \sum_{j=0}^{N} \frac{1}{2^{j+1}} a_j (X^*_0,U^*_0).
		\label{eq:zeta-by-the-N+1-bits}
	\end{equation}
	is the closest to the PO in height.
	Therefore, its ribbon traps the initial point $(X^*_0,U^*_0)$.
	See \fref{fig:step-ribbon}.
\end{theorem-non}
To see this, consider the amount of misfit between
$\zeta^{F}_{\mathrm{PO}}$ and $\zeta^{F}_N$.
It is nothing but the contribution truncated away from the string
(\ref{eq:po-sequence}) under the coarse-graining to level $N$. Therefore,
\begin{equation}
	\delta \equiv \zeta^{F}_{\mathrm{PO}} -\zeta^{F}_N
	=\sum_{j=N+1}^{\infty} \frac{1}{2^{j+1}} a_j (X^*_0,U^*_0)
	\label{eq:misfit}
\end{equation}
and we find $|\delta| \le {1}/ {2^{N+1}}$.
On the other hand, the difference of height
between neighboring steps is $\Delta_N={1}/{2^N}$.
Therefore,  the selected step is the closest.\footnote{%
	\ This statement holds
	irrespective of whether the tiling by ribbons is proper or not.
} 
At the large $N$ limit, the misfit $\delta$ vanishes and
the PO asymptotically sits on the selected step,
whether the enclosing ribbon shrinks or not.$\blacksquare$

\subsection{One-time map}
\label{subsec:2-3:one-time-map}
We now investigate the one-time map from a rectangle $ D_0 $ on a certain PSS
onto $ D_1 $ on the next PSS.
\begin{eqnarray}
	\begin{array}{cccc}
		{\cal F}:  &             D_0     &     \longrightarrow &       D_1 \\
		&\rotatebox{90}{$\in$} &                             & \rotatebox{90}{$\in$} \\
		& (X_0,U_0)    &     \longmapsto      & (X_1,U_1).
	\end{array}
	\label{eq:one-time-mapF}
\end{eqnarray}
This is a challenging problem, because the AKP flow involves
both hyperbolic and elliptic singularities. Separation as well as
blow up inevitably occur and a brute force numerical integration only is not sufficient to grasp the feature.
However, the map restricted to the collision manifold, $\left.{\cal F}\right |_I  $,
is worked out by Gutzwiller in \cite{gutzwiller77}.
The Hamiltonian (\ref{eq:2d-hamiltonian}) is rewritten
using polar coordinates ($r, \psi$) and $(\chi,
\vartheta$) for both the coordinates and momenta;
\begin{eqnarray}
	x&=r \cos{\psi},~~~u&=\sqrt{\mu} e^\chi \cos{\vartheta},\nonumber\\		
	y&=r \sin{\psi},~~~v&=\sqrt{\nu} e^\chi \sin{\vartheta}.
	\label{eq:doublepolar}
\end{eqnarray}
The kinetic energy is $K = e^{2\chi}/2$ and under the canonical choice of energy
$H=-1/2$ it follows
\begin{equation}
	\label{eq:r-and-chi}
	r={2}/(1+e^{2\chi}).
\end{equation}
Slowing down the orbit by $dt'=e^{3\chi} dt$, and taking the limit $\chi \ra \infty$, the equation of motion
is reduced to an autonomous form
\footnote{\
	The collision occurs at $\chi=\infty$.
	This is equivalent to the blow up technique used by Devaney \cite{devaney78}
	to remove the singularity due to the collision and introduce the collision manifold. See also McGehee \cite{mcgehee74}.
}
\begin{eqnarray}
	\frac{d \vartheta}{dt'}&=-
	\left(\sqrt{\mu}\cos{\vartheta}\sin{\psi}-\sqrt{\nu}\sin{\vartheta}\cos{\psi}
	\right),\nonumber\\
	\frac{d \psi}{dt'}&=-2
	\left(	\sqrt{\nu}\cos{\vartheta}\sin{\psi}-\sqrt{\mu}\sin{\vartheta}\cos{\psi}
	\right).
	\label{eq:autonomous-flow}
\end{eqnarray}
This gives the map
\begin{eqnarray}
	{\cal M}:(\vartheta_0,\psi_0)  \ra (\vartheta_1,\psi_1),
	\label{eq:one-time-map-polar}
\end{eqnarray}
where $(\vartheta_0,\psi_0)$ and $(\vartheta_1,\psi_1)$
parameterize respectively the initial and final $\mathrm{I}$.
($\psi_0$ and $\psi_1$ are either 0 or $\pi$, because $y=0$ on PSS.
$\psi_0$ is taken to be 0 as a choice of the fundamental initial domain).
Thus all we need to obtain $F_I$ in terms $(X,U)$ is
properly pulling back and pushing forward (\ref{eq:one-time-map-polar});
\begin{eqnarray}
	\xymatrix{
		{\cal F}|_I:& (X_0,U_0)  \ar@{|->}[r]^{*}  &  (\vartheta_0,\psi_0)  \ar@{|->}[r]^{\cal M}
		&  (\vartheta_1,\psi_1)  \ar@{|->}[r]^{*}  &  (X_1,U_1)	},
	\label{eq:pull-back}
\end{eqnarray}
where the map with the starred arrow should be calculated by
\begin{eqnarray}
	X &=& 2 ~\sign (\cos \psi)\cos^2\vartheta,~~U= \sign(\cos \vartheta) \frac{\pi\sqrt{\mu}}{2},
	\label{eq:boundary-relation}
\end{eqnarray}
which is valid on $\mathrm{I}$.
The above outline to obtain ${\cal F}_I$  is
substantiated in \ref{appendix:a}.
A careful examination is necessary because firstly the initial domain of the map ${\cal M}$
divides into three regions $(1, 2, 3)$ due to the separation by hyperbolic singularities
(see the flow in \fref{fig:separation})
and furthermore the pulling back and push forward procedure in (\ref{eq:pull-back})
introduces additional criticality (when $\vartheta_0$ passes through $\pi/2$, or,
when $\vartheta_1$ passes through $3\pi/2$).
Consequently,  the collision manifold $\mathrm{I}$ is divided eventually into
five regions $1, 2A,2B,2C, 3$ and we have to consider
how each is mapped by ${\cal F}_I$.
\input separation-table.tex
This issue is discussed in \ref{appendix:a}
and the result is summarized in \tref{tab:separation-table} above.
\Fref{fig:boundary-map} is graphical representation of it.
\input boundary-map.tex
\noindent
We observed that (i) the boundary map $\left.{\cal F}\right |_I $ acts on the
boundaries of $(++)$ and $(+-)$ separately, sending them to the left and right respectively,
and that (ii) $\left.{\cal F}\right |_{(+-)} $ rotates the boundary of $(+-)$ by one sub-region.

The virtue of the collision manifold analysis is that $\left.{\cal F}\right |_I $ extends
to the interior and helps to grasp the behavior of the internal map $\F$---the point
emphasized by Devaney \cite{devaney78}.
\input one_time_map_scheme3d.tex
In \fref{fig:one_time_map_scheme3d},
we combine \fref{fig:boundary-map}
with the numerically calculated interior map.
Indeed above two characteristics $\left.{\cal F}\right |_I $
clearly controls the interior map.
The middle regions $(+-)$ and $(-+)$ swaps the order to become $(-+)'$ and $(+-)'$ in the image
corresponding to  the separation property (i) of $\left.{\cal F}\right |_I $.
Remarkably, the distortion of the orthogonal lattice in $D_0$
shows that there exist focus points in the interior map, for instance,
$v^1_{++}$ and $v^1_{+-}$, which are respectively the image of the sides of the separator curve $C$.
\begin{eqnarray*}
	C^0_{++} &\ra v^1_{++},\\
	C^0_{+-} &\ra v^1_{+-}.
\end{eqnarray*}
For the notation of critical objects, see \ref{appendix:b}.
Furthermore, the rotation property (ii) implies that the vertical lines
connecting $2A$ and $2B$ in $D_0$ must be bent to connect the images
$2A'$ and $2B'$ both in the bottom boundary of $D_1$.
This produces wing-shaped curves in $D_1$ showing folding property in ${\cal F}$.
The wings contract to a single point $c^0_{+}$, and it is as well a focus
point to which the $X_0=0$ ends of horizontal lines in $(+-)$, namely ${\cal I}_{+}$, are mapped;
\begin{eqnarray*}
	{\cal I}_{+} & \ra c^1_{+}.
\end{eqnarray*}
We investigate these phenomena accurately in \ref{appendix:b}.
The occurrence of blow-up and contraction may appear ad-hoc, but
it is quite systematic. This can be seen if we follow in \tref{tab:separation-table} the seven parts in the boundary of $(+-)$
(and their image) counter-clockwise;
\begin{equation}
	\label{eq:perimeter-invariance}
	\begin{array}{ccccccccccccc}
		v^0_{+-}&\!\ra \!&2A           &\!\ra \!   &{\cal I}^0_{+}&\ra &2B    &\ra &       c^0_{-}     &\ra & 2C    &\ra &C^0_{+-} \\
		0       &    &\ell_v	   &       &\ell({\cal I})&    &\ell_c&    &       0      &    &\ell_h-\ell_c&    &   \ell(C) \\
		\wedge  &    &       &     &  \vee &    &     &    &\wedge   &    &       &    &  \vee  \\
		C^1_{+-}&\!\ra \!&2A'    &\ra  &c^1_{+}&\ra &2B'  &\ra &{\cal I}^1_{-}&\ra & 2C'   &\ra & v^1_{+-} \\
		\ell(C)    &    &\ell_h-\ell_c&       &   0      &    &\ell_c&    &    \ell({\cal I})     &    & \ell_v   &    &  0.
	\end{array}
\end{equation}
Here $\ell(C)$ and $\ell({\cal I})$ denote respectively the length of
the separator curve $C$ and ${\cal I}$. ($\ell({\cal I}) = \pi\sqrt{\mu}$, $\ell_{v,h,c} =2\cos^2\vartheta_{v,h,c}$).
The blow-up and contraction of a part are specified by
$\wedge$ and $\vee$ symbols respectively.
Note $(+-)$ and $(+-)'$ are congruent each other, because $(+-)=P(T((+-)'))$,
see \eref{eq:congruence-B}. Indeed the blow-up and contraction are involved symmetrically equal-times
so that the perimeters of $(+-)$ and $(+-)'$ are intriguingly kept the same.
\input c-curve.tex
The reason why a curve is contracted into a point
and a point is blown up to a curve is as follows.
In \fref{fig:c-curve}, we show the result of
our collision parameter analysis, which follows Gutzwiller \cite{gutzwiller88a,yellowbook}.
The collision orbits are created as a one parameter family.
parameter $A$. The first short time interval is analytically calculated and
continued by numerical integration.
The positions of the first arrivals on the Poinca\'e surface
\noindent
forms a curve in the $(X_1,U_1)$-rectangle developed
by the parameter A. This is nothing but the separator curve $C$.
This is a typical sample of the blow-up process.
Note that one can reversely track back by numerical integration
(with slow down) starting from $C^1_{+-}$,
but it is difficult to follow their motion after closely reaching the upper end
of ${\cal I}$.

Continuing numerically the collision trajectory for many crossings of PSS,
the stable and unstable manifold are produced as shown in
Gutzwiller Fig.2 in \cite{gutzwiller77} and Fig.26 in \cite{gutzwiller88a}.
In a way, our approach and collision analysis are complementary to each other.
We focus on finite $N$ ribbons, while the collision analysis focuses
on inter-ribbon-ribbon curves. That is, the longitudinal line separating
two ribbons are nothing but the stable/unstable manifold, because
the code change is just induced by the collision.
The difference is that, while it is a tough problem in the collision parameter
analysis to tell which hierarchy a line is subject to, our approach has the
advantage of a simple book-keeping. Hence it can be easily used to locate
PO directly, and helps to resolve the coexistence of stable and unstable POs
of the same code. We discuss these points in detail below.

\section{Creation Mechanism of Proper Tiling by Ribbons}
\label{sec:3-DSS-creation}
\subsection{Proper Tiling by Ribbons of the Initial value Domain $D_0$. Proof Starts:}
We prove below that
\begin{theorem-non}
	At any $N$, the ribbons are tiled properly
\end{theorem-non}
by mathematical induction with respect to $N$.
The $N=0$ case is special and the one-time-map scheme in \fref{fig:one_time_map_scheme3d} is responsible for $N \ge 1$.
The future case is considered; past case goes similarly.
In the proof it is clarified how level $N+1$ tiling is
created from level $N$, and this in turn shows how a non-shrinking
ribbon can appear in a special case.

\subsubsection*{$N=0$:} The $N=0$ tiling is determined by $a_0=\sign(X_0)$.
The ribbons are $(-)$, $(+)$ from left and right,
and the step height $\zeta^F_{N=0}$ is respectively
$-\frac{1}{2},~\frac{1}{2}$. Therefore, the tiling is proper.
Here, the part ${\cal I}^0_{\pm}$ of the collision manifold ( \fref{fig:one_time_map_scheme3d}) takes the role of the separator.
\subsubsection*{$N=1$:} Now, $a_1$ is determined by the separator curves
$C^0_{+\pm}$ and $C^0_{-\pm}$ in the one-time map ${\cal F}$.
As the result, the height distribution over the ribbons
is
\begin{equation}
	\begin{array}{ccccc}
		\zeta^{N=1}& \!\!\!\!-\frac{3}{4}&\!\!\!\!-\frac{1}{4}&\frac{1}{4}&\frac{3}{4}\\
		                                        &\uparrow &\uparrow &\uparrow   &\uparrow\\
	\mathrm{Ribbon}:(a_0,a_1): &      (--)   &    (-+)    &   (+-)       &(++)\\
                                              &L     \cdots &  \cdots   &  \cdots     &\cdots   R
    \end{array}
    \label{eq:N=1_of_induction}
\end{equation}
Each ribbon extends from top to bottom ($U_0=B$ to $-B$). Thus, the tiling at $N=1$ is also proper.

\subsubsection*{$N$ to $N+1$:}~~Let us show that,
if the level $N$ tiling of $(X_{0},U_{0})$ is proper, then the level $N+1$ tiling is also proper for $N\ge1$.
The former is determined by \eref{eq:zeta} by $(N+1)-$bit binary sequence $\left(a_0, a_1,\cdots, a_N\right)$.
Let us make the initial value dependence of $\zeta^N$ via $\F$ explicit, that is,
\begin{eqnarray}	
	\zeta^N(X_0,U_0) &= \sum_{j=0}^{N}	
	\frac{\left(s \circ \F^j\right)\left(X_0,U_0\right)}{2^{j+1}},  ~~(X_0,U_0) \in D_0
	\label{eq:zeta-explicit}
\end{eqnarray}
where $s$ is a function
\begin{eqnarray*}
	\label{eq:s-symbol}    s(X,U) \equiv \sign{(X)}.
\end{eqnarray*}
In the same way, $\zeta^{N+1}(X_0,U_0)$ is by definition calculated from the$(N+2)-$bit
sequence $\left(a_0, a_1,\cdots, a_N, a_{N+1}\right)$.
But the last bit
\begin{eqnarray*}
    a_{N+1}\left(X_0,U_0\right) =  \left(s \circ \F^{N+1}\right) \left(X_0,U_0\right)
\end{eqnarray*}
can be only calculated via the full $\F^{N+1}$; it cannot be obtained by simply combining the one-time map
$\F$ with the given level $N$ height function
$\zeta^{N}(X_0,U_0)$.
One may resort to a numerical calculation, but then the analytic
understanding of the system is lost.
A crack of nutshell is to call up the inverse of ${\cal F}$, and map once {\it backward}
from $D_0$ onto $D_{-1}$ to obtain
\begin{equation}
	a_{-1} =\sign (X_{-1})=\left(s\circ\F^{-1}\right)\left (X_0,U_0\right).
	\label{idea}\end{equation}
Then,  one can obtain the necessary $(N+2)-$bit sequence
$\left(a_{-1}, a_0, a_1, \cdots,a_{N}\right)$ and then examine if the corresponding level $N+1$ height function
$\zeta^{N+1}(X_{-1},U_{-1})$ gives a proper tiling.
It is actually a tiling on $D_{-1}$, but, from time-shift symmetry, it is equally a tiling on $D_{0}$.
Let us calculate it explicitly;
\begin{eqnarray}
	\zeta^{N+1} \left( X_{-1},U_{-1} \right)   & \equiv
	\sum_{j=0}^{N+1}\frac{s\circ\F^j\left( X_{-1},U_{-1} \right)}{2^{j+1}}\nonumber \\
	& = \frac{1}{2} s\left(X_{-1},U_{-1}\right)+
	    \sum_{j=0}^{N} \frac{s\circ\F^j \left(\F (X_{-1},U_{-1}) \right)}{2^{j+2}}  \nonumber \\
    & = \frac{1}{2}\sign\left(X_{-1}\right) +\frac{1}{2}\zeta^N ( X_0,U_0),
    \label{eq:master-formula}
\end{eqnarray}
where, at the second line, $j \ra j-1$ is applied for the sum variable,
and the third line is obtained by \eref{eq:zeta-explicit}.
Now, to prove  (\ref{eq:master-formula}) keeps the properness of tiling, we have clarify how
$\Finv$ acts on the ribbons on $D_0$.

The level $N$ tiling is made of altogether $2^{N+1}$ ribbons.
Let us label them by $k$ from left to right so that the left and right-half of $D_0$
are tiled respectively as\footnote{~By time reversal symmetry (\ref{appendix:c}),
the tiling of $D^{+}_{0}$ and $D^{-}_{0}$ are symmetric each other under
$(X, U)\ra (-X,-U)$.}
\begin{align}
	\begin{split}
		D^{-}_{0}  &  = \sum_{ k \in \Lambda_{-}} T_k,  ~~\Lambda_{-}\equiv \{ 1 ,\cdots 2^{N} \},\\
		D^{+}_{0}  &  = \sum_{ k \in \Lambda_{+}}T_k,  ~~\Lambda_{+}\equiv \{ 2^N+1,\cdots 2^{N+1}\}.
	\end{split}
\label{eq:D0tiling}
\end{align}
The height $\zeta^{N}$ has a common value for any $(X_0,U_0)$ within a ribbon.
Hence, it can be regarded as a class function considering a ribbon as an equivalent class
of initial points with the same step-height: we write the height of a ribbon $T_k$ as $\zeta^{N}(T_k)$.
The properness of level $N$ tiling is $
\zeta^{N}(T_1)< \zeta^{N}(T_2) < \cdots <\zeta^{N}(T_{2^{N+1}})$ and
allowing the extended use of a function on a set  $\left(f(\{ x_1,\cdots,x_N\})=\{f(x_1),\cdots,f(x_N)\}\right)$
to a class function,
\begin{align}
\begin{split}
\zeta^N [D^{-}_0] &\equiv \{\zeta^N(T_k)\}_{k \in \Lambda_{-}} \subset (-1,0), \\
\zeta^N [D^{+}_0] &\equiv \{\zeta^N(T_k)\}_{k \in \Lambda_{+}} \subset (0,+1).
\end{split}
\label{eq:D0range}
\end{align}
\input proof.tex
\noindent
In \fref{fig:proof}, the scheme of ${\cal F}^{-1}$ is laid over the level $N$ tiling of $D_0$.
To account for the backward time shift ($D_{1,0} \ra D_{0,-1}$) from \fref{fig:one_time_map_scheme3d},
critical objects are respectively renamed as$v^{-1}_{\pm+}$ and $C^0_{\pm+}$.
But, labels for four regions are kept so that $(+\pm)' \equiv \F[(+\pm)]$
(and $(+\pm) \equiv \Finv[(+\pm)']$).

\subsection{DSS creation mechanism}
\label{subsec:DSS creation mechanism}
We consider ribbons $T^k, k\in \Lambda_{+}$ in $D^+_0$.
\begin{enumerate}
	\item    Each ribbon $T_k$ is divided by the separator curve $C^0_{\pm +}$ into sub-parts
	$a_k \subset  (++)$ and $b_k \subset (-+)$,
	and by the separation property of $\Finv$,
	$\Finv (a_k) \subset  (++) \subset D^{+}_{-1}$ and
	$\Finv (b_k) \subset  (-+)  \subset D^{-}_{-1}$.
	Because $C^0_{\pm+} \ra v^{-1}_{\pm+}$, all $\Finv (a_k)$ ($\Finv (b_k)$) have top point
	at $v^{-1}_{++}$ (bottom point $v^{-1}_{-+}$) and,
	as a whole,  tiles $(++)$ [$(-+)$].  Every ribbon is elongated
	to full height (from $U_0=-B$ to $+B$), and the number of ribbons is in this way  doubled.
	\item    Because ${\cal F}^{-1}$ is orientation-preserving, the order of ribbons are maintained
    region by region. Therefore, the properness of tiling  of $(\pm+)'$ inherits to $(\pm+)$.
	\label{item:orientation-preserving-tiling}
    \item    Because ${\cal F}^{-1}$ is area-preserving, it holds (denoting the area of $T_k$ as $S(T_k)$))
	\begin{align}
	    \begin{split}  S(T_{k})  & \equiv S(a_{k}) + S(b_{k}) =S({\cal F}^{-1}(a_{k}) ) + S({\cal F}^{-1}(b_{k}) )\\
	                             & \implies ~S(T_{k}) \ge S({\cal F}^{-1}(a_{k}) ), S({\cal F}^{-1}(b_{k}) ).
	\end{split}
	\label{eq:area-diminishing}
	\end{align}
	But the height of ribbons ${\cal F}^{-1}(a_{k})$, ${\cal F}^{-1}(b_{k})$, and $T_k$
	are all equal,
	\eqref{eq:area-diminishing} implies that the created ribbons are in general finer than their parent.
	This is case (A) referred in the introduction.
	However, there is a remarkable exception;
	\begin{align}
	    S({\cal F}^{-1}  (a_{k}) )=0 ~\implies S(T_{k}) = S({\cal F}^{-1}(b_{k}) ).
	\label{eq:area-non-diminishing}
	\end{align}
	This is the case (B), leading to the Broucke-type stable PO in a non-shrinking ribbon.
	This point is further examined in \sref{subsec:4-2:advent-of-stable-periodic-orbits} below.
\end{enumerate}

\subsection{Step-Height Distribution and End of the Proof}
\label{subsubsec:3-3-3:step-heights_distribution}
Applying  $\zeta^{N+1}$ in \eref{eq:master-formula} on the region $(a_0,a_1)\subset D_{-1}$
as an extended class function (acting every ribbon inside the region $(a_0,a_1)$), we obtain
\begin{eqnarray}
	\zeta^{N+1}[(a_0,a_1)] =\frac{1}{2} s[(a_0,a_1)] +\frac{1}{2} \zeta^{N}[\F[(a_0,a_1)]]
	\label{eq:master2}
\end{eqnarray}
For all ribbons inside $(a_0,a_1)$, the sign of $X_{-1}$ is $a_0$, so the first term is simply $ a_0/2 $.
(The term $a_{-1}$  in \eref{idea} implies $\sign(X_{-1})$).
In the second term,
\begin{eqnarray*}
\F[(a_0,a_1)] &  \subset
\begin{cases}
	D^+_0 &\text{if} ~a_1=+1\\
	D^-_0  &\text{if} ~a_1=-1,
\end{cases}
\end{eqnarray*}
and via \eref{eq:D0range},
\begin{eqnarray*}
\zeta^N(\F[(a_0,a_1)]) & \subset
\begin{cases}	(0,1)   & \text{if} ~ a_1=+1\\	
				(-1,0)  & \text{if} ~ a_1=-1,
\end{cases}
\end{eqnarray*}
or,in one line,
\begin{eqnarray}
	\zeta^N(\F[(a_0,a_1)]) &  \subset	\frac{a_1}{2}  + \left( -\frac{1}{2},\frac{1}{2} \right).
\end{eqnarray}
Therefore, \eref{eq:master2} gives
\begin{eqnarray} \zeta^{N+1}[(a_0,a_1)] \subset \frac{a_0}{2}  +
\frac{a_1}{2^2} + \left( -\frac{1}{4},\frac{1}{4} \right).
\label{eq:master3}
\end{eqnarray}
The distribution of height is
\begin{equation}
\begin{array}{ccccc}
\zeta^{N+1}\subset&
        \left(-1,-\frac{1}{2}\right)&\left(-\frac{1}{2},0\right)&\left(0,\frac{1}{2}\right)& \left(\frac{1}{2},1\right) \\
        &  \uparrow         & \uparrow              &\uparrow             &     \uparrow   \\
        D_{0}~\text{region} &   (--)                &   (-+)      &(+-)   &       (++)         \\
        & L\cdots&\cdots&\cdots        &\cdots R
\end{array}
\label{eq:global}
\end{equation}
The ribbon tiling inside each region is proper due to item (ii) of the creation mechanism,
and now \eref{eq:global} shows the range distribution over regions is also proper.
Therefore, the level $N+1$ tiling is proper. $\blacksquare$\\
A remark: We see clearly that, dropping the last term $(-1/4,1/4)$, it gives the basic height distribution of
$N=1$ tiling in \eref{eq:N=1_of_induction}.

\section{Stable and Unstable Periodic Orbits in AKP}
\label{sec:4:stable-and-unstable-po}
\subsection{Use of Ribbons to Locate the
Initial Point of a Periodic Orbit}
\label{subsec:4-1:location-of-po}

In \sref{subsec:2-2:symbolic-coding-and-devil's-staircase-surface} we have
shown that, for a given periodic binary code in \eref{eq:po-sequence},
the initial point $(X^*_0,U^*_0)$ of corresponding PO should be enclosed in
the level $N$ ribbon, whose step has the height of the first $N+1$ bits of the code
and this holds at any $N$. The same is true for the past case.
And future and past tiling are in general transverse each other (see \ref{appendix:c}).
Therefore,  the initial point $(X^*_0,U^*_0)$ should be inside the junction of
responsible future and passed ribbons.
Furthermore, in \sref{sec:3-DSS-creation}, we have proved that the
tiling is properly ordered by the height. Therefore, it is possible to locate the appropriate ribbon
by its height and we can locate PO-initial-point $(X^*_0,U^*_0)$ inside a
junction of corresponding future and past ribbons with no mistake.

But now, one must carefully examine
whether at large $N$, all ribbons shrink to vanishing width, case (A),
or some ribbon escapes from shrinking. case (B).  See (iii) in \sref{subsec:DSS creation mechanism} above.

Case (A) appeals naive intuition---some part of a ribbon is definitely chopped away
by the separator, it would then loose the area.  As the elongation keeps its height as before,
it would become thinner.  At large $N$, the junction enclosing the $(X^*_0,U^*_0)$ would
converge to a point and the initial point would be singled out\footnote{~
And even the uniqueness of the PO with a given code would be proved.
This argument is indicated as a possibility in \cite{yellowbook}.}.
\Fref{fig:crossing-of-future-past-ribbons} in fact corresponds to this case.
\input crossing-of-future-past-ribbons.tex
However, even though case (B) looks pathological, it really occurs
and it is the origin of Broucke's stable orbits in AKP.
Here, each ribbon is indeed chopped by the separator
and resultant two parts are elongated back to the full length.
But, it can occur that one part can have full area and the other none after the chopping.
This can occur when the latter has already shrunk to a line segment!
The separator can then only cut out measure zero area, and the survivor remains with finite area.
This is so subtle, but the consequence is crucial. It gives a room for a stable PO survives in AKP.
We are amazed that such an exception leads to physically basic phenomenon.

\subsection{Advent of Stable Periodic Orbits}
\label{subsec:4-2:advent-of-stable-periodic-orbits}

\subsubsection{Stable periodic orbits and non-shrinking ribbon}
\label{stable_periodic_orbits_and_non_shrinking_ribbon}
Now, let us show how the non-shrinking ribbon
occurs taking the case of  Broucke's PO as a good example.
Broucke reported two {\it stable} POs, and here we take the shorter one;
it is rank $n=3$ with the code $(+--+--)$.
We call it `Broucke's stable PO3-6.\footnote{\ %
The {\it unstable} periodic orbit with the same code had been given an identification number `-6'
among rank 3 distinct POs by Gutzwiller in his PO classification scheme \cite{gutzwiller81}.}
We show in \fref{fig:broucke_evolution_PO3-6} how the
future ribbons evolve with the increase of $N$.
\input broucke_evolution_PO3-6.tex
The top row shows tiling $T_{6}, T_{12}, T_{42}$. In each,
three particular ribbons are shown,
namely the ribbon
which encloses the initial point $(X_0^*,U_0^*)$
of the Broucke's PO (we call it Broucke's ribbon and give a mark `B'),
and two neighboring ones.
Remarkably only the Broucke's ribbon survives, while
other diminishes rapidly with width $\sim 1/2^N$.
In the first column $T_{6}\ra T_{7}\ra T_{8}$
($N(\mathrm{mod}~6)=0,1,2$),
we clearly observe the process that
(i) each ribbon is chopped by the separator curve
into two parts, (ii) one is  sent to the left and the other to the right
half of the rectangle, and
(iii) each is stretched between upper and lower boundary.
As a result, duplication of full-height ribbons occurs,
each finer than its parent roughly by half.
Now, in the second column starting $T_{12}$,
the separator curve is chopping out only very fine ribbon near the bottom!
Further in the third column, the separator is now inactive---
chopping out only a line segment. {\it The non-shrinking ribbon is
protecting itself from shrinkage by changing  its tail to a line at early stage. }
The same occurs for the past Broucke's ribbon.
Therefore, the overlap of Broucke's future and past ribbons
constitutes finite-size domain around the initial point  $(X_0^*,U_0^*)$,
and any orbit starting from a point inside the neighborhood evolves
producing the same code with the Broucke's PO forever in both the future and the past.
The Broucke's PO in the center is a stable PO in itself.

\subsubsection{The bifurcation process $U\ra S+U'$; threshold behavior}
\label{subsubsec:threshold behavior}
\vspace{5mm}
Now let us follow the decrease of the anisotropy by increasing $\gamma$ and investigate how
the Broucke stable PO3-6 comes out.
As shown in \fref{fig:broucke-transition}, there is a threshold $\gamma_{th}$
where the unstable PO ($U$) changes into the stable one ($S$)
following the advent of non-shrinking ribbons enclosing $S$.
At the same time, a new unstable PO ($U'$) is born. Thus the stable PO emerges in a bifurcation process
$U\ra S+U'$.
\input broucke-transition.tex
\noindent
All of the POs ($U,S,U'$) in the process are symmetric under the $Y$ transformation
\begin{equation*}
  Y: x\ra x, ~y\ra -y.
\end{equation*}

The bifurcation proceeds in the following way.
\vspace{5pt}
\\
\noindent
(1) Below the threshold, both future (F) and past (P) ribbons
have shrunk to curves at the asymptotic $N$ and they mutually
cross each other at a single point on the $U_0=0$ line\footnote{ \ %
Precisely, as in \fref{fig:broucke_evolution_PO3-6},
a PO (and F, P ribbons enveloping it) evolve periodically in $D_0$, but
at $T_N$ ($N \equiv 0 \pmod  {2n}$), the maximum overlap comes at $U_0=0$.
}.
$U_0=0$ implies $p_x=0$, and the orbit perpendicularly crosses the $x$-axis at, say,
$(X_0,U_0)=(X_0^*,0)$. Thus, the orbit is symmetric under $Y$-transformation.
Now, an infinitesimal shift from $(X_0^*,0)$ leads to the slip off
from the crossing point of F and P ribbons;
thus disables the orbit to repeat the code of PO. This is the way the PO ($U$) is unstable
in terms of ribbons below the threshold.\vspace{10pt} \\ \noindent
(2) Right on the threshold, the F and P ribbons
become tangent each other at $U_0=0$. \vspace{10pt} \\ \noindent
(3) Above the threshold the ribbons stop shrinking and they start extending an overlap around $U_0=0$. See \fref{fig:broucke_location}.
\input broucke-location.tex
\noindent
Inside the overlap, the orbit can repeat the
sequence in both future and past. The previously unstable orbit ($U$) becomes stable ($S$)
remaining at $U_0=0$. Therefore $S$ is also $Y$-symmetric.
The initial point of the new-born PO ($U'$) locates at the corner of the overlap
so that a slight shift again (as was the case of $U$) leads to the slip off
from the overlap. It realizes instability
still keeping periodicity in this way. It is remarkable that it is also $Y$-symmetric
even though $U_0$ is not vanishing now.
This is no contradiction since $U_0=0$ is a sufficient condition for the PO to be
$Y$-symmetric but it is not a necessary condition.
Indeed, we show below that $U'$ belongs in a different symmetry class (self-non-retracing) from that of $U$ and $S$ (self-retracing).
We add that there is no other PO of the same code within the overlap.
For detail, see \ref{appendix:e}.

\subsubsection{Lyapunov exponents}
\label{subsubsec:Lyapunov_exponents and periods}
The change of the maximum Lyapunov exponent
of Broucke PO is shown in the bifurcation diagram in
\fref{fig:broucke-bifurcation-diagram}.
\input broucke-bifurcation.tex
We can now follow the transition process through $\gamma=0.1$---$0.8$.
We observe the followings.
\begin{itemize}
\item[(1)]
Above $\gamma_\mathrm{th}$,
both stable and unstable orbit of the same code co-exist.
As the associated orbit profiles clearly shows,
the PO in the stable branch is self-retracing (the same with the unstable PO below $\gamma_\mathrm{th}$),
while that in the unstable branches self-non-retracing respectively.
That is, the bifurcation proceed in the process
\begin{equation}
   U(R) \ra S(R)+U'(NR),
\label{eq:transition-scheme}
\end{equation}
where, R and NR stands for self-retracing and self-non-retracing\footnote{ \
We write R for self-retracing (rather than SR) to avoid confusion
with `S' for a stable PO.}.
See \fref{fig:broucke_location} for the location
of their initial values in the rectangle.
\item[(2)]
Below and above threshold, $\lambda_{max} \propto |\gamma-\gamma_\mathrm{th}|^{1/2}$
gives a good description.
\item[(3)]
As discussed in \ref{subsubsec:threshold behavior},
the initial point $(X_0^*,U_0^*)$ of the unstable PO ($U'$)
above threshold locates at the edge of the overlap of $F$ and $P$ ribbons
(\fref{fig:broucke_location}).
$U^*_0$ also exhibits the typical threshold behavior
$U^*_0 \propto (\gamma -\gamma_\mathrm{th})^{1/2}$.
\item[(4)]
It is interesting to note that the period of the periodic orbits are insensitive to the transition.
The period of $U$ as a function of $\gamma$ below $\gamma_{th}$ smoothly continues to
that of $S$ above $\gamma_{th}$. This may be natural since both POs are self-retracing,
but the period of the self-non-retracing one ($U'$) is also degenerate in very good approximation.
\end{itemize}

\subsection{Orbit Symmetry Consideration}
\label{subsec:symmetry-consideration}
\vspace{3mm}
\subsubsection{Three classes of $Y$-symmetric orbits}
\vspace{3mm}
Since all involved POs in the bifurcation process
$U(R) \ra S(R)+U'(NR)$ are $Y$-symmetric, let us now
focus on the $Y$-symmetric POs. By simple topology and symmetry consideration,
we can obtain an overview on this bifurcation process.
Firstly, we prepare two keys.
(1) A retracing ($R$) PO should be distinguished from a non-retracing  ($NR$) PO.
A $NR$-PO is simply a closed curve and homotopic to $S^1$,
but, in $R$-PO, the particle is going back and forth
on the same curve connecting two turning points.
To account for this specific feature, let us say $R$-PO is homotopic to {\it squashed} $S^1$.
 See \fref{fig:S1-vs-squashed-S1}.
\input S1-vs-squashed-S1.tex
\noindent
(2) $n_\perp$, the number of perpendicular crossing of a $Y$-symmetric PO
with the heavy $x$-axis, must be either 2 or 0.
This is because an orbit with odd $n_{\perp}$ cannot be closed while $n_{\perp}=4, 6,\cdots$
can close but in disconnected loops.

With above preparation, we can prove a remarkable fact:
\begin{theorem-non}{}
Any $Y$-symmetric periodic orbit is subject to one of the following three classes;
\begin{eqnarray}
&~(a) ~~  \mathrm{ R~~~ with ~~} n_\perp=2,\nonumber\\
&~(b) ~~  \mathrm{ NR~ with ~~} n_\perp=0, \label{eq:symmetry-classification}\\
&~(c) ~~  \mathrm{ NR~ with ~~} n_\perp=2. \nonumber
\end{eqnarray}
\end{theorem-non}
The key of proof is to consider
how to realize with the $Y$ symmetric orbit an appropriate $n_\perp$ for the topology
(retracing or non-retracing).
See \fref{fig:y-sym-classification}(a-c).
\begin{itemize}
\item[(i)]
For retracing PO to be $Y$-symmetric, at least one perpendicular
crossing of the $x$-axis must be included.
But, being a squashed $S^1$, just a single perpendicular crossing
already saturates $n_\perp=2$. This crossing is multiplicity 2 in itself.
Crossings other than it are X-type junction each with multiplicity 4.
This is class (a).
\item[(ii)]
On the other hand, for a non-retracing PO, it can be $Y$-symmetric even without perpendicular crossing;
either $n_\perp=0$ (b) or $n_\perp=2$ (c). \\
In (b), all the crossings are X-type junction, each with multiplicity 2.\\
In (c), all the crossings are X-type but for two distinct perpendicular crossings each with multiplicity one.
$\blacksquare$
\end{itemize}

\input y-sym-classification.tex
Having proved the classification of $Y$ symmetric POs, let us reconsider Broucke transition
$U(R)\ra S(R) + U'(NR)$, and try to understand
why the pattern of the bifurcation is in this way in the light of the threshold behavior
and the classification theorem in four steps.

\begin{itemize}
\item[(i)] $S$ must be class (a) (meaning self-retracing): ~ Under the decrease of anisotropy, the orbit is stabilized and $S$ is born.
As found in \fref{fig:broucke-transition}, it is just born at  the threshold
$\gamma= \gamma_{\mathrm{th}}$, when the future and
past ribbon become tangent each other; hence the location of the initial point $(X^*_0,U^*_0)$ is in the middle of the maximum ribbon overlap at $U_0^*=0$ $(p_{\perp}=0)$.
This means perpendicular emission from the heavy $x$ axis.  Hence, $S$ must be
$Y$ symmetric.  Besides, the classification requires that either $n_\perp=2$
or $0$. Since at least there is one perpendicular crossing exists,
$n_\perp$ must be 2 and the class of $S$ is either (a) or (c).
In (a), the crossing is self-retracing and multiplicity 2, while in (c), two separated perpendicular crossings, each multiplicity one, are necessary.
But, as we see in \fref{fig:broucke-transition}, at the threshold, there is no room for the separation and (c) is excluded. Therefore, $S$ should be in class (a), which means $S$ is self-retracing.
\item[(ii)] $Y$-symmetry:~The transition $U \ra S + U'$ is induced by the configuration change of respective F and P ribbons at the threshold, and, as ribbons carry the same code before and after the transition, the PO transition must be code-preserving.  As shown by Gutzwiller \cite{gutzwiller81},
the code of PO dictates the symmetry of PO, so that $U$ and $U'$ must be
also $Y$-symmetric, once $S$ is understood to be $Y$ symmetric.
Therefore, we can apply the classification theorem of $Y$ symmetric PO
to all of them.

\item[(iii)] $U$ is also class (a) (self-retracing):~
$Y$-symmetry of $U$ is argued above [(2)], but also directly seen
in \fref{fig:broucke-transition}. It locates at the crossing of the F and P ribbons at $U_0^*=0$, and, just the same reason as [(1)], $U$ is $Y$-symmetric and in class (a). As discussed at \fref{fig:broucke-transition}, the $U(R)$/$S(R)$ transition
here just corresponds to,  in term of ribbons,  the configuration change
single-point/finite-width overlap.

\item[(iv)] $U'$ is class (b) and non-self-retracing:~
As $U$ is now understood to be in class (a), we consider
in \fref{fig:topology-change-yoko}
all possible routes, via which the initial $U(R)$ in class (a)
may change its feature through the transition.
\input topology-change-yoko.tex

Route [I] is $(a)\ra (a)$, but changes from unstable to stable.
This is just $U(R)\ra S(R)$, and just corresponds to the above
single-point/finite-width overlap. The initial value remains $U_0^*=0$
and only with small change in $X_0^*$ occurs.
Now, let us discuss the route $S(R) \ra U'$.
($U'$ must be $Y$ symmetric, as the transition is code-preserving).
It can be in principle either via [II] or [III].
In [II], the self-retracing property is broken, and the $n_{perp}=2$
crossing becomes $X$-crossing. Then, non-vanishing $U_0$ is
generated quickly after $\gamma_{th}$.  This is just in accord with the threshold behavior of $U'$ ($U^*_0 \propto (\gamma -\gamma_\mathrm{th})^{1/2}$);
unstable with initial value at the edge of the overlap of F and P ribbons.
(On the other-hand, if the route were via [III], retracing-property is broken
in such a way that rapid $\Delta X_0$ is created without $\Delta U_0$. This is totally against the threshold behavior of ribbons).
Thus, the route must be via [II] and $U'$ must be $U'(NR)$
\end{itemize}
The above is a {\it  a post-diction} on the Broucke' transition
and summarized in \fref{fig:y-sym-classification}(d). It is just the
transition as it is.
Two remarks are in order;
\begin{itemize}
  \item[(1)]
            For a PO in class (a), the total number of crossings of the $x$-axis (i.e. the length)
            is $4 n +2$, $4n$ comes from $n$ cross-junction with multiplicity 4
            and 2 from a single perpendicular crossing with multiplicity 2.
            The rank of a PO is half of its length;
            thus, the rank of class (a) PO must be odd ($2n+1$).
            For this reason,  we conjecture that
            {\it the Broucke-type transition, associated with
            the non-shrinking ribbon, will occur in $Y$ symmetric odd rank PO. }\footnote{%
                   \ Precisely, the tangency of the F and P ribbons at $U_0=0$ implies
                    $Y$-symmetry of involved POs, and $n_\perp=2$, but the possibility
                    of the pre-PO ($U$ in the Broucke's case) being in class (c) is not logically excluded.
                    Our preliminary result is that from $n=3$ up to $n=23$ and $\gamma<8/9$,
                    there is only one advent of non-shrinking ribbons at every odd rank starting from (a).
                    We are trying to consolidate this issue.
            }
  \item[(2)]
            We have observed route III transition for a PO in rank 15, where $\gamma >9/8$.
            See \fref{fig:PO15-1-location} below.
\end{itemize}

\section{Application}
\label{sec:5-applications}
\subsection{The two-dimensional AKP PO search}
\label{subsec:5-1:algorithm}

In order study the POs in this system, especially to challenge the uniqueness issue,
we first of all have to search out the POs {\it exhaustively}.
By exhaustively we mean avoiding any limitation on the PO, irrespective of its symmetry and whether it is
unstable or stable. The recent search by Contopoulos et al. \cite{contopoulos05}
was of this type but it is limited to perpendicular emission from the heavy axis ($p_{0,y}=0$). Then,
choosing the initial position $x_0$ on the heavy axis ($y_0=0$)
also fixes the momentum $p_{0,x}$ via energy conservation; hence
a one-parameter shooting varying $x_0$ only is sufficient. Their analysis yielded important information
especially on the existence of stable PO in AKP including relatively high rank PO. However, the limitation is rather severe;
it limits the PO only to the $Y$-symmetric ones. Furthermore, it cannot detect the class (b), $n_\perp=0$, PO ($NR$)
which is created as $U'(NR)$ after the bifurcation $U(R) \ra S(R)+U'(NR)$ . To be exhaustive,
we organize our search as follows.

\begin{itemize}
\item[(1)]
The basic flow of the analysis is based on specifying the code of a PO at a certain anisotropy.
For instance, if the code $(+--+--)$ is specified, the routine searches out Broucke' PO-36;
$U(R)$ only if $\gamma<\gamma_\mathrm{th}$, but both $S(R)$ and $U'(NR)$ if $\gamma>\gamma_\mathrm{th}$.
One can be sure that there is no more PO of this code within the setup resolution.
\Fref{fig:broucke-transition} and \ref{fig:broucke_location}(a-c) are the outcome of the code request $(+--+--)$. \\
\item[(2)]
Given the requested code of PO, we now take advantage of the PO trapping
mechanism. That is, the properness of the tiling of $D_0$ by ribbons
guarantees that
$(X^*_0,U^*_0)$ should be inside the base ribbon of the step whose
height $\zeta^F_N(a)$ is calculated by (\ref{eq:zeta-by-the-N+1-bits}).
We use $N=48$ for the maximum precision in the double precision calculation.

\item[(3)]
The above {\it asymptotic} ribbon with $N=48$ may be,
if it subjects to case (A), extremely narrow ($\Delta X \sim O(1/2^{48})$)
and the neighboring steps may be almost degenerate in heights.
On the other hand, if it is in case (B), it may retain some finite width.
Here we proceed protectively; rather than directly tackling the asymptotic ribbon,
we focus on level $2n$ ($\ll 48$) ribbon, which should embody the asymptotic one
by the properness of the tiling\footnote{%
~This corresponds to the relaxed ribbon
with boundaries ${\cal B}_L$ and ${\cal B}_R$
in \fref{fig:broucke-chi2-contours}.}.
The interval of the ribbon,  $(X_0^\mathrm{min},X_0^\mathrm{max})$
at the mesh points of $U_0 \in (-B,B)$ can be calculated
by a bi-section method to satisfy
\begin{eqnarray}
\zeta^F_{2n}(X_0^\mathrm{min},U_0)\leq &\zeta^F_{N}(a)& \leq \zeta^F_{2n} (X_0^\mathrm{max},U_0)
\label{eq:ribbon_width}
\end{eqnarray}

Now the search area is reduced from vast $D_0$ to
a level $2n$ ribbon extending from $U=-B$ to $U=B$.

\item[(3a)]
At this step, we check whether the target ribbon shrinks or not by
inspection of its full profile $U_0 \in (-B,B)$.

\item[(4)]  Now the shooting for a PO inside the level $2n$ ribbon---
find a point that makes the misfit
\begin{equation}
\chi^2 (X_0,U_0) \equiv (x_\mathrm{F}-x_0)^2+y_\mathrm{F}^2+
(p_{x,\mathrm{F}}-p_{x,0})^2+(p_{y,\mathrm{F}}-p_{y,0})^2
\label{eq:chi2misfit}
\end{equation}
between the initial and final $2n$-th crossing of the $x$-axis
vanishing\footnote{
    ~$y_\mathrm{F}\equiv 0$ by definition. $(X_0,U_0)$ in the left-hand side fixes the initial point
    in the Cartesian coordinates, and the integration of orbit
    for the calculation of right-hand side
    is done in $x,y,p_x,p_y$, with slow down around the origin when close-encounter
    with the origin is involved.
}.
~~Practically we have firstly searched for $X_0^*$, which minimizes $\chi^2$, at every $U_0$.
This gives a function $X_0^*(U_0)$ over the interval $(-B,B)$, and
through it, we can regard $\chi^2$ as a function of $U_0$. That is,
\begin{eqnarray*}
 \chi^2(U_0)&\equiv& \chi^2\left(X_0^*(U_0), U_0\right).
\end{eqnarray*}
Thus, the two-parameter search is effectively reduced to one-parameter one.

\item[(4a)]
In case (A), $\chi^2(U_0)$ should be a convex function of $U_0$ with a single bottom at $U_0^*$
with vanishing $\chi^2$ within numerical error.
Then $(X_0^*,U_0^*)$ is the wanted for initial position\footnote{%
~In this case, bottom-search by a tri-section method on
$\chi^2(U_0)$ is vital. See our earlier report \cite{arob14}.
}.
In case (B), $\chi^2(U_0)$ turns out multi-bottomed vanishing at each bottom.
To deal with both possibilities, graphical analysis of the full profile of $\chi^2(U_0)$ is
performed.
\item[(5)]
We combine the information from the past ribbon on top of the above procedure.
\end{itemize}

With step (3a) and (4a), the search is organized not to miss the violation of uniqueness.
Below, we report two applications;a detailed case study of a high-rank PO
($\gamma=0.85-0.93$) and an exhaustive PO search
regarding the uniqueness issue at high anisotropy ($\gamma=0.2$).


\subsection{The case study of bifurcations of a high-rank PO ($n=15$).}
\label{subsec:5-2:po15}
The investigation of this PO is motivated by the longest PO (length 30)
among the stable POs reported by Contopoulos et al.\cite{contopoulos05}.
However, the code of their PO is not given and there is a possibility of
miss-identification, considering the exponentially large number of POs
at such high rank. Thus,
the description below may be better read in its own right irrespective to the motivation \footnote{%
    ~ We read $x_0$ by eye from their Figs.4 and 6,
    and confirmed reproduced orbit closes, below bifurcation threshold,
    at the stated 30-th crossing after slight adjustment.
    But,  passing the threshold,  it gradually fails to close. Their analysis is limited to $p_{x,0}=0$,
    and there is a possibility of error, that observed PO is the same with
    ours ($U(R) \ra U'(NR)+S(R)$), where $U'(NR)$ is $p_{x,0} \neq 0$.
    We here quote our initial values $(X_0,U_0)$ with sufficient digits for reproduction.
    At $\gamma=0.87$,  there is only $U$:(0.11465, 0), and
    at $\gamma=0.88$, there are both $S$:(0.099342, 0) and $U'$: (0.10021, 0.050793).
}.
The code of our PO15 is
\begin{eqnarray}
    (+-+-+-+--+-+-+-)^2.
\label{eq:PO15}
\end{eqnarray}
We find it bifurcates twice; $U \ra S+U'$ and then
$S \ra S'+S''$ as seen in \fref{fig:PO15-1_bifurcation}.
The bifurcation pattern is summarized in \fref{fig:PO15-bifurcation-scheme}  using the PO class
(\fref{fig:y-sym-classification}) and bifurcation route (\fref{fig:topology-change-yoko}).
As the orbit at this high rank is very complicated, let us examine
the bifurcation as a challenge to the general theory consideration in
\sref{sec:4:stable-and-unstable-po}.
%
\input PO15-1_bifurcation.tex
\vspace{-0.5cm}
\input PO15-1-bifurcation-scheme.tex
\noindent
(1) The first bifurcation $U \ra S+U'$: ~
All $U$, $S$, $U'$ are $Y$-symmetric and our classification can be applied.
\Fref{fig:PO15-1_bifurcation} shows that
the initial value of $U$, $S$ is $U_0^*$,
while the $U_0^*$ of $U'$ is rapidly created after the bifurcation.
This leads us to infer $U \ra S$ is via route [I]
($(a)\ra (a)$)
and $U \ra U'$ is via route [II] ($(a)\ra (b)$).
This means the bifurcation is $U(R) \ra S(R)+U'(NR)$;
just the same with Broucke's PO3-6 and PO5-40.
The ribbon structure in \fref{fig:PO15-1-location} consistently shows
that the bifurcation is exactly caused along with the advent of non-shrinking ribbon.
$(X_0^*,U_0^*)$ of $U'$ locates at the edge of the overlap, while that of $S$
is at the center of the overlap with $U_0^*=0$.
\input PO15-1-location.tex
Now, the orbit profiles are given in \fref{fig:PO15-1-a}.
As orbits are length 30, they look at a glance as a cloud
but from the density, apparently $S$ is self-retracing and
$U'$ is non-retracing. One can also pin-point the tuning-point
at the kinematically boundary in the latter. The close inspection
is more intriguing;  $S(R)$ has a single $n_2$ point,
while $U'(NR)$ has two perpendicular crossings, each with multiplicity one.
This is just the multiplicity prediction from the classification (\ref{eq:symmetry-classification})
and \fref{fig:topology-change-yoko}.
\input PO15-1_a.tex

\noindent
The second bifurcation $S \ra S'+S''$:~
This is a new case, which occurs in the very low anisotropy regime ($\gamma_2 > 8/9$),
where the flow (\ref{eq:full-polar-eom}) has no longer hyperbolic singularities and
the existence of unstable PO at an arbitrary code is no longer guaranteed \cite{devaney78}.
The rapid initial value variation is in the $X$ direction.
Thus, we infer that $S''$ is produced via root [III], which means it is in class (c)
and the bifurcation should be $S(R) \ra S'(R)+S''(NR)$.
Indeed, the initial positions of $S'$ and $S''$ in \fref{fig:PO15-1-location} are
horizontally aligned. Horizontal because the variation is in the $X_0$ direction,
and two initial points for $S''$ because $S''$ is in class (c) ($n_\perp=2$).
(See the two multiplicity-one crossings in [III] in \fref{fig:topology-change-yoko}).
\noindent
\input PO15-1_b.tex
This can be directly verified in the orbit profile in \fref{fig:PO15-1_b}(c).
Let us note a further success of the theory.  It tells $U'$ must be $NR$ but it can be
hardly seen from the orbit since the orbit is almost doubled and looks as though
self-retracing. However, the twice magnification in \fref{fig:PO15-1_b}(a) verifies
it is indeed $NR$.

Summing up, the PO15 first bifurcation follows precisely the theory classification, and the theory
also explains the second bifurcation which embodies class (c) PO.


\subsection{Search of all distinct POs (rank $n \le 10$)
at high anisotropy $\gamma=0.2$}
\label{subsec:5-3:exhaustive}

Here we report our exhaustive search for the POs up to rank
$n=10$ (length $2n=20$) at the high anisotropy ($\gamma=0.2$).
As noted in the introduction,
it was previously considered that at such high anisotropy
all POs are unstable and isolated, mainly based on the early
numerical analysis \cite{yellowbook,gutzwiller81}.
On the other hand, recently some possibility has been expressed
in \cite{contopoulos05} that the existence of ample stable orbits
in AKP may indicate stable orbits even at high anisotropy.
Therefore, we here revisit high anisotropy AKP armed by our two-parameter shooting algorithm
which embodies steps (3a) and (4a) above 
so that it is sensitive to the possible violation of uniqueness.

\subsubsection{Distinct periodic orbits and distinct binary code}
To challenge this problem, firstly let us consider the counting of {\it distinct} POs.
The Hamiltonian (\ref{eq:2d-hamiltonian}) is symmetric under discrete symmetry transformation
$X\!:\!x(t)\ra -x(t)$,
$Y\!:\!y(t)\ra -y(t)$ and
${\cal T}\!:\!t \ra -t$.
Thus, any partner orbit, generated from one PO by the symmetry transformations
is again a solution of the equation of motion and a respectable PO.
But, they have the same shape, period and stability exponent
and it is legitimate to put them into an equivalent class. Two POs
are distinct only when they belong to different classes.
If a PO is in itself has none of the symmetries, $2^3$ orbits belongs to the same class
and the class has degeneracy $\sigma_{\mathrm{sym}}=2^3$. On the other hand, if the PO is
self-symmetric under some transformation, then it does not generate a different partner,
and the degeneracy is halved for each self-symmetry\footnote{~
If a PO is non-self-retracing, $\cal{T}$ produces degenerate pairs
orbiting in opposite-direction each other, but if self-retracing, $\cal{T}$ is immaterial---after
it amounts to simply the freedom of the choice of one among two turning points
as the starting point.}.
There are ten symmetry types of PO, five for a non-self-retracing (NR) and
five for a self-retracing (R) PO. The degeneracy factor for each symmetry-type is
listed at the fourth row of \tref{tab:symmetry-counting}.
\input symmetry-counting.tex
\noindent
The $O$ type among $NR$ is newly found at rank 7 and 9, see discussion below.
The number of distinct POs in a given symmetry class listed in the following rows
are predicted (under the uniqueness assumption) by counting {\it distinct} sequences.
The prescription is given by Gutzwiller in detail and the code table up to rank $n=5$ is
given in Table I of \cite{gutzwiller81}. Our \tref{tab:symmetry-counting}
is its extension to $n=10$. (Explicit sequences take pages and only the number
of sequences is tabulated). Now, let us briefly recapitulate the prescription.
First, a rank $n$ PO is represented by a length $2n$ binary code;
if the orbit traverses the heavy axis upwards $n$ times, then it must traverse
also $n$ times downwards to come back the initial point.
Next, the symmetry of a PO corresponds to the rule of its code.
For instance, if a PO is $X$ symmetric, its code must satisfy a rule $X: a_{2n-i-1}=-a_i$ \cite{gutzwiller81}.
Thus, as all POs are divided into classes by the equivalence under symmetry transformation
in order to scrutinize distinct POs, the $2^{2n}$ binary sequences at rank $n$ should be divided
into classes using symmetry rules. However, to reach distinct sequences,
it is not sufficient to divide by the equivalence under $X$, $Y$, ${\cal T}$ rules,
but one must also divide by the equivalence of sequences under the code-shift operation
${\tau}$, that is, the freedom
of choosing the starting bit among the cyclic binary code. At this final step,
there is a slight subtlety. If the rank $n$ is prime, the code-shift symmetry simply amounts
to $n$-fold degeneracy. (It is $n$ rather than $2n$, because, by definition,
the starting bit must be chosen from crossings with $p_y>0$).
However, it must be noted that the set of rank $n$ POs includes $m$ times
repetition of a lower rank primary orbit of rank $p$, where $p$ a prime divisor $p=n/m$.
In such a case, the code-shift degeneracy reduces to $p$.
Dividing out the code-shift degeneracy taking account this,
one eventually reaches the distinct sequences.
\input symmetry-counting-with-tshift.tex
\Tref{tab:symmetry-counting-with-tshift} shows how to organize this counting.
Uniqueness of a PO for a given code
means one and only one distinct PO for each distinct code.
Under this assumption, the bottom row of \tref{tab:symmetry-counting-with-tshift}
gives the number of distinct PO prediction in \tref{tab:symmetry-counting}.

\subsubsection{The results:uniqueness holds at $\gamma=0.2$ }
Now, we describe how the checker  (3a) and (4a) in the algorithm worked.
\begin{itemize}
  \item[(3a)]
    Always the ribbon of the code shrinks at $\gamma=0.2$ as $\sim 1/2^{2n}$.
    Specifically for $n=10$, the maximum width,
    $\max \left(  X_0^\mathrm{max}(U_0)-X_0^\mathrm{min}(U_0)\right)$
    (see (\ref{eq:ribbon_width})) over the interval $U_0 \in (-B,B)$
    does not exceeds $10^{-7}$ for any one of the tested 13648 codes. (See \tref{tab:symmetry-counting})
\item[(4a)]
    Always the chi-squared (misfit) curve is convex with a single bottom
    with negligible value. It is mostly under $10^{-20}$ (the best value is $\sim 10^{-25}$)
    and the worst is $\sim 10^{-2}$ only for a few POs.
\item[(5)]
    The bottom ${(X_0}^*,{U_0^*}$) agrees without exception
    with the crossing point of the future and past {\it lines}. (At large $N$
    ribbons are reduced to lines in the scrutiney at $\gamma=0.2$).
\end{itemize}
Subsequent stability test has proved that all the PO are unstable.
Therefore, we conclude that within the above resolution,
there are only unstable POs and that the PO is unique for any given code up to $n=10$
at $\gamma=0.2$.

\subsubsection{A new symmetry type $O$ and other PO samples}

Let us comment on the new symmetry class ($O$ type). This type is not considered
in Gutzwiller \cite{gutzwiller81}. We are firstly embarrassed when we cannot
satisfy by the search result the sum rule that number of the PO
should be $2^{2n}$ when devision by the symmetry equivalence is removed.
It may have indicated the violation of uniqueness even though we are working
at the highly chaotic region $\gamma=0.2$. Eventually we were able to locate
the reason as due to this $O$-type; there is only one at rank $n=7$ and seven
at rank 9.  After counting correctly them, the sum rule is satisfied, and
the uniqueness holds at $\gamma=0.2$. See \tref{tab:symmetry-counting}.
The $O$ type orbit is neither $X-$ nor $Y-$symmetric,
but symmetric under the transformation $O:x \ra -x, ~ y \ra -y$.
The observed ones are all self-non-retracing and ($\sigma_{\mathrm{sym}}=2^2$).
The symmetry class number $5$ is assigned, $10$ is reserved for the possible
occurrence of O-type retracing PO at $n \ge 11$.

\input typical-pos-at-n9-Otype.tex
Further examples are taken from 13648 distinct unstable POs
at rank 10.
\input typical-pos-at-n10.tex
The last one is $Y$-symmetric and $NR$, then our classification tells
it is in class (c). Indeed close investigation reveals it has two perpendicular
crossings, each multiplicity one as the class (c) PO should.

\subsection{A Verification of the Gutzwiller's Action Formula}
\label{subsec:5-4:action}
In Gutzwiller's periodic orbit theory,
the semi-classical description of the quantum density of state is given by
\begin{equation}
  g(E)\equiv\sum_i\frac{1}{E-E_i+i\epsilon}
  \approx
  {g}_{POT}(E)\equiv -\frac{i}{\hbar}
\sum_{\Gamma\in\mathrm{POs}}\frac{T_{\Gamma,0}}{2\sinh\left(
\lambda_\Gamma/2\right)}
e^{iS_\Gamma(E)/\hbar-i\pi\nu_\Gamma/2},
\label{eq:trace-formula}
\end{equation}
where the sum is over all classical periodic orbits.
Each PO is designated by $\Gamma$;
$S_\Gamma$, $\lambda_\Gamma$, $\nu_\Gamma$
are respectively action, Lyapunov exponent,
and number of conjugate points of $\Gamma$,
while $T_{\Gamma,0}$ is the period of the primary cycle of $\Gamma$.
From the scaling property of AKP, the action behaves
as $S_\Gamma=\Phi_\Gamma/\sqrt{-2E}$
with a constant $\Phi_\Gamma$ characteristic to each PO.
In the endeavor to calculate ${g}_{POT}(E)$ in (\ref{eq:trace-formula}),
Gutzwiller first introduced the symbolic coding of the PO
in order to put the sum under control. The next step was
to express the constant $\Phi_\Gamma$ by the binary code of the PO.\footnote{
   \    $T(E)=\frac{\partial S}{\partial E}
        =S/(-2E)$ and for our canonical choice $E=-1/2$, T=S, and
        $T_{\Gamma,0}$ was obtained by dividing it by $n$.
        For Lyapunov exponent, a `crudest approximation'
        $\lambda=n \lambda_0 \sim 1.5 n$ was adopted in \cite{gutzwiller80}.
}
~~For this issue he introduced the following empirical formula for $\Phi(a)$ of
a periodic orbits of rank $n$ (length $2n$) with
the primary code $a=( a_0, a_1, \cdots, a_{2n-1}) $
\begin{eqnarray}
\label{eq:action-appox}
    \Phi(a) = 2 n \tau \cosh \left(\frac{\lambda_A}{2} \right)-
    \frac{1}{2} \tau \sinh \lambda_A
    \sum^{2n}_{i=1} \sum^{+\infty}_{j=-\infty} a_i a_j e^{-\lambda_A |j-i|}.
\end{eqnarray}
This formula has only two parameters, $\tau$ and $\lambda_A$,
 but it was found that, for $n=5$, this formula can provide
 $\Phi$ of any PO in remarkable agreement with the measured value.\footnote{%
    \   For the family  of rank $n$ PO,
        the $\Phi_\mathrm{max}$, $\Phi_\mathrm{min}$ and $\braket{\Phi }$ predicted
        by (\ref{eq:action-appox}) are easily calculated as
        $\Phi_\mathrm{min}=0$, and (at large $n$)
         $\Phi_\mathrm{max}=2n \tau$, $\braket{\Phi }=2n\tau (1+e^{-\lambda_A})$.
 }
~~ We have performed the test of this formula using the
 action of rank $n=10$ PO obtained by our high accuracy
 measurement. In one of the tests, we first determined
 the two parameters by fit to the action data of 44 distinct POs
 in the rank $n=5$ family. ($\gamma=0.2$ [$\mu^2=5$]). The result  is
 \begin{equation}
 \begin{array}{ccc}
    \tau=   &  2.871    & (2.8844)\\
    \lambda_A= & 0.6014  & (0.622
     )
 \end{array}
 \label{eq:parameters}
 \end{equation}
 Quoted figures in the parenthesis are from \cite{gutzwiller80} for
 the Silicon anisotropy ($\mu^2=4.80$). Now, we test
 whether this formula with the parameters fixed at the values
 in (\ref{eq:parameters}) can endure the extrapolation to
 $n=10$ family. In \fref{fig:action_test}
 we show the result as a scatter plot.
\input action_test.tex
 At $n=10$ there are 13648 distinct POs.  Each point
in the plot represents one of them. it can be seen that
the prediction works for the wide range
of the action constant value extending up to 50.
We show in \tref{tab:msd} the MSD defined as
\begin{equation}
    \mathrm{MSD}=\frac{
    \sum_{i \in \mathrm{rank n distinct POs}}
     \left(S_i^{Data}-S_i^{Th.}\right)^2}{\mathrm{number of distinct POs}}.
\end{equation}
\begin{table}[!h]
\caption{\label{tab:msd}
        The MSD. The formula (\ref{eq:action-appox})
        (with parameter $\tau$ and $\lambda$ fixed at level $n=5$) can describe all $n$ up to 10
        almost with equal accuracy at $n=5$.
        }
\begin{indented}
\item[]
\begin{tabular}{ccccccccccc}
\br
rank n & 1&   2    &      3        &4       &\underbar{5}        &6        &7          &   8     &   9      &10\\
\mr
\# PO  & 2&   4    &      8        &18   &44        &122    &362      &1162   &3914   &13648\\
MSD& 0&0.00965&0.0229&0.0379&0.0500&0.0521&0.0499&0.0494&0.0506&0.0536\\
\br
\end{tabular}
\end{indented}
\end{table}

 As clearly seen the marvelous success of the formula (\ref{eq:action-appox})
 continues up to $n=10$.

 Let us briefly discuss the implication of this success.
Following Gutzwiller, let us integrate the density of state up to $E$
\cite{gutzwiller80}.
Then, we obtain
\begin{equation}
    \int^E dE' D_\mathrm{POT}(E')=\sum_n \sum_{a}\frac{1}{2n
    \sinh\left(\lambda_\Gamma/2\right)}
    e^{-s \Phi(a)},
\label{eq:spin-sum}
\end{equation}
where $a$ is the code of the PO, and a dimensionless variable
$s$ is introduced by
\begin{equation}
    s \Phi = -i S/ \hbar.
\end{equation}
By this `Wick rotation', the quantum formula is mapped to statistical formula
--- the grand partition function of a statistical spin system.\footnote{
\ Under an approximation
$2\sinh\left(\lambda_\Gamma/2\right)
\approx \exp{(\lambda_\Gamma/2)}$, the Lyapunov exponent can be
traded to a chemical potential.}
Under this transformation, the action formula (\ref{eq:action-appox})
corresponds to an Ising spin chain, which has only two body
interaction with exponential decay. Now that the validity of
(\ref{eq:action-appox}) has been confirmed even up to $n=10$,
it may be stated that AKP semi-classical quantum theory
at the high anisotropy region is dual to the above spin theory.
\section{Conclusion}
\label{sec:6-conclusion}
In this paper, we have fully used the ability of symbolic coding in AKP
and considered level $N$ devil's staircase surface and the tiling of the initial value
domain by ribbons introduced by the surface.
We have proved the properness of the tiling by ribbons
from the creation mechanism of $N+1$
from $N$ tiling, which clarifies how the non-shrinking ribbon can emerge.

Our key points are as follows;
\begin{itemize}
\item[(i)] Non-shrinking ribbon emerges when future and past asymptotic curve
becomes tangent each other at $U_0=0$ at threshold anisotropy.
\item[(ii)] The bifurcation is $U(R) \ra S(R) + U'(NR)$.
Initial point $(X_0*, U_0*)$ of stable PO $S$  locates in the overlap of
future and past ribbons, while the unstable PO $U'$ locates at the edge of the overlap.
\item[(iii)] From topology and symmetry consideration,
we have explained the above bifurcation scheme, and we give a conjecture
that the stable PO occurs in the $Y$-symmetric rank odd self-retracing orbit.
A case study of high-rank PO15 supports it.
\item[(iv)]
An exhaustive PO search verifies the uniqueness conjecture holds
at high anisotropy $\gamma=0.2$  (if newly found type $O$ orbit is accounted for).
The Gutzwiller's action approximation formula works amazingly for all POs
up to rank 10 at this anisotropy.
\end{itemize}

The topology and symmetry approach developed in this paper
gives a frame to constrain the bifurcation, but, we admit that
a direct analytical understanding of the advent of non-shrinking ribbon
is most wanted for.
Relatedly, it is tempting to look at this bifurcation from quantum side---separating of $S$ and $U'$ from quantum data,
using inverse chaology  as {\it a quantum prism}.
Previously, we could successfully extract low rank unstable PO data
(including Lyapunov exponents) from AKP quantum spectrum \cite{ptep14}.
To tackle with the bifurcation is a serious challenge;
it requires higher order correction in $\hbar$ as well as taking satellite POs
into account \cite{ozorio87, main99,schomerus97}.
Work in this direction is in progress.
\ack
This paper is an outcome of  a course of several years' work in AKP; both KK and KS
wrote their Ph. D. theses on their contribution. Taking this occasion, we would like to thank
people who gave us useful suggestions and encouragement at various stages of the work.
We thank our colleagues, S. Ishihara for useful discussion, T. Nakajima for his enthusiasm
in the search of non-shrinking ribbons at the later stage, E. N. Bjerrum-Bohr,
P. Damgaard, W. Ochs for unfailing encouragement.
We thank people working on dynamical systems,  especially M. Saito, M. Shibayama,
and K. Tanikawa for their interest and various suggestions.
Finally, we would like to thank  M. Gutzwiller for his kind correspondence in 2008 (just once).
We have send him one of our papers at early stage,
and after answering our questions, he wrote that [he is] ``a great admirer of figures $\cdots$'', and
kindly guided us to \cite{gutzwiller81}, encouraging to publish our results.
We regret that we can no longer send this completed paper to him and ask for his criticism.
This work is partially supported by Research Project Grant (B) by Institute of Science and Technology,  Meiji University.

\appendix
\section{The  Boundary One-time Map ${\cal F}_I$}
\label{appendix:a}
In terms of double polar coordinates (\ref{eq:doublepolar})
the equation of motion given by the Hamiltonian (\ref {eq:2d-hamiltonian})
is
\begin{eqnarray}
	\frac{d \chi}{dt}=&-\frac{\left(
	1+e^{2 \chi}
	\right)^{2}}{4e^{\chi}}\left(
	\sqrt{\nu}\cos{\vartheta}\cos{\psi}+\sqrt{\mu}\sin{\vartheta}\sin{\psi}
	\right),\\
	\frac{d \vartheta}{dt}=&-\frac{\left(
	1+e^{2 \chi}
	\right)^{2}}{4e^{\chi}}\left(
	\sqrt{\mu}\cos{\vartheta}\sin{\psi}-\sqrt{\nu}\sin{\vartheta}\cos{\psi}
	\right),\\
	\frac{d \psi}{dt}=&-\frac{1}{2}e^{\chi}\left(
	1+e^{2\chi}\right)
	\left(
	\sqrt{\nu}\cos{\vartheta}\sin{\psi}-\sqrt{\mu}\sin{\vartheta}\cos{\psi}
	\right).
	\label{eq:full-polar-eom}
\end{eqnarray}
We consider the limit $\chi\ra \infty$;
this corresponds to $r\ra 0$, because $r={2}/(1+e^{2\chi})$
under the choice $H=-1/2$.
We also slow down the trajectory by a transformation of the
time variable  $dt'=e^{3\chi} dt$ to remove the singularity.
Then the first two equations become autonomous form (\ref{eq:autonomous-flow})
and the $\chi$-equation decouples (in fact $\chi$ remains
$\infty$ for any finite $t'$), and  (\ref{eq:autonomous-flow}) describes the evolution of
$\vartheta$ and $\psi$ in the limit $\chi\ra \infty$
\cite{gutzwiller77}.

The one-time map, ${\cal F}$ in (\ref{eq:one-time-mapF}),
describes how $D_0$
is mapped onto $D_1$.
Because the limit $\chi\ra \infty$ restricts
$D$ to the collision manifold $I$ (\ref{eq:backbone}),
 (\ref{eq:autonomous-flow}) embodies all necessary information to find
restricted one-time map $\left.{\cal F}\right|_I:I\ra I$.
Only a slight complication is that
$\left.{\cal F}\right|_I$ is a map $(X_0,U_0)\mapsto (X_1,U_1)$,
while (\ref{eq:autonomous-flow}) gives ${\cal M}:
(\vartheta_0 ,\psi_0)\rightarrow(\vartheta_1 ,\psi_1)$.
Thus we have to convert ${\cal M}$ to $\left.{\cal F}\right|_I$
as in (\ref{eq:pull-back}) going back and forth
between the different parametrization of the same point on $I$.
From (\ref{eq:xu-transformation}) and (\ref{eq:r-and-chi})
we obtain for finite $\chi$
\begin{eqnarray*}
	X&=2\cos{\psi}\frac{\cos^{2}\vartheta+e^{-2\chi}}{1+e^{-2\chi}},~~
    U&=\sqrt{\mu} \arctan\left(e^{\chi} \cos \vartheta\right).
\end{eqnarray*}
With $\chi\ra \infty$, we observe
\begin{eqnarray*}
  X & \ra 2 \cos \psi \cos^{2} \vartheta, ~~
  U & \ra \sqrt{\mu} ~\sign \left(\cos \vartheta \right)
	\left(\frac{\pi}{2}-\frac{1}{e^{\chi}\left|\cos \vartheta\right|}\right).
\end{eqnarray*}
Therefore, except for the critical case $\theta=\pi/2$,
we can use the following relation for the necessary conversion;
\begin{eqnarray}
  X & = 2 ~ \sign (\cos\psi) \cos^2\vartheta,  ~~
  U & = \frac{\pi\sqrt{\mu}}{2}\sign(\cos\vartheta),
\label{eq:boundary-relation-appendix}
\end{eqnarray}
where for $X$ we have used the fact that, on the Poincar\'e section $y=0$,
either $\psi=0$ (the case $X>0$) or $\psi=\pi$ (the case $X<0$).
This is (\ref{eq:boundary-relation}) in the text.

To extract ${\cal F}_I$ from ${\cal M}$, it is best to follow \cite{gutzwiller77}.
Fix $\psi_0=0$ (that is, choose $X_0>0$),
and let $\vartheta_0$ increase from $0$ to $\pi$. (This is sufficient
thanks to the symmetry of the system. )
This makes $(X_0,U_0)$ circulates around the boundary of the
half-Gutzwiller rectangle. (See \fref{fig:boundary-map}).
On the other hand, follow the stream line in \fref{fig:separation}
until it reaches the next PSS,
that is, until it crosses either $\psi_1=0$ or $\psi_1=\pi$.
In this way, one can read off the final ($\vartheta_1, \psi_1$) for ${\cal M}$
and, via (\ref{eq:boundary-relation-appendix}), one gets $(X_1,U_1) \in I$
for ${\cal F}_I$.

Now let us study the flow in \fref{fig:separation}.
It has two types of singularities for $\gamma < 8/9$;
\vspace{-0.5cm}
\input separation-fig.tex
\vspace{-1cm}
\begin{align}
   &(a)~ \mathrm{Elliptic} : \sin{\vartheta}=\sin{\phi}=0
              ~ [(\vartheta,\phi)= (m,n)\pi,~m,n\in \mathbb{Z}] \\
   &(b)~ \mathrm{Hyperbolic} : \cos{\vartheta}=\cos{\phi}=0
              ~ [(\vartheta,\phi)= (m+\frac{1}{2},n+\frac{1}{2})\pi,~m,n\in \mathbb{Z}]
\label{eq:lattice}
\end{align}
By the shift of $\pi$ either in $\vartheta$ or $\psi$,
the linearized matrix and its eigenvalues change the sign.
Therefore, a source ($E_{+}$) and a sink ($E_{-}$)
locate alternatively at every $\pi$ on the lattice (a).
One the other hand, on the lattice (b),
hyperbolic singularity $H_{v}$ and $H_{h}$ locate alternatively,
where $H_{v}$ ($H_{h}$) attracts the trajectory vertically (horizontally).

The flow divides the initial domain into
three regions, depending on whether
the next crossing of the $y-$axis occurs
with $X_1>0$ ($\psi_1=0$) or with $X_1<0$ ($\psi_1=0$).
The division is determined by two critical angles $\vartheta_v$
and $\vartheta_h$ \cite{gutzwiller77}. The three regions are as follows.
\begin{itemize}
\item[\bf{1}]:
    $0 \leq \vartheta_0 < \vartheta_{v}$; the stream line is repelled by $H_{v}$ to the left,
    and then attracted into $E_{-}$ rotating counter-clockwise.
    $0  \geq \vartheta_1 >  -\vartheta_{h}$ and $\psi_1=0$.
\item[\bf{2}]:
    $\vartheta_{v} < \vartheta_0 < \pi-\vartheta_h$; it is repelled by both $H_{v}$ and $H_{h}$,
    and reaches $\psi_1=\pi$. Thus,
    $\pi+\vartheta_h  < \vartheta_1 <  2\pi-\vartheta_{v}$ and $\psi_1=\pi$.
\item[\bf{3}]:
    $\pi-\vartheta_h < \vartheta_0 \leq \pi$; the repulsion by $H_{h}$ acts to the right
    and  the rotation by $E_{+}$ is clockwise.
    $\pi+\vartheta_v > \vartheta_1 \geq  \pi$ and $\psi_1=0$.
\end{itemize}
For the map ${\cal M}$ (the $\vartheta\psi-$representation), this division is sufficient.
However, the restricted map $\left.{\cal F}\right|_I$ (the $XU-$representation) uses the conversion
via (\ref{eq:boundary-relation-appendix}).
This introduces further criticality at $\vartheta_0=\pi/2$
and $\vartheta_1=3\pi/2$ and, as the consequence,
region 2 is divided into three sub-regions for ${\cal F}_I$.

 We should add  that the most important for the above derivation of
 ${\cal F}_I$ is the order of initial critical angles
\begin{eqnarray}
   \vartheta_v < \pi/2 < \pi - \vartheta_c < \pi - \vartheta_h.
\label{eq:angle-order}
\end{eqnarray}
(The order of final critical angles
$
  \pi+\vartheta_h < \pi+ \vartheta_c < \pi+ \vartheta_h < 3\pi/2 < 2\pi - \vartheta_v
$
is then guaranteed by the time-reversal symmetry).
The order \eref{eq:angle-order} follows from the distribution of the singularities in
(\ref{eq:lattice}) for $\gamma <8/9$, and the numerical
confirmation is shown in \fref{fig:variation-of-critical-angles}.
\input variation-of-critical-angles.tex

\section{One-time Map: Blow-up and Contraction}
\label{appendix:b}
We use below symbols for critical objects to help the book-keeping.
All in the domain of $\cal{F}$ have naturally superscript `0',
and those in the image `1'.
The side of a separator curve $C$ has double subscript common to its adjacent region.
A focus point is expressed by a letter ($v,h$ and $c$), taken from the subscript
of the critical angle ($\vartheta_v,\vartheta_h$ and $\vartheta_c$),
by which the $X$-coordinate of the point is determined.
For instance, $C^0_{++}$ is the side of a separator in $D_0$ and adjacent to $(++)$,
and $v^1_{++}$ is a focus point in $D_1$, adjacent to $(++)'$, and
$X_v=2\cos^2\vartheta_v$. See \tref{tab:separation-table}.

Let us start by examining how ${\cal C}^0_{++}$ is mapped by $\cal{F}$,
since ${\cal F}$ on $(++)$ is relatively simple without rotation.
\Fref{fig:one_time_map_red_beak} shows how sets of vertical and horizontal line segments (set b and c), meeting at the separator ${\cal C}^0_{++}$ are mapped by ${\cal F}$.
We observe every set turns into a {\it `beak'} with its tip
at $v^1_{++}$, while the full vertical line (a) not touching
${\cal C}^0_{++}$ is simply distorted.
\input one_time_map_red_beak.tex
\vspace{-0.5cm}
\noindent
Thus we find a rule of contraction
 \begin{equation}
  {\cal F}:C^0_{++}\longmapsto v^1_{++}.
 \label{eq:contraction-rule-red}
\end{equation}
By considering a move of vertical line segment
(and its image in $(++)'$) until it reaches $v^0_{++}$,
we find a rule of blow up;
 \begin{equation}
  {\cal F}: v^0_{++}  \longmapsto C^1_{++}.
 \label{eq:expansion-rule-red}
\end{equation}
 (\ref{eq:contraction-rule-red})
and
 (\ref{eq:expansion-rule-red})
are time-reversal pair.
Now, let us proceed to the $C^0_{(+-)}$.
It is adjacent to $(+-)$ and $\left.{\cal F}\right |_I $
rotates the boundary $2$ as seen in \fref{fig:boundary-map}.
We clarify in \fref{fig:one_time_map_blue_beak}
how the rotation affects the interior map.
\begin{itemize}
\item[(i)]
${\cal F}$ maps each set of vertical and horizontal line segments ($p,~q,\cdots$)
meeting at $C^0_{+-}$ into a {\it beak} with its tip at a point $v^1_{+-}$.
Therefore,
\begin{equation}
{\cal F}:C^0_{+-}\longmapsto v^1_{+-}.
\label{eq:contraction-rule-blue}
\end{equation}
\end{itemize}
%
\input one_time_map_blue_beak.tex
\noindent
This just corresponds to the contraction rule (\ref{eq:contraction-rule-red}).
On the other hand, the body of the beak reflects the rotation as follows.
\begin{itemize}
\item[(ii)]
The left-side of the beak (dashed line) always connects the $v^1_{+-}$
and another critical point $c^1_{+}$.
And it is the image of the line segment
horizontally connecting ${\cal C}^0_{+-}$ to ${\cal I}^0_{+}$
(the $X>0$ side of the enter line of the collision manifold $\mathrm{I}$ in $D_0$).
Therefore, we find a contraction rule
\begin{equation}
  {\cal F}:{\cal I}^0_{+}\longmapsto c^1_{+}.
  \label{eq:I-to-c}
\end{equation}
\item[(iii)]
The right-side of the beak, on the other hand, is the image of the vertical line segment
connecting ${\cal C}^0_{+-}$ to either $2B$ or $2C$.
In the former, it is simply a long vertical curve. In the latter,
it is a short {\it hook}, connecting $v^1_{+-}$ and a point in $2C'$, now in the upper
boundary due to the rotation.
\end{itemize}
Now, let us investigate the vertical line segments ($a,b,c,\cdots$) connecting
$2A$ to $2B$.
\begin{itemize}
  \item[(iv)]
  By the rotation, $2A'$ and $2B'$ are both in the bottom boundary
  of $(+-)'$. Thus,  $a',b',c',\cdots$ form {\it wing}-like curves---
  folding, induced by the rotation.
  The limit $a,b,c,\cdots\ra {\cal I}_{+}$ corresponds to
  $a',b',c',\cdots \ra c^1_{+}$.
  Comparing Limits of both code, we find again
  ${\cal F}:{\cal I}^0_{+}\longmapsto c^1_{+}$.
\end{itemize}
\begin{itemize}
\item[(v)]
Reversing the previous code, let us consider ${\cal I}_{+},c,b,\cdots\ra a$.
The image is a code, starting from a single point ${c^1}_{+}$, passing through enlarging wings,
and the limit is a cusped curve enveloping all wings. See \fref{fig:one_time_map_blue_beak}(b).
Here occurs a blow up of the end-point of $a$, namely
\begin{equation}
   v_{+-}^0 \longmapsto C^1_{+-}.
\label{eq:blow-up-rule-vC1}
\end{equation}
\end{itemize}
Remarkably, we observe in \fref{fig:one_time_map_blue_beak}(b) that the contraction
and blow-up come in a pair, keeping the perimeter length of the boundary.  (See more in
\eref{eq:perimeter-invariance}).

\section{Transverse Crossing of Future and Past Tiling}
\label{appendix:c}
Let us first note a remarkable fact that $D_0$ and $D_1$
consist of altogether 8 sub-regions, but that
there are actually only two distinct shapes. Half are congruent to $(++)$
in $D_0$ and the other half to $(+-)$. See \fref{fig:congruence}.
\input congruence.tex
\vspace{-0.5cm}
This comes from the invariance of AKP under time-reversal and  parity transformations.
First the time reversal changes the direction of momentum while keeping the
coordinate values.
Namely,
\begin{equation*}
T: (X,U) \ra (X, -U)
\label{eq:time-reversal-1}
\end{equation*}
This in particular implies that an orbit with the initial value
$(X_0,U_0)$ and evolving backwards in time is the same with that
starting with $(X_0, -U_0)$ and evolves forward in time.
Therefore, it follows for general $N$ that
\begin{equation}
    \zeta^N_P(X_0,U_0)=\zeta^N_F(X_0,-U_0)
\label{eq:time-reversal-2}
\end{equation}
as a relation between the future and past height functions.
Now, just as the sub-regions $A,\ldots,D$ are the ribbons of
{\it future} height functions $\zeta^F_{N=1}$, so
$\bar{A},\ldots,\bar{D}$ are the ribbons of the past function $\zeta^P_{N=1}$,
because $(a_0,a_1)=(+-)$ for ${\cal F}$ is equivalent to ($a_0$, $a_1$ interchanged) $(-+)'$ for ${\cal F}^{-1}$
and so on.
Therefore, it follows from time reversal symmetry that
\begin{equation}\label{eq:time-reversal-3}
\bar{A}\equiv A,~\bar{B}\equiv B,\bar{C}\equiv C,\bar{D}\equiv D
~(\mathrm{mod} ~U \ra -U).
\end{equation}
This is abbreviated below as $\bar{A} \equiv T(A), \cdots$.
Next, the symmetry under the parity transformation
\begin{equation*}
\vec{x} \ra -\vec{x},~~\vec{p} \ra -\vec{p}
\label{eq:parity-1}
\end{equation*}
implies that an orbit starting from $(X_0,U_0)$
and its partner starting from $(-X_0,-U_0)$ evolve
keeping the relation $X(t)=-\tilde{X}(t)$ (and hence
$a_n=-\bar{a}_n$). Therefore,
\begin{equation*}
\zeta^F_N(X_0,U_0)=-\zeta^F_N(-X_0,-U_0),~~
\zeta^P_N(X_0,U_0)=-\zeta^P_N(-X_0,-U_0)
\label{eq:parity-2}
\end{equation*}
and we find that
\begin{equation}
A \equiv P(D), B\equiv P(C), \bar{A} \equiv P(\bar{D}), \bar{B}\equiv P(\bar{C}),
\label{eq:parity-3}
\end{equation}
where $P$ acts on $D_0$ as well as $D_1$ as $P:~X \ra -X,~ U\ra -U$.
Finally, (\ref{eq:time-reversal-3}) and (\ref{eq:parity-3}) together
the eight sub-regions $\bar{A},\ldots,\bar{D}$, and
$\bar{A},\ldots,\bar{D}$ are grouped into two classes;
\begin{eqnarray}
  A &\equiv T(\bar{A}) \equiv P(D) \equiv P(T(\bar{D})) ,   \label{eq:congruence-A}
 \\
  B &\equiv T(\bar{B}) \equiv P(C) \equiv P(T(\bar{C})).    \label{eq:congruence-B}
\end{eqnarray}
Now, the direction of increasing height of the future $N=1$ surface is shown by an arrow
in \fref{fig:congruence}. This is nothing but the $N=1$ fact in the proof
of the properness of ribbon tiling by mathematical induction. Now, by the above
symmetry relation, the arrow for the past $N=1$ surface is $U \ra -U$ mirror
of the future arrow. Therefore, the future ribbons and past ribbons are transverse
each other.
This inherits to higher level ribbons, as the level $N\ra N+1$
proceeds keeping the properness of ribbon tiling. The sole
exception is where future and past ribbons become tangent each other
as seen in \sref{sec:3-DSS-creation}.

\section{Transverse Chopping and Longitudinal Splitting}
\label{appendix:d}
If one is content with just comparing the location
of the new and previous ribbons, it is simple;
each of previous ribbon {\it longitudinally} splits into two finer ones
and just that.
The height $\zeta_N^F(X_0,U_0)$ of level $N$ step
is calculated by (\ref{eq:zeta}) from $a_j, j=0,\cdots, N$
and $\zeta_{N+1}^F(X_0,U_0)$ has additional last bit $a_{N+1}$
which contributes $\pm \Delta_{N+1}$ ( $\Delta_{N+1}=1/2^{N+1})$.
Here the sign depends on $(X_0,U_0)$ via $\F^{N+1}$.
But, as we proved, the tiling at $N+1$ is proper
Therefore, the level $N$ ribbon longitudinally splits into two finer ribbons
in such a way that the left of split-line gets $-\Delta_{N+1}$ and right gets $\Delta_{N+1}$.
See \fref{fig:ribbon-longitudinal-splitting}.
Note, in the case of non-shrinking ribbon, the longitudinal split-line
occurs on  the boundary of a level $N$ ribbon.
\input ribbon-longitudinal-splitting.tex
\vspace{-0.5cm}
The level $N+1$ ribbons are created by
chopping the $N$ ribbons {\it transversely}
into two parts by the (fixed) separator curve
and mapping each into different half of rectangle
with elongation from $-B$ to $B$ as shown in \sref{sec:3-DSS-creation}.
So, if one picks some ribbon at large $N$,
and wishes to trace back from which part of initial domain it comes,
it requires tremendous task. See \fref{fig:iteration-tree}.
\input iteration-tree.tex

\section{Uniqueness of PO within the Overlap}
\label{appendix:e}
As for the unstable PO3-6, the initial position is remarkably very close to the edge of
the overlap. This is just as it should; the unstable
PO must be, to be a PO, on the union of future and past ribbons of its code,
and yet, to be unstable, {\it should not be} much inside the junction.
\input broucke_chi2_contours.tex
This is a subtle point, since the exact corner is homo-clinic point and
cannot be a periodic point. The fact is that the unstable PO3-6 turns out extremely close to the corner
as shown in \fref{fig:broucke-chi2-contours}.
We thank Tanikawa and Shibayama pointing out this issue of homo-criticality
and mentioning that this kind of close proximity often occurs.
The unstable PO3-6 is non-self-retracing (`NR').
There is no other PO of the same code on the union of the future and past ribbons
as is also clear in \fref{fig:broucke-chi2-contours}.
We should add that we have observed stable satellite in the Broucke's island.
The detail is under investigation.

\end{document}

%% file: gutzwiller-lips.tex
\begin{figure}[!h]
\begin{center}
	\includegraphics[width=14cm]{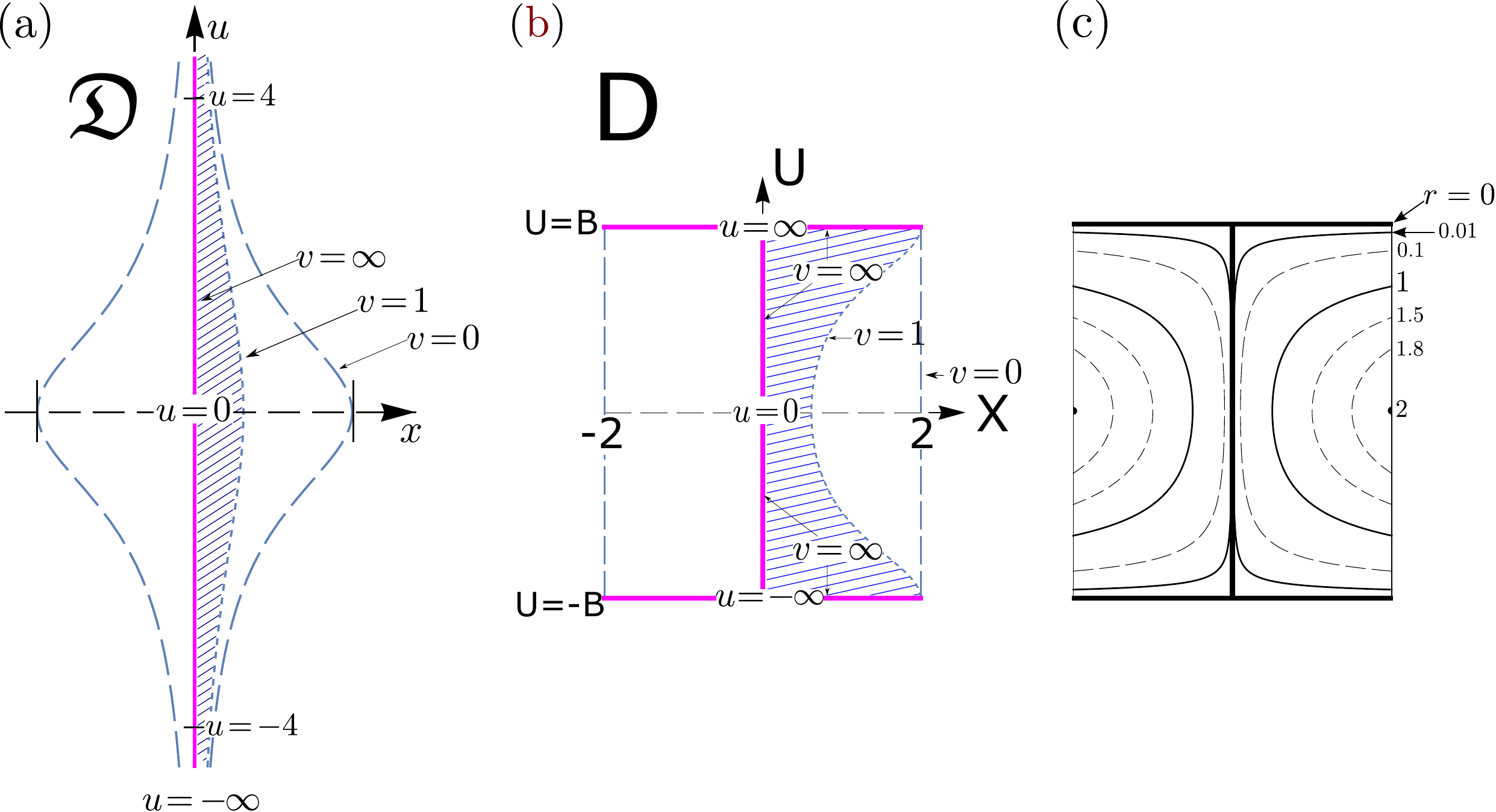}
\caption{
(a) The initial domain $\mathfrak{D}$ on the Poincar\'e surface of section ($y=0$).
(b) Gutzwiller rectangle D.
I-shaped {\it backbone} (solid line) is the collision manifold ($r=0$).
Graphs are drawn to scale (at $\gamma=0.2,~B=2.349$).
The map $\mathfrak{D}\rightarrow$ $D$ preserves area. For instance, hatched regions
($-\infty<u<\infty, 1 \le v $) in $\mathfrak{D}$ and $D$ have the same area.
 (c) Equi-r contours in $D$ approach the backbone at $r \rightarrow 0$.
}
\end{center}	
\label{fig:gutzwiller-lips}
\end{figure}

%% file: coarse-grained-dss.tex
\begin{figure}[!h]
	\centering
	\includegraphics[width=12cm]{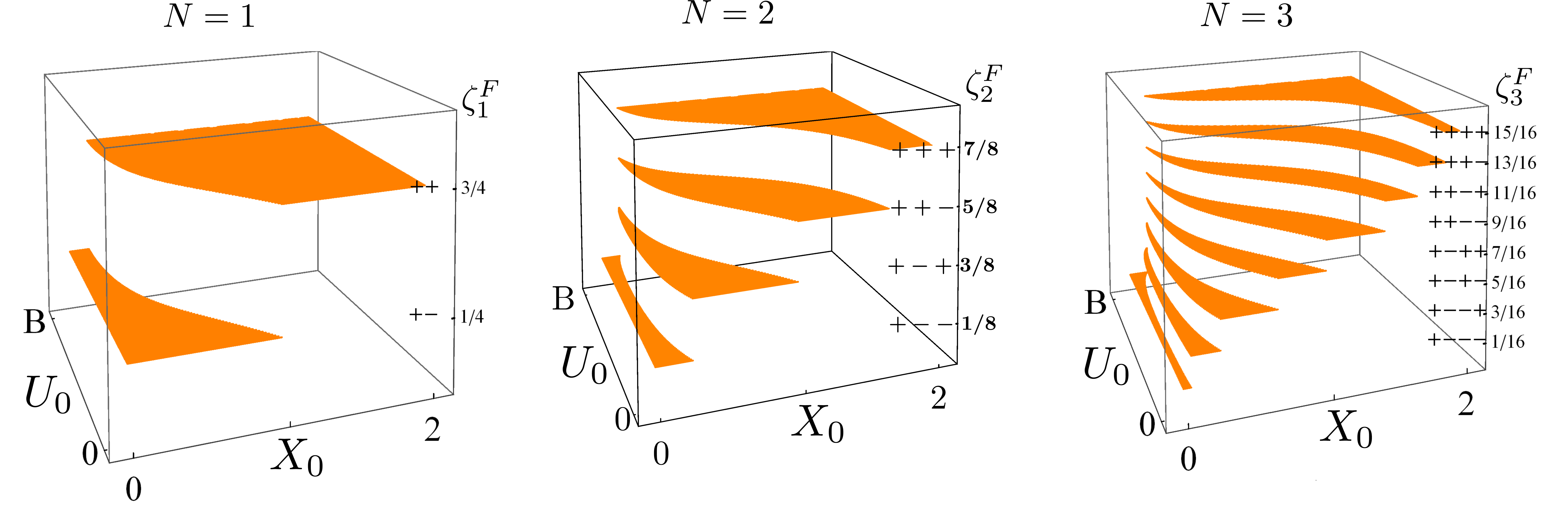}
	\caption{The future DSS with $N=1,2,3$. (Only the first quadrant part shown).
                    The level $N$ surface consists of altogether $2^{N+1}$ steps with heights $\zeta_N^F$ in $(-1,1)$.}
   	\label{fig:coarse-grained-dss}
\end{figure}

%% file: step-ribbon.tex
\begin{figure}[!h]
	\begin{center}
	\includegraphics[width=5cm]{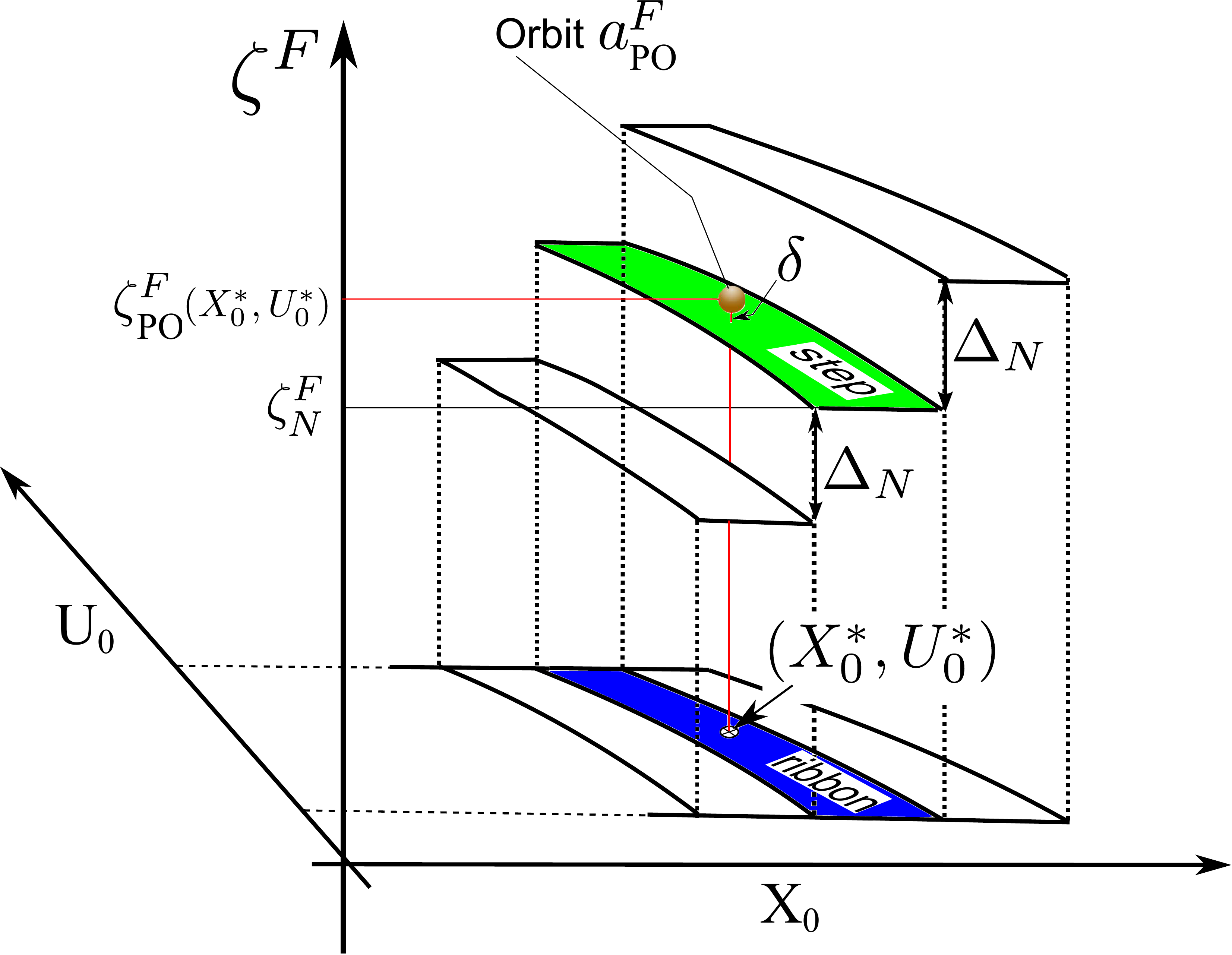}
	\end{center}
	\caption{Each step of DSS stands on its own {\it ribbon}.
            A PO is illustrated by a ball on its initial point $(X^*_0,X^*_0)$ and has a
            height (\ref{eq:zetaPO}). Also shown are the closest step (green), and
            its base ribbon (blue) enclosing the initial point $(X^*_0,X^*_0)$.
	}
	\label{fig:step-ribbon}
\end{figure}

%% file: separation-table.tex
\begin{table}[!h]
\caption{\label{tab:separation-table}
Regions 1, 2, 3 are separated by hyperbolic singularities $H_v$ and $H_h$;
$\psi_0=0$ by definition but $\psi_1=0,\pi,0$ respectively for 1,2,3.
Region 2 is sub-divided by criticalities at $\vartheta_0=\pi/2$ and
$\vartheta_1=3\pi/2$. $X_{v,h,c}=2 \cos^2(\vartheta_{v,h.c})$ and $B=2\pi\sqrt{\mu}/2$.}

\begin{indented}
\item[]
\begin{tabular}{cccccrcr}
\br
&$\vartheta_0(\psi_0\equiv 0)$ & $\vartheta_1$ & $\psi_1$ & $X_0$ & $U_0$ & $X_1$ & $U_1$\\
\mr
1  & $0 \rightarrow \vartheta_v$ & $0\rightarrow -\vartheta_h$& 0 & $2\rightarrow X_v$& $B$ & $2\rightarrow X_h$ & $B$ \\
\mr
$\Updownarrow$ & $\vartheta_v$& \multicolumn{4}{c}{Separation by  $H_v$} && \\
\mr
2A & $\vartheta_v\rightarrow \frac{\pi}{2}$ &$\pi + \vartheta_h\rightarrow\pi+\vartheta_c$ & $\pi$ & $X_v\rightarrow 0$ & $B$ & $-X_h\rightarrow -X_c$ & $-B$\\

\multicolumn{1}{c}{} & $\frac{\pi}{2}{}^{\P}$ &  ($\pi+\vartheta_c$) & & \\

2B & $\frac{\pi}{2}\rightarrow\pi-\vartheta_c$ & $\pi+\vartheta_c\rightarrow\frac{3\pi}{2}$ & $\pi$ & $0\rightarrow X_c$& $-B$ & $-X_c\rightarrow 0$ & $-B$ \\

\multicolumn{1}{c}{} & ($\pi-\vartheta_c$) & $\frac{3\pi}{2}{}^{\clubsuit}$&  & \\

2C & $\pi-\vartheta_c\rightarrow\pi - \vartheta_h$ & $\frac{3\pi}{2}\rightarrow 2 \pi - \vartheta_v$ & $\pi$& $X_c\rightarrow X_h$&$-B$& $0\rightarrow-X_v$&$B$\\
\mr
$\Updownarrow$ & $\pi-\vartheta_h$ &\multicolumn{4}{c}{Separation by  $H_h$} &&\\
\mr
3  & $\pi-\vartheta_h\rightarrow\pi$ & $\pi + \vartheta_v\rightarrow\pi$ & $0$&$X_h\rightarrow 2$& $-B$ & $X_v\rightarrow 2$ & $-B$\\
\br
\end{tabular}
\item[] $^{\P}$~ $\vartheta_0$ crosses $\frac{\pi}{2}$ from below inducing $U_0:B\rightarrow -B$.
\item[] $^{\clubsuit}$~ $\vartheta_1$ crosses $\frac{3\pi}{2}$ from below inducing $U_1:-B\rightarrow B$.
\end{indented}
\end{table}

%% file: boundary-map.tex
\begin{figure}[!h]
	\centering
	\includegraphics[width=10cm]{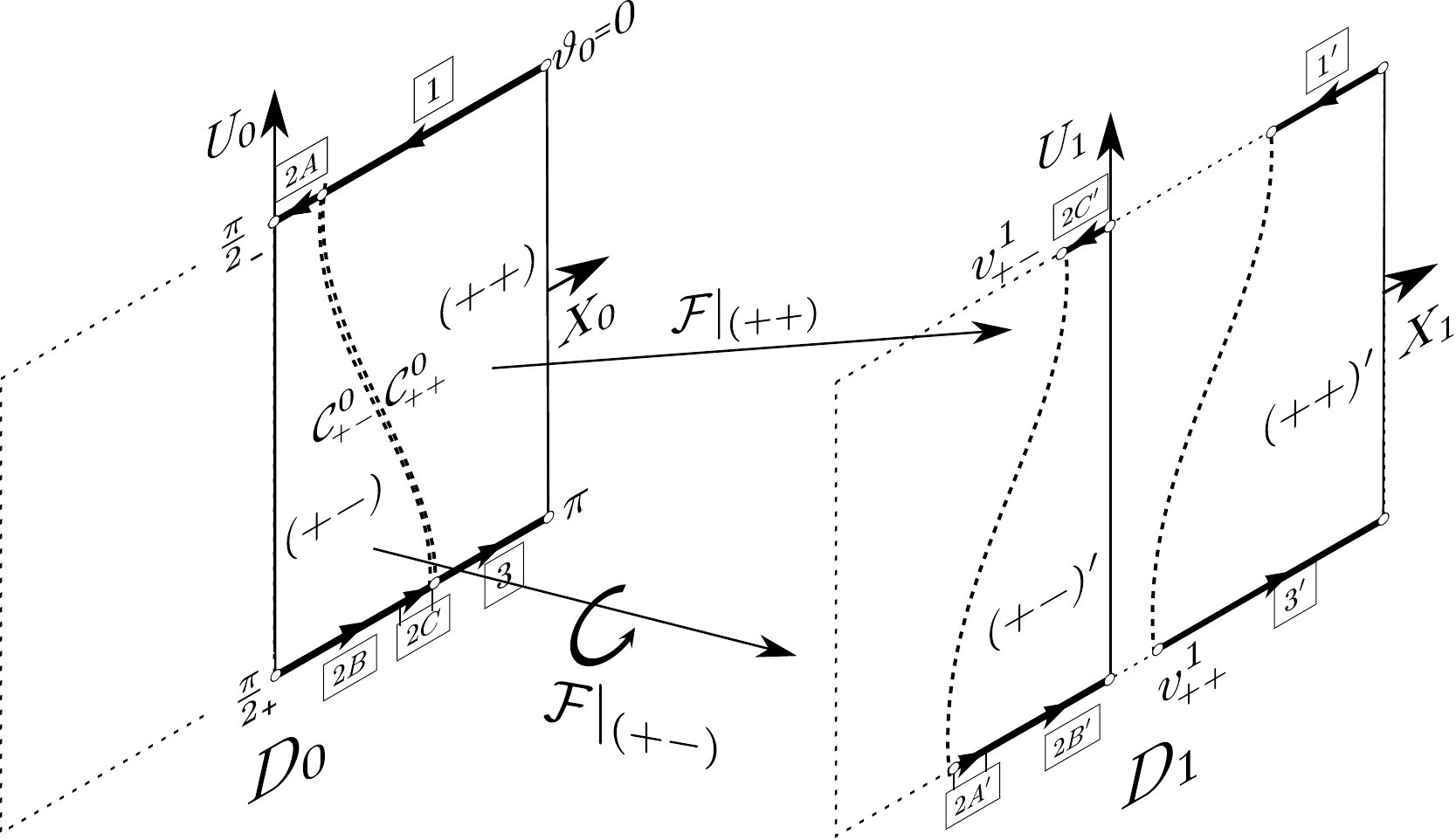}
	\caption{
Boundary map $\left.{\cal F}\right |_I$
given by \tref{tab:separation-table}.
Full map ${\cal F}$ sends $(++)$ and $(+-)$ to the left and right
respectively; besides, the boundary of $(+-)$ is `rotated'.
$C^0_{+\pm}$ are mapped not onto boundary curves of $(+\pm)^\prime$, but into
the end points $v^1_{+\pm}$ respectively.
}
\label{fig:boundary-map}
\end{figure}

%% file: one_time_map_scheme3d.tex
\begin{figure}[!h]
	\begin{center}
	\includegraphics[width=16cm]{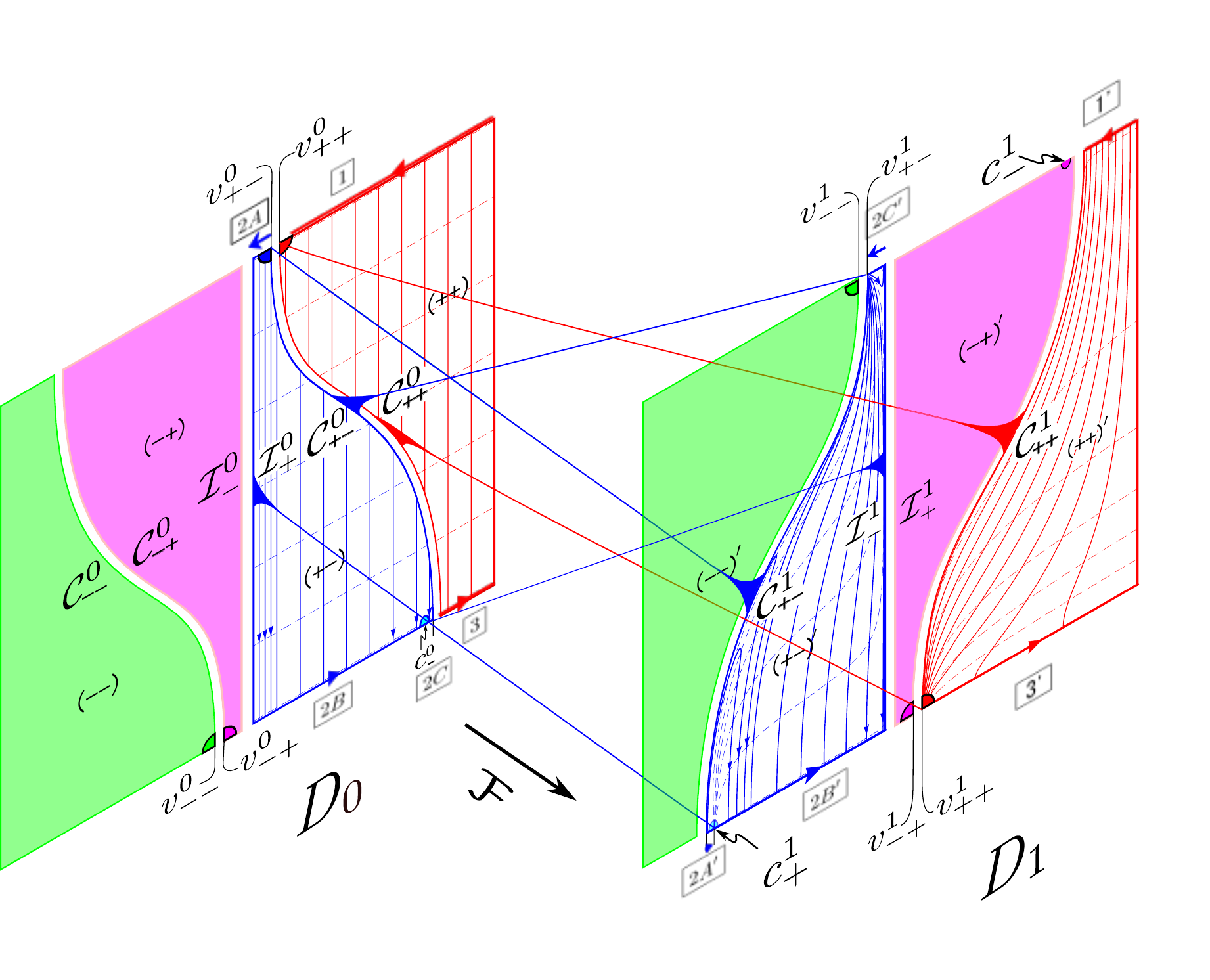}
	\end{center}
    \caption{\label{fig:one_time_map_scheme3d}
            Critical maps (\ref{eq:contraction-rule-red})---(\ref{eq:blow-up-rule-vC1}).
      Connectors of criticalities are shown only on the boundaries of
      $(++)$ and $(+-)$. $\gamma=0.2$ and to scale. }
\end{figure}

%% file: c-curve.tex
\begin{figure}[!htb]
\begin{center}
	\includegraphics[width=7cm]{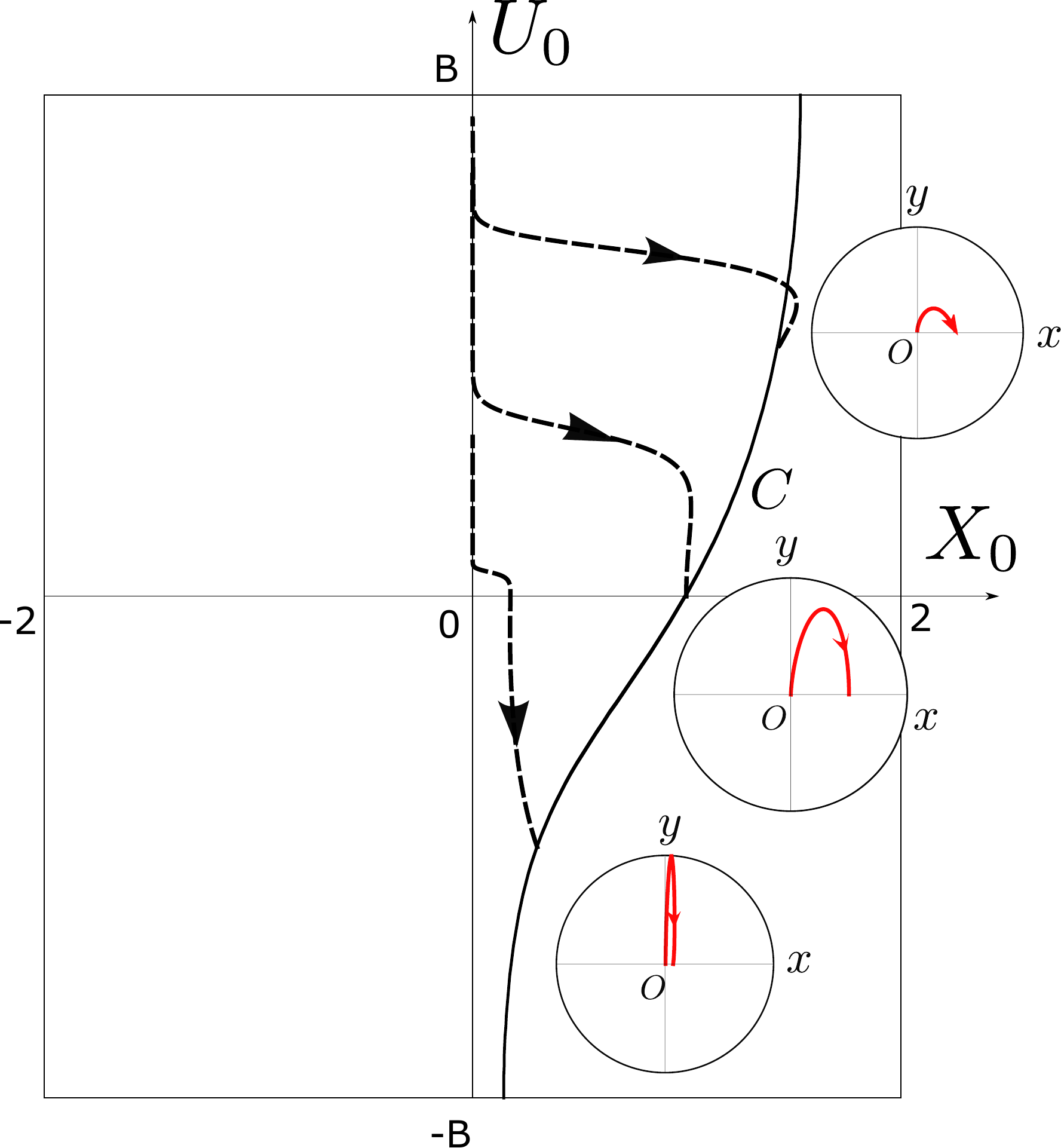}
	\end{center}
	\caption{
     Three collision orbits in circles,  and their projection in $(X,U)$ plane (dashed curves).
     The latter reach the separator curve $C$ at the first arrival
     at the PSS.}
\label{fig:c-curve}
\end{figure}

%% file: proof.tex
\begin{figure}[!h]
	\centering
	\includegraphics[width=16cm]{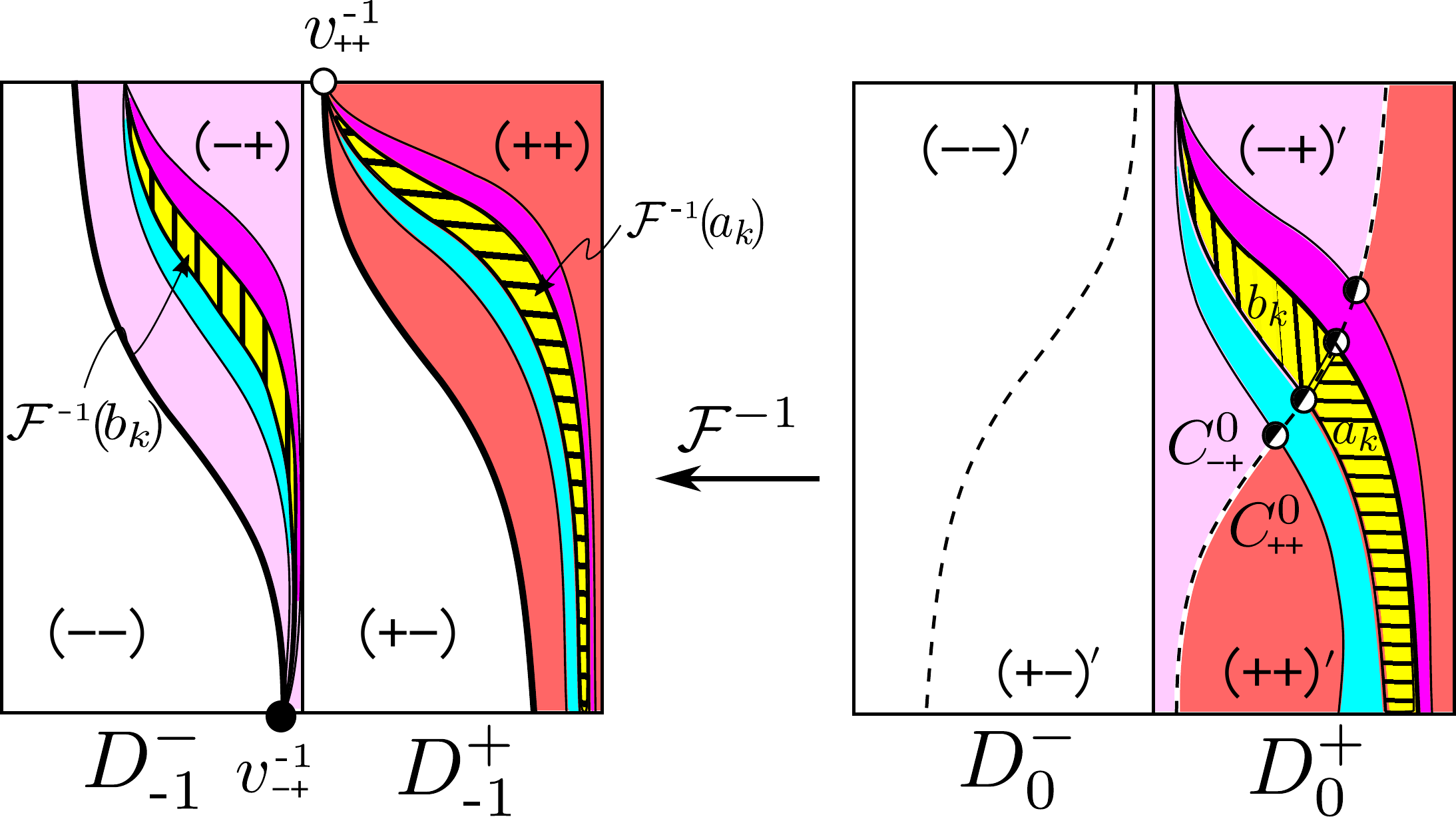}
	\caption{The separator $C^0_{\pm+}$ in ${\cal F}^{-1}$ (\fref{fig:one_time_map_scheme3d})
	chops a ribbon in $D_0^{+}$. Since $C^0_{\pm+} \rightarrow v^{-1}_{\pm+}$,
    $a_k$ ($b_k$) is mapped backwards  into $(++)$ [$(-+)$] forming a  full height ribbon
    with vertex  $v^0_{++}$ ($v^0_{-+}$); number of ribbons are doubled.
	Meant to illustrate a general case, but data with $\gamma=0.2$, $N=3$ is used.}
	\label{fig:proof}
\end{figure}

%% file: crossing-of-future-past-ribbons.tex
\begin{figure}[!ht]
	\centering
\includegraphics[width=17cm]{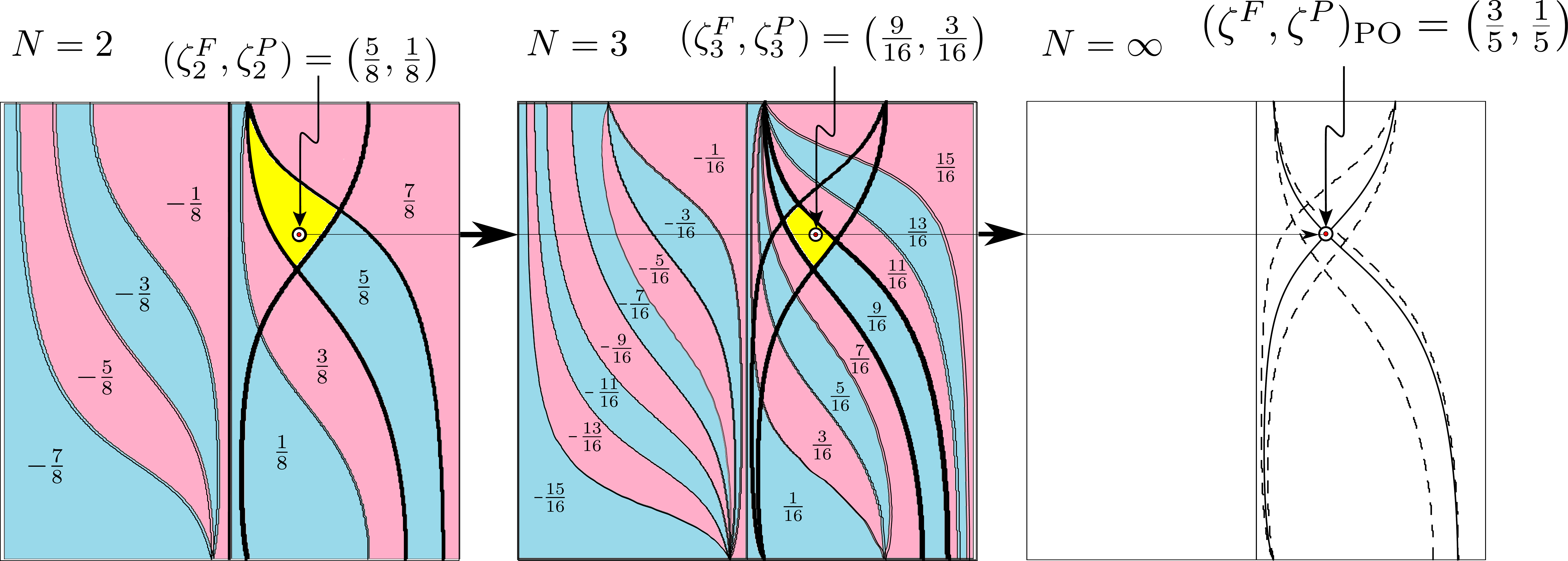}
	\caption{Ribbon-evolution at $\gamma=0.2$.
All future ribbons and one past ribbon relevant for the PO are shown, and the overlap
is highlighted. The PO (shown by circle dot) is code $(++--)$, (rank 2, Id=3).
$(\zeta^F, \zeta^P)=(3/5,1/5)$ by (\ref{eq:zetaPO}). The overlap is
with heights $(5/8,1/8)$  ($(9/16,3/16)$) at $N=2$ ($N=3$) by (\ref{eq:zeta}).
}
	\label{fig:crossing-of-future-past-ribbons}
\end{figure}

%% file: broucke_evolution_PO3-6.tex

\begin{figure}[!ht]
	\centering
	\includegraphics[height=15cm]{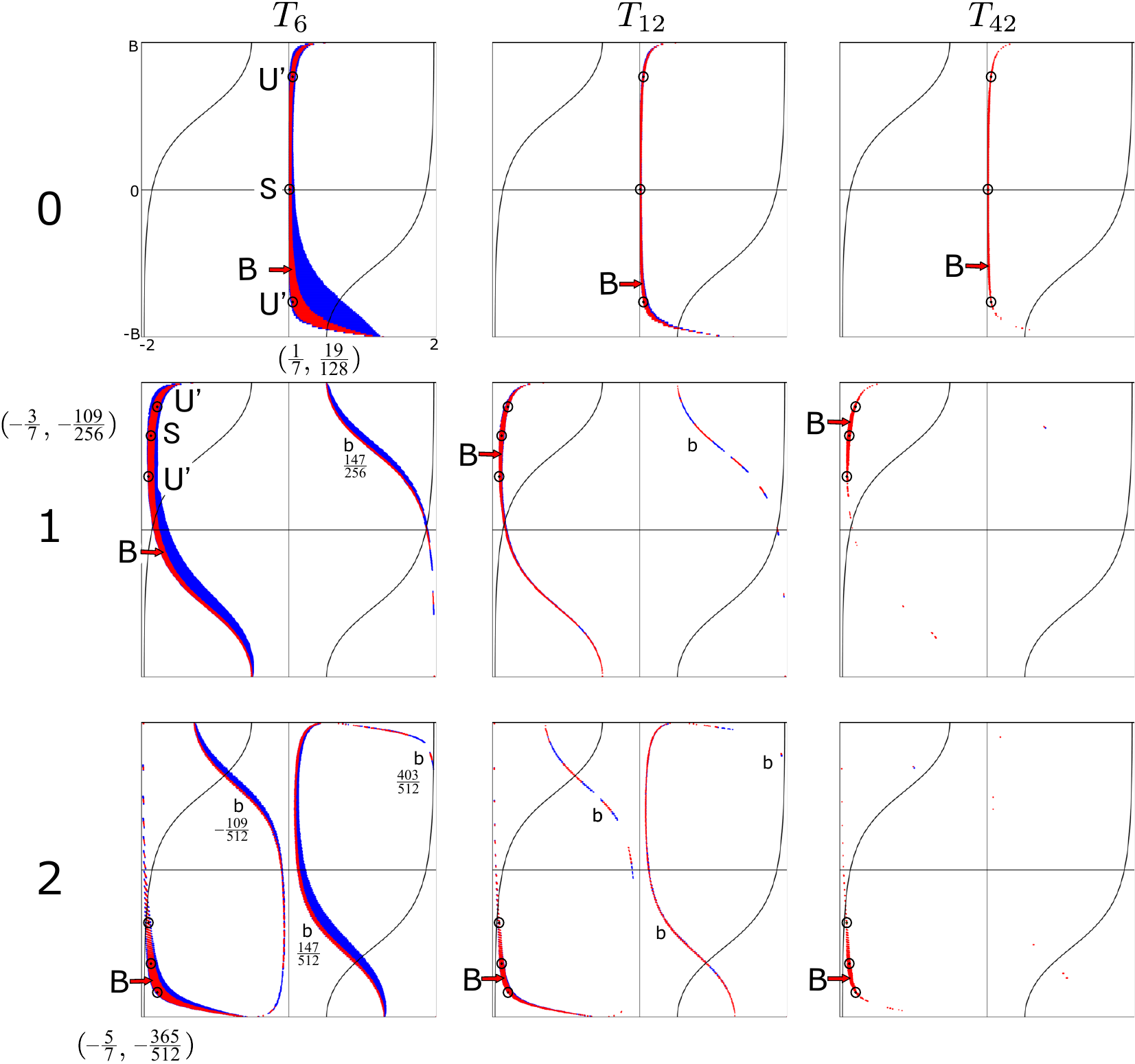}
	\caption{Evolution of future ribbons of Broucke's PO3-6.
    Snap shots of tiling $T_N$ at $N=(6, \cdots, 44)$ with $T~(\mathrm{mod}~6)$.
    $\gamma=0.611$.   Circle-dots: ($S$) and unstable ($U')$ PO.
    `B': the ribbon enclosing them (`b' chopped away parts). `B' remains non-shrinking, while
    adjacent-side ribbons vanish quickly.
    Heights  $\left(\zeta^{F}_{\mathrm{PO}}(X^*_0,U^*_0), \zeta^{F}_N(a)\right)$ as
    given by (\ref{eq:zetaPO}), (\ref{eq:zeta-by-the-N+1-bits}).
    $\zeta^{F}_N(a)$ can be also calculated by \eref{eq:master2},
    e.g. comparing $T_6$ and $T_8$,
    $-\frac{365}{512}=-\frac{1}{2}+\frac{1}{2}\left(-\frac{1}{2}+\frac{1}{2}\cdot\frac{19}{128}\right)$.
    Initial positions of POs in $T_0$ are for canonical reading of code as $(a_0,\cdots,a_5)=(+--+--)$;
    those in $T_N$ are given by acting $({\cal F}^{-1})^N$.
	}
	\label{fig:broucke_evolution_PO3-6}
\end{figure}

%% file: broucke-transition.tex
\begin{figure}[!ht]
    \centering
    \includegraphics[width=14cm]{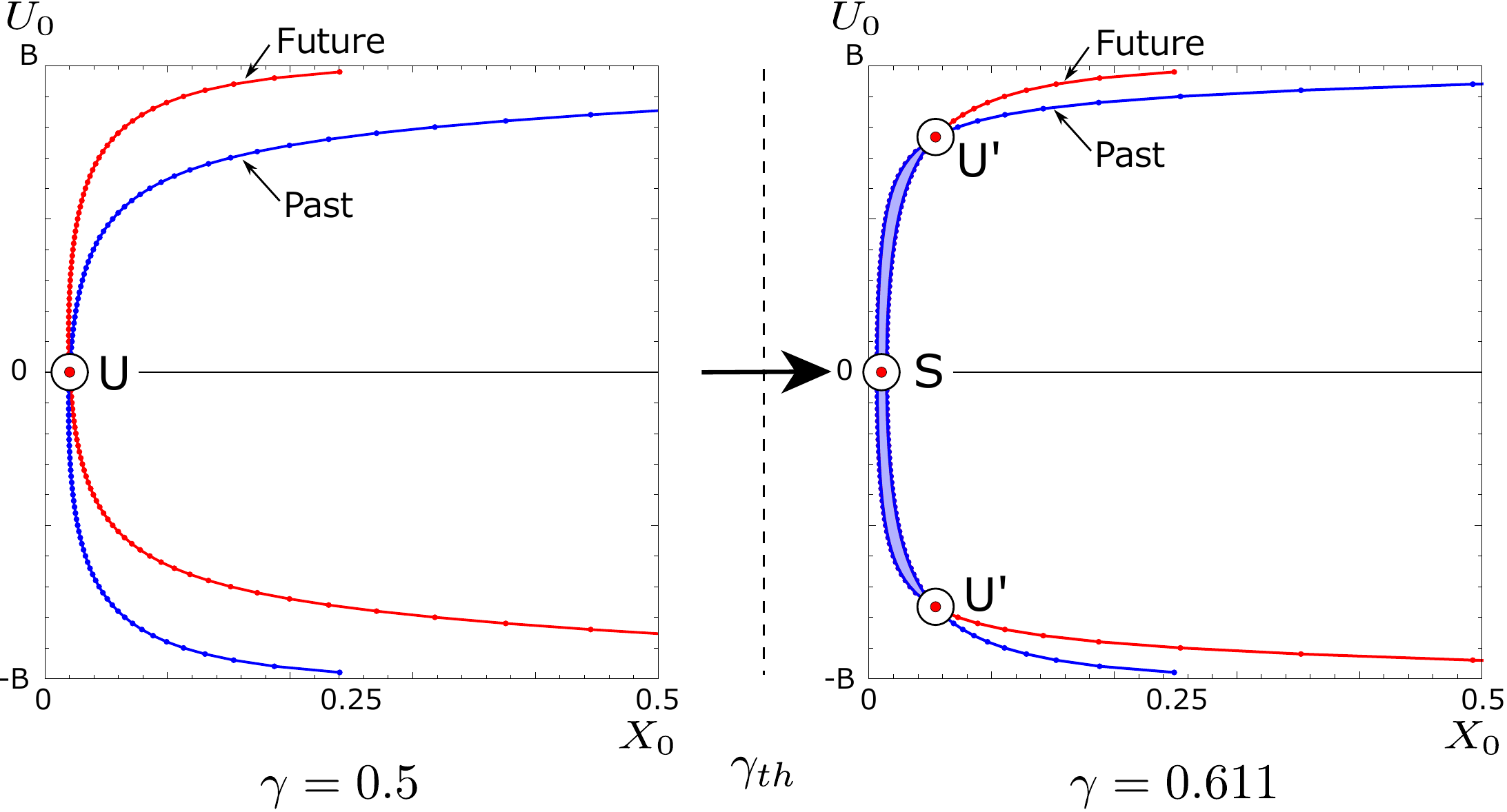}
        \caption{The transition process $U\rightarrow S+U'$.
   The ribbons are calculated at asymptotic $N$ ($N=48 $).
    Left:$\gamma =0.5$.  Right:$\gamma =0.611$.
     ($\gamma_\mathrm{th}=0.572350895$ for the Broucke's PO3-6).
      The two $U'$s are the appearance of the same PO
     at different time slices ($T=45$ and $T=48$).
     }
     \label{fig:broucke-transition}
\end{figure}

%% file: broucke-location.tex
\begin{figure}[!ht]
\begin{center}	
\includegraphics[width=16cm]{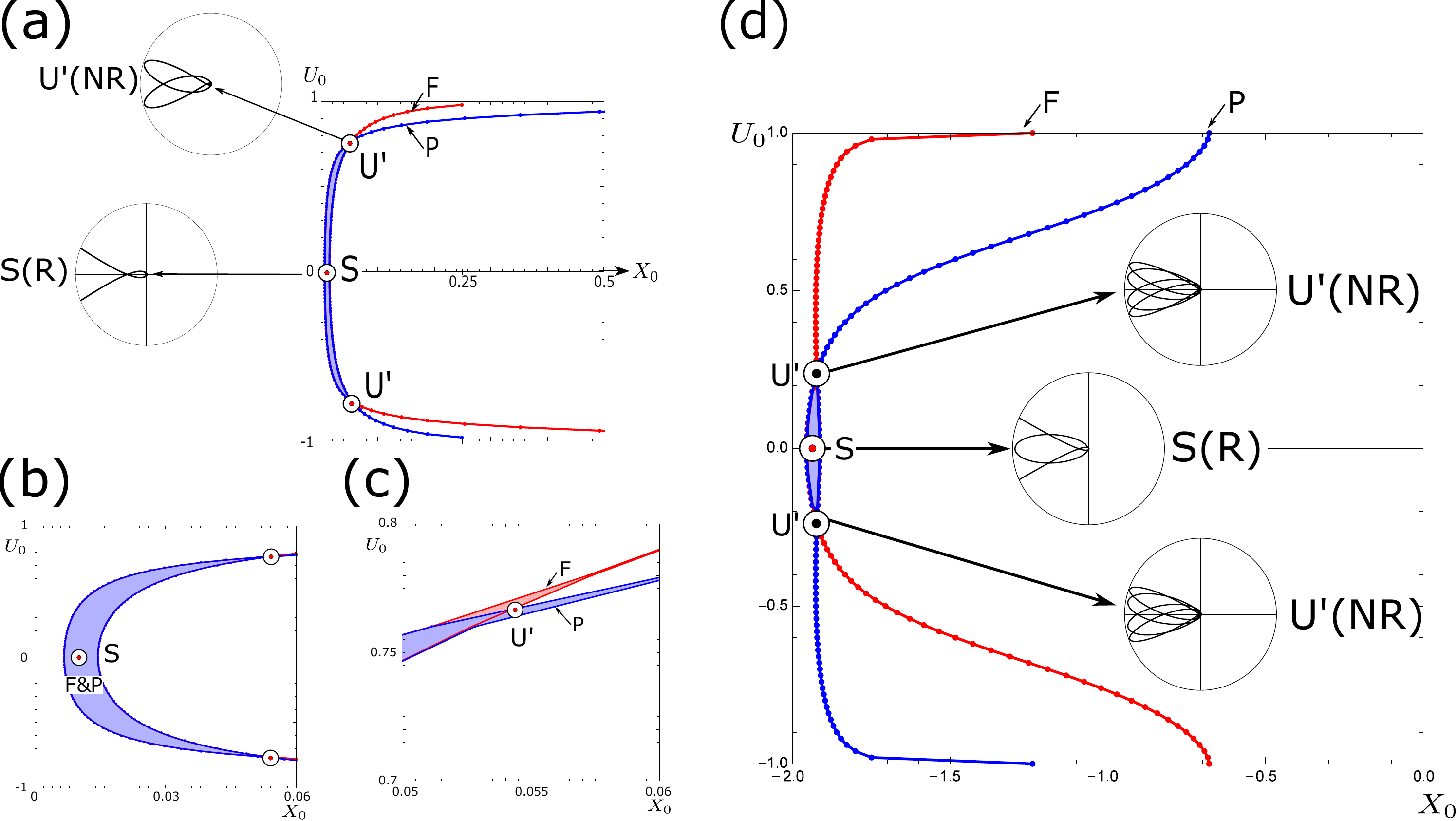}
\end{center}
\caption{\label{fig:broucke_location}
    Broucke PO bifurcates as $U(R) \rightarrow S(R)+ U'(NR)$.
    Initial points of Broucke's POs after the bifurcation
    are shown in the overlap of future (red) and past (blue) ribbon at level $N=48$.
    (a-c) PO3-6 ($\gamma=0.611)$, (d) PO5-40 ($\gamma=0.77$).
    $S(R)$ and $U'(NR)$ respectively locate in the maximum overlap and at
    edge of the overlap. See magnified figures (b), (c) for PO3-6.
    Insets show N and NR orbit profiles.
	}
\end{figure}

%% file: broucke-bifurcation.tex
\begin{figure}[!h]
\includegraphics[width=16cm]{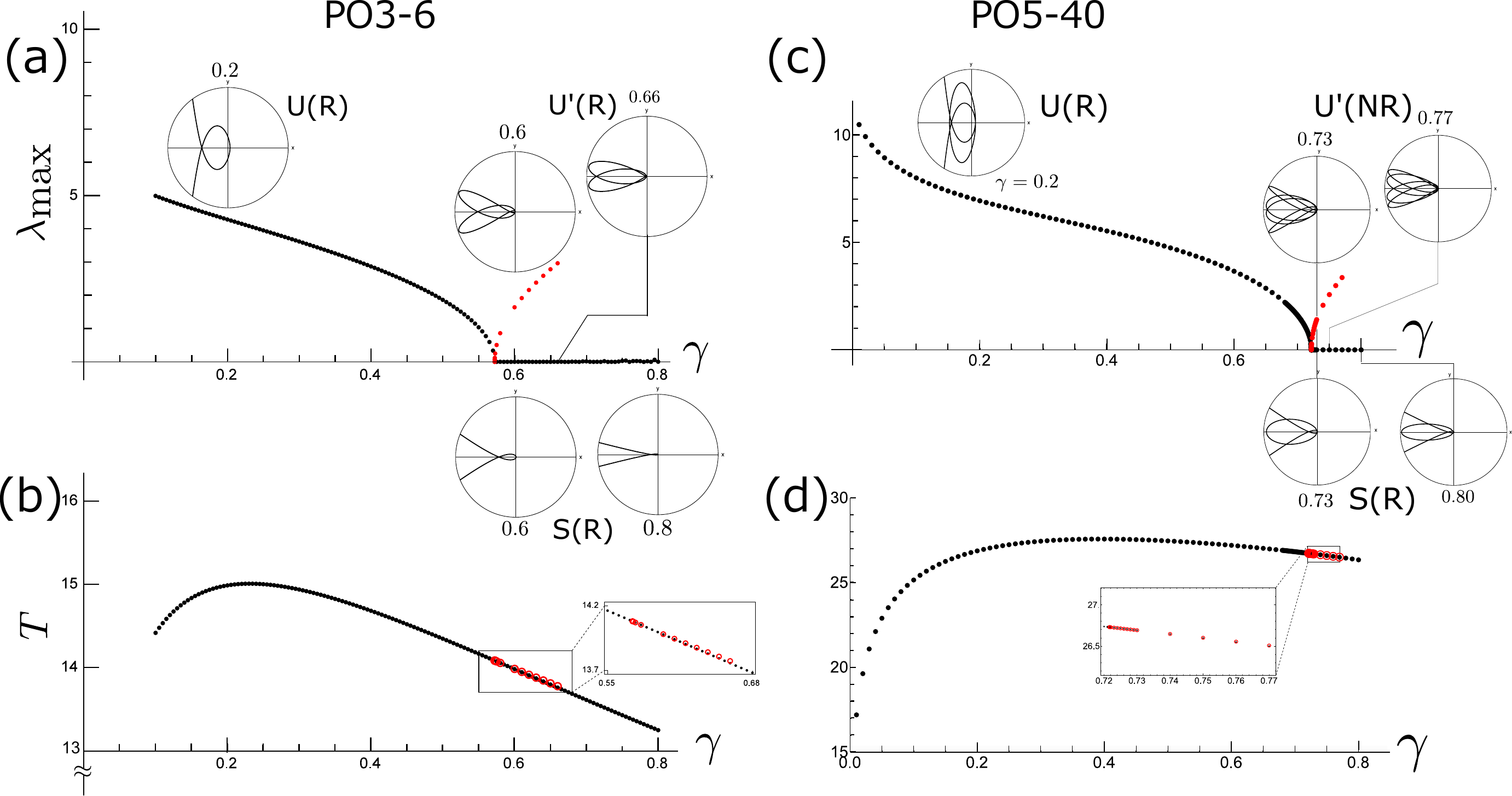}
\caption{
\label{fig:broucke-bifurcation-diagram}
Bifurcation of (a-b) PO3-6 and  (c-d) PO5-40.
Both $U(R) \rightarrow S(R)+ U'(NR)$.
The retracing orbit, perpendicularly emitted
from the heavy axis changes from stable to unstable, keeping self-retracing property. From the maximal Lyapunov
exponents $\lambda_\mathrm{max}$, (a) $\gamma_\mathrm{th}=0.5723$,
(c) $\gamma_\mathrm{th}=0.7216$.
A new $U'$ is born and it is $NR$.
(See \fref{fig:broucke_location} for the coexistence of stable and unstable PO).
Remarkably period continuously changes through the threshold (see (b),(d)),
and, above threshold, periods of $S(R)$ and $U'(NR)$ are extremely degenerate
despite their distinct orbit profiles.
}
\end{figure}

%% file: S1-vs-squashed-S1.tex
\begin{figure}[!h]
    \begin{center}
    \includegraphics[width=12cm]{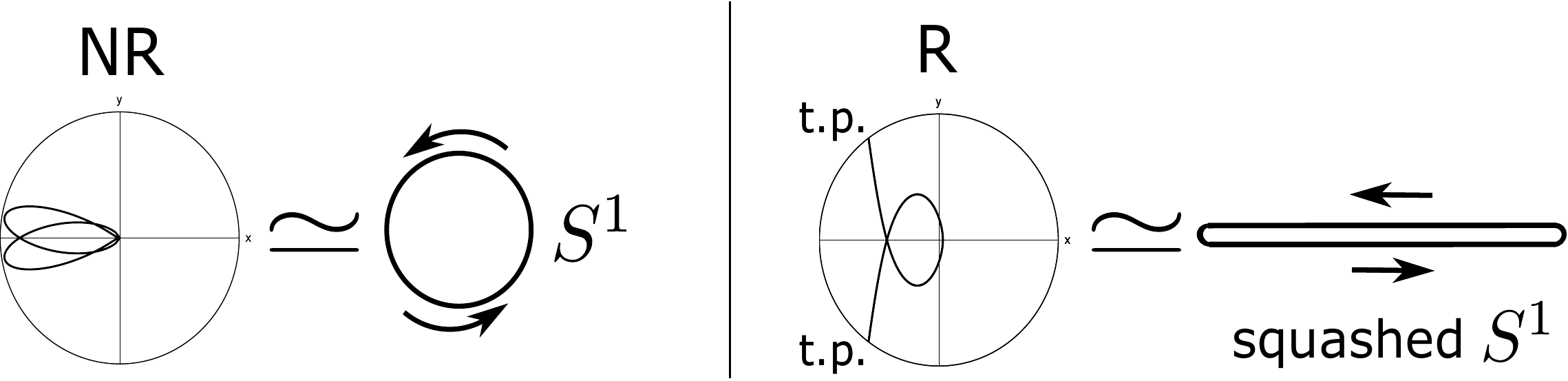}
    \end{center}
    \caption{
    A self-non-retracing PO (such as $U'$ in PO3-6) is homotopic to $S^1$, while
    a self-retracing PO ($U$ and $S$) is considered to a {\it squashed} $S^1$.
    In the right, each t.p. denotes a turning point.}
    \label{fig:S1-vs-squashed-S1}
\end{figure}

%% file: y-sym-classification.tex
\begin{figure}[!h]
    \centering
    \includegraphics[width=12cm]{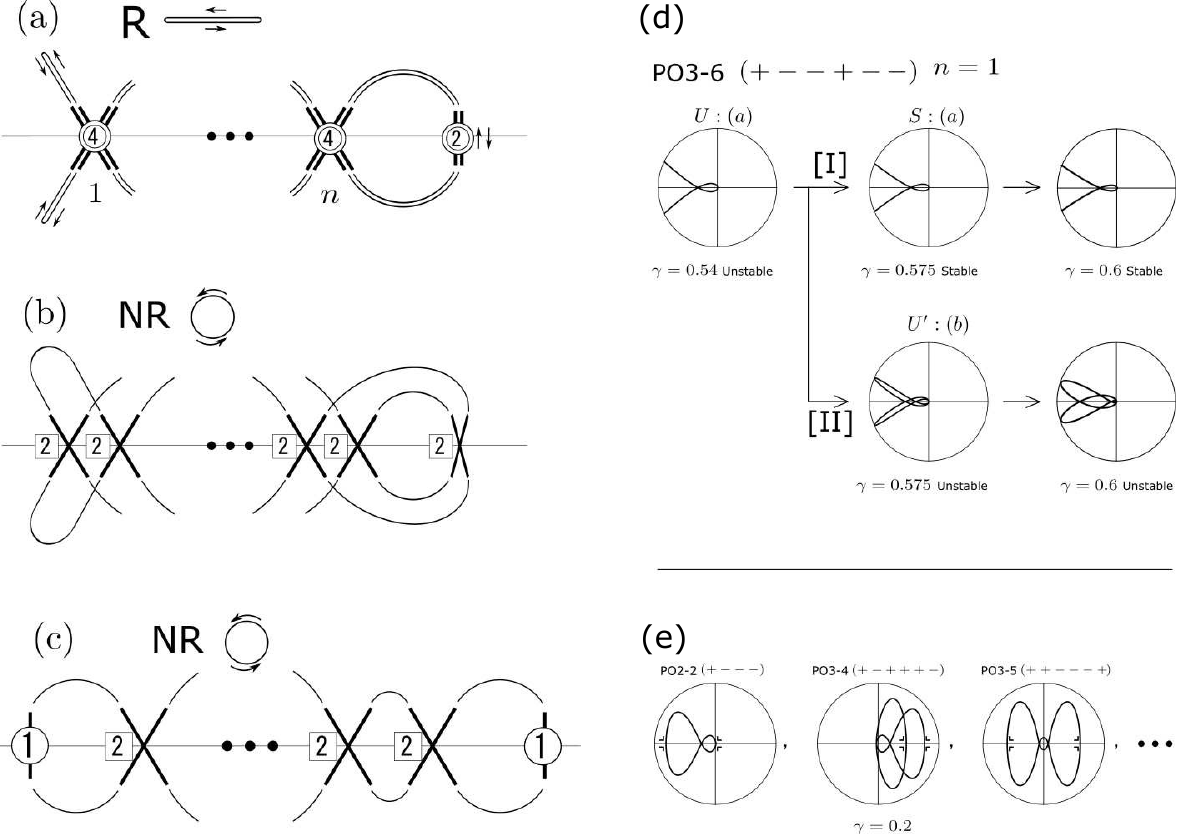}

    \caption{(a),(b),(c) Three classes of $Y$-symmetric PO.  Crossing multiplicities are shown.
     In (c) two perpendicular crossings (multiplicity one)  are set at ends for easy identification.
     (d) Broucke's PO3-6 bifurcation
     $U(R) \rightarrow S(R)+U'(NR)$. $U \rightarrow S$ are stability transition within class(a) (route[I]),
     $U\rightarrow U'$ is $(a)\rightarrow (b)$ (route [II]).
     (e) POs in class (c) ($\gamma=0.2$, all unstable).  PO15-1 $S''(NR)$ is also in class (c) (\fref{fig:PO15-1_b}).
        }
    \label{fig:y-sym-classification}
\end{figure}

%% file: topology-change-yoko.tex
\begin{figure}[!h]
	\centering
	\includegraphics[width=15cm]{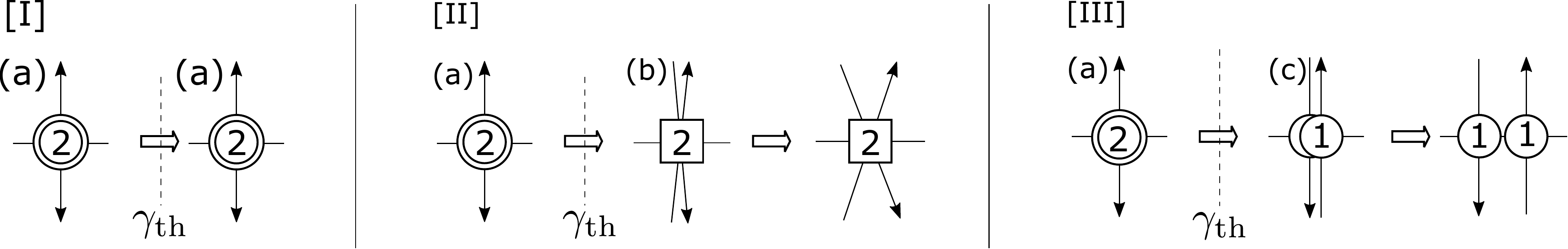}
\caption{Three routes of transition of $Y$-symmetric PO with decreasing anisotropy.   From class $(a)$ ($n_\perp =2$),
    [I] it remains in $(a)$ but changes from unstable to stable.
    [II] $(a) \rightarrow (b)$ ($n_\perp=2 \rightarrow 0$).
    Non-vanishing $U_0$  is generated quickly after $\gamma_{th}$.
    [III] $(a) \rightarrow  (c)$ No change of $n_\perp$, but multiplicity
    at crossing changes as $2\rightarrow 1+1$. $X_0$ changes rapidly after
    $\gamma_{th}$.
    }
	\label{fig:topology-change-yoko}
\end{figure}

%% file: PO15-1_bifurcation.tex
\begin{figure}[!h]
	\centering
	\includegraphics[width=15cm]{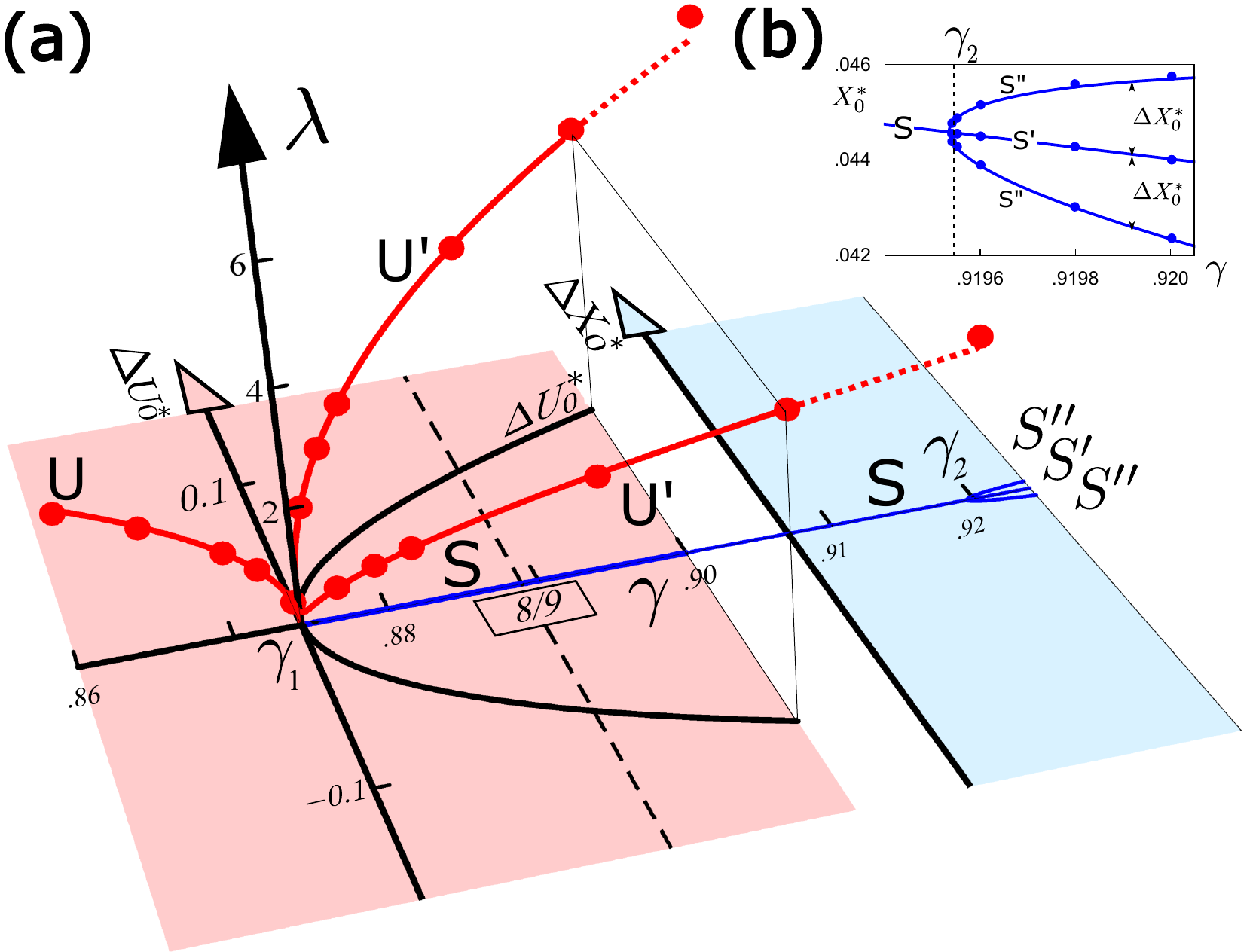}
	\caption{
 (a) A rank 15 PO bifurcates twice; $U(R) \rightarrow S(R)+U'(NR)$ and then $S(R)\rightarrow
 S'(R)+S''(NR)$. The coordinates are $(\Delta U_0^*,  \gamma, \lambda)$
and $(\Delta X_0^*, \gamma, \lambda)$ respectively.
Curves are from a simple fit
$  \lambda_\mathrm{max}
  =   a \theta(\gamma_1-\gamma) \sqrt{\gamma_1-\gamma}
   +  b \theta(\gamma-\gamma_1) \sqrt{\gamma-\gamma_1},
   (\gamma_1, a, b)\!=\!(0.8743,22.36,30.87)$
and $U_0^* \!=\!\pm c\sqrt{\gamma-\gamma_{1,U}},
 (\gamma_{1,U},c)\!=\!(0.8744,0.6822)$
give good fit with $| \gamma_1 - \gamma_{1,U}| \le ~O(10^{-4}$).
(b) magnifies the stable PO's bifurcation near the threshold.
The fit is $
 \Delta X_0\!=\!c \sqrt{\gamma-\gamma_2},(\gamma_2,c)\!=\!(0.9195,0.07818)$.
 }
\label{fig:PO15-1_bifurcation}
\end{figure}

%% file: PO15-1-bifurcation-scheme.tex
\begin{figure}[!ht]
	\centering
    	\includegraphics[width=10cm]{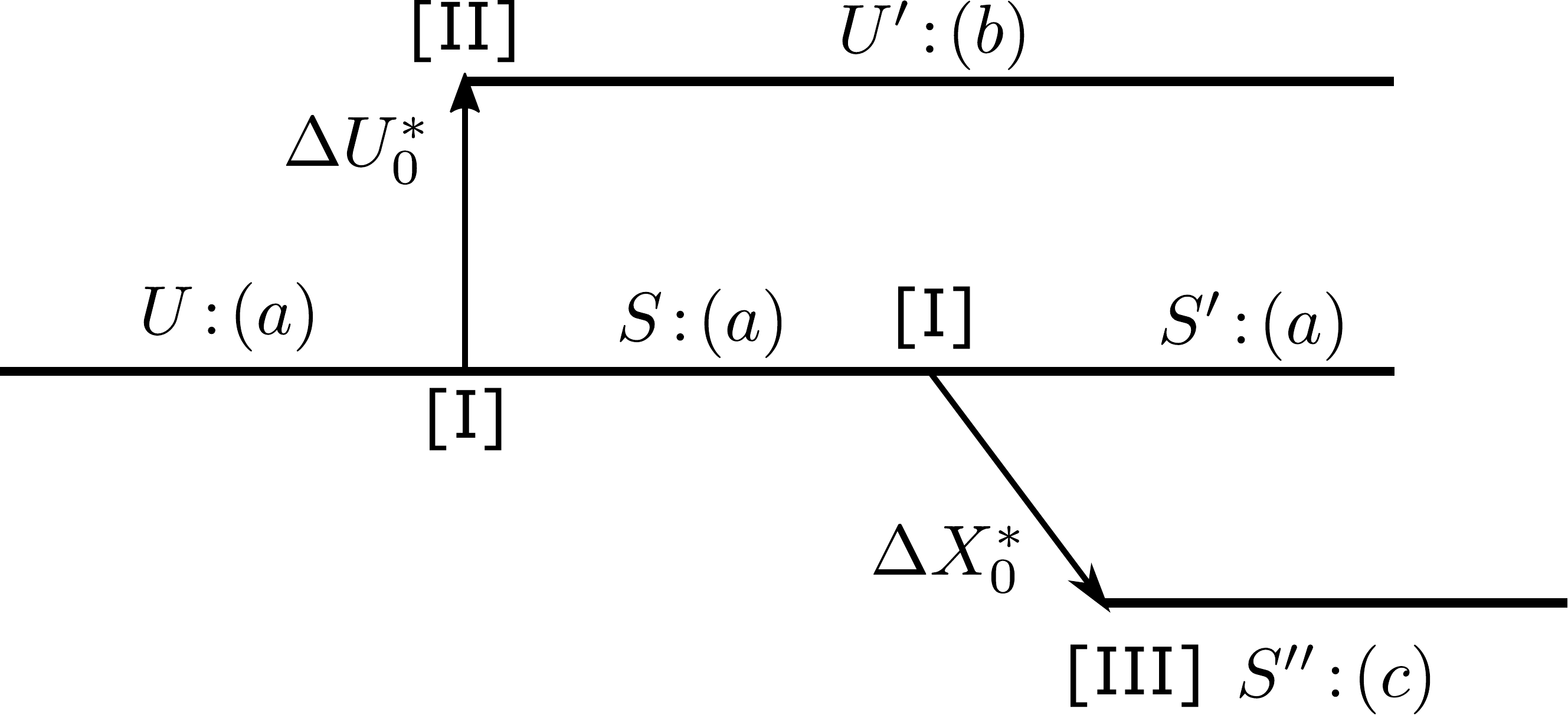}
    	\caption{Bifurcation scheme of the PO15. }
\label{fig:PO15-bifurcation-scheme}
\end{figure}

%% file: PO15-1-location.tex
\begin{figure}[!h]
	\centering
  \includegraphics[width=14cm]{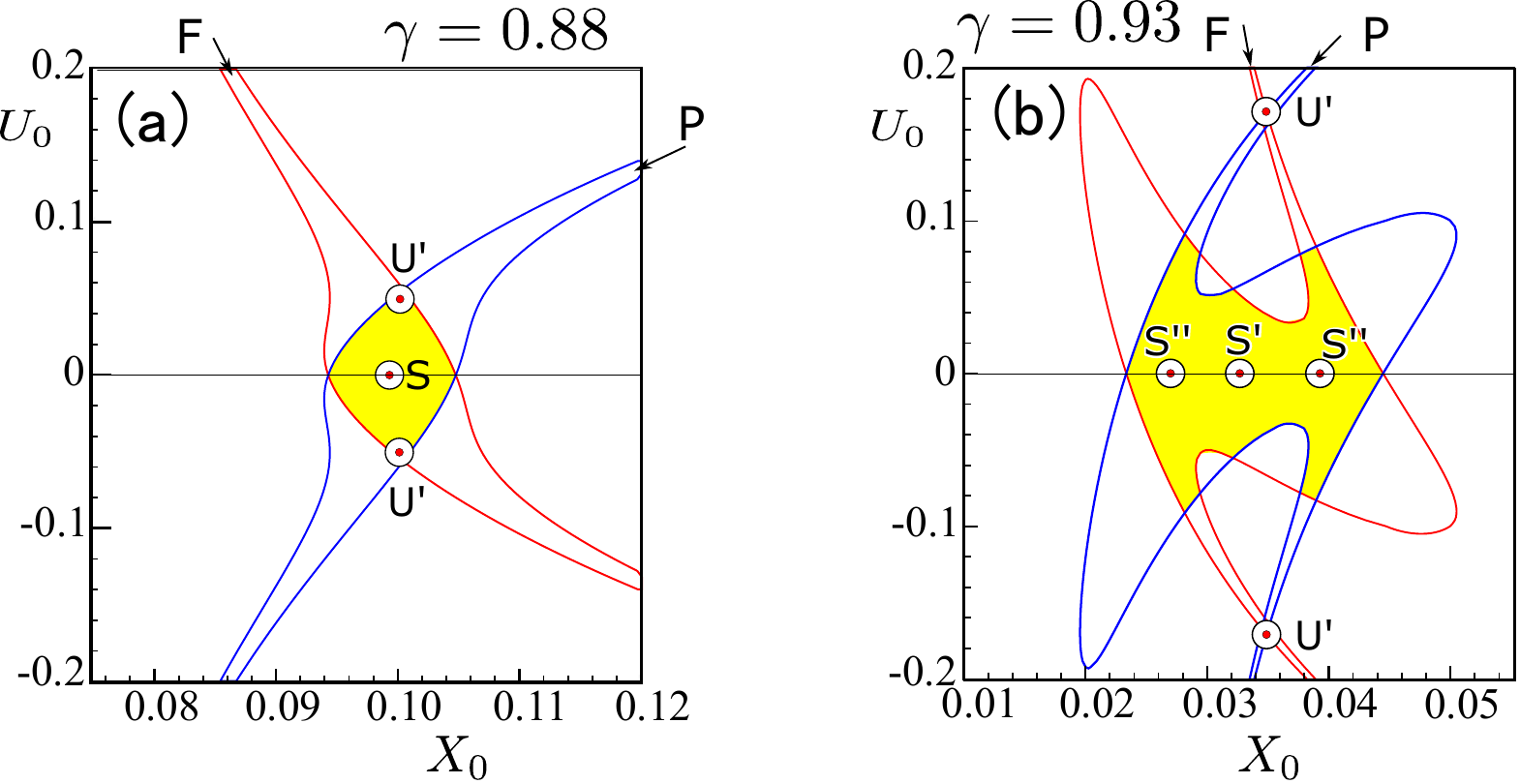}
	\caption{Initial points of a PO15
(a)  after $U \rightarrow S+ U'$,
(b) after $S\rightarrow S'+ S''$.
}
	\label{fig:PO15-1-location}
\end{figure}

%% file: PO15-1_a.tex
\begin{figure}[!h]
	\centering
	\includegraphics[width=16cm]{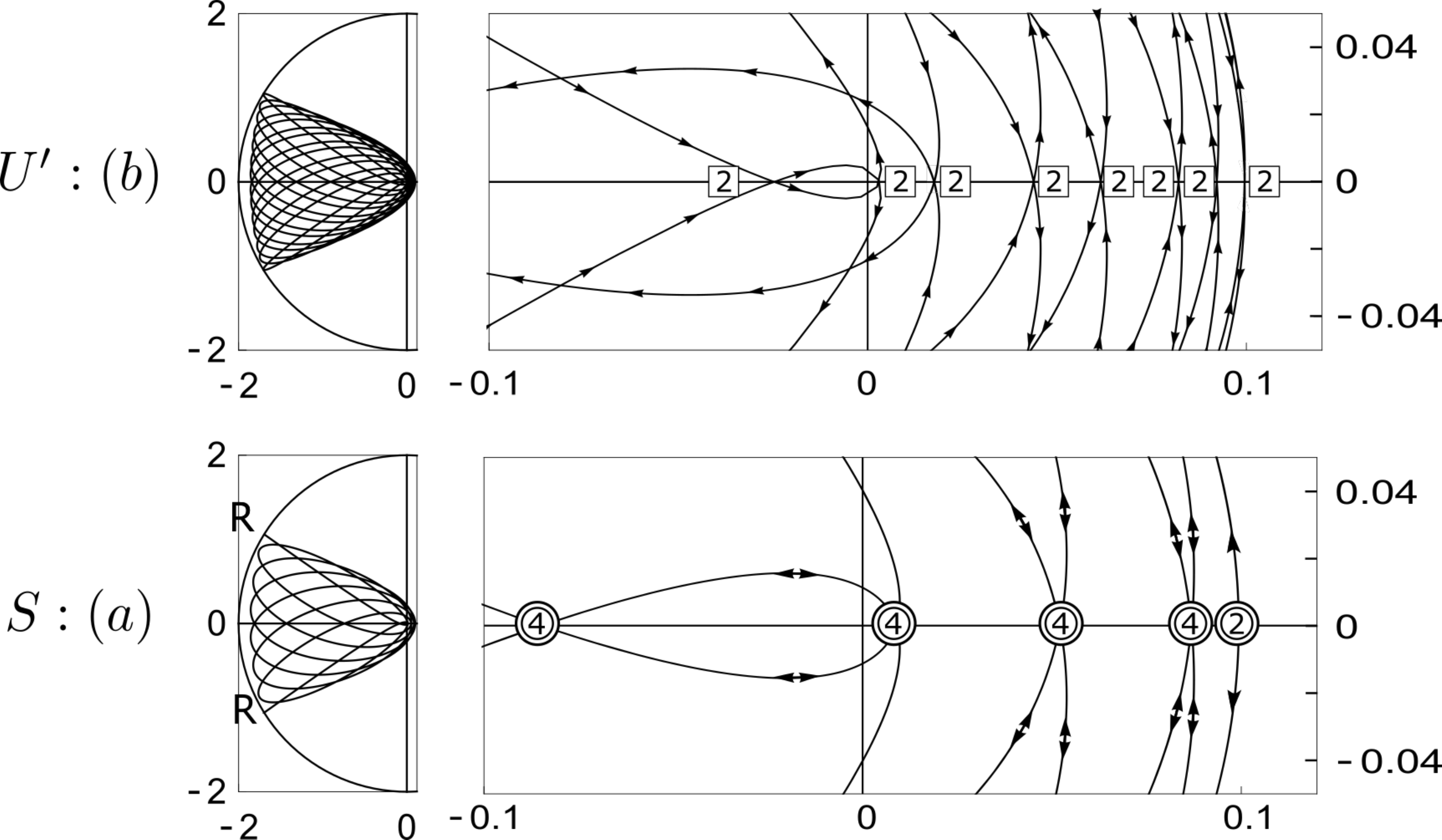}
	\caption{A rank 15 PO. $\gamma=0.88$.
Crossing multiplicity shows $U'(NR)$ ($S(R)$) is class
(b) [(a)].
 }
	\label{fig:PO15-1-a}
\end{figure}

%% file: PO15-1_b.tex
\begin{figure}[!h]
	\centering
	\includegraphics[width=16cm]{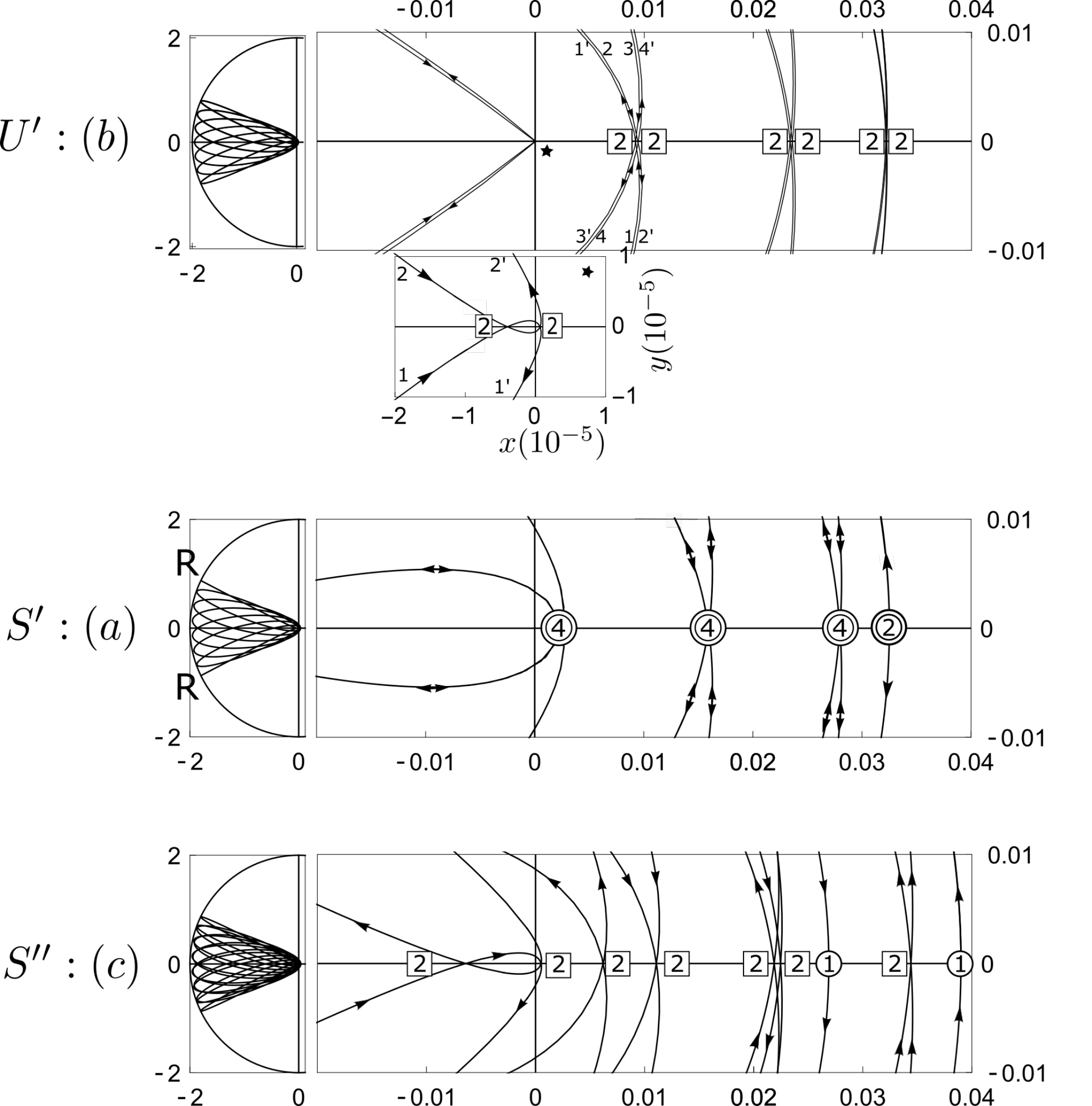}
	\caption{
$\gamma=0.93$.
The number of crossings tells $S'(R)$ [$S''(NR)$]  in class (a) [(c)].
 }
\label{fig:PO15-1_b}
\end{figure}

%% file: symmetry-counting.tex
\begin{table}[!h]
\caption{\label{tab:symmetry-counting}The number of distinct POs predicted under uniqueness assumption
in each of ten symmetry classes indexed by $k$. The fourth row lists the degeneracy
$\sigma_{\mathrm{sym}}(k)$.}
\begin{indented}
\item[]
\begin{tabular}{c  rrrrr  rrrrr  r}
\br
& \multicolumn{5}{c}{Non-self retracing: NR} & \multicolumn{5}{c}{Self-retracing: R} & \\
\ns \ns
& \crule{5}&\crule{5}\\
& ---  &  $X$  &  $Y$  &  $XY$  & $O$  & --- &  $X$ &  $Y$  & $XY$ & $O$ & \\
\mr
Sym.Type $k$: & 1 & 2 & 3& 4& 5& 6& 7 & 8 & 9 & 10 & \\
$\sigma_\mathrm{sym}(k)$: & 8 &   4 &   4 &  2 & 4 &  4  &  2 &  2 & 1 & 2 & \\
\mr
Rank $n$ &&&&&&&&&&&total \#\\
\mr
 1 &     0 &   0 &   0 &  1 & 0 &  0  &  0 &  1 & 0 & 0 &     2 \\
 2 &     0 &   0 &   1 &  1 & 0 &  0  &  1 &  1 & 0 & 0 &     4 \\
 3 &     0 &   1 &   2 &  2 & 0 &  1  &  0 &  2 & 0 & 0 &     8 \\
 4 &     2 &   2 &   7 &  1 & 0 &  2  &  3 &  1 & 0 & 0 &    18 \\
 5 &    12 &   6 &  12 &  4 & 0 &  6  &  0 &  4 & 0 & 0 &    44 \\
 6 &    57 &  14 &  30 &  2 & 0 &  13 &  4 &  2 & 0 & 0 &   122 \\
 7 &   232 &  28 &  57 &  8 & 1 &  28 &  0 &  8 & 0 & 0 &   362 \\
 8 &   902 &  62 & 127 &  1 & 0 &  58 & 11 &  1 & 0 & 0 &  1162 \\
 9 &  3388 & 120 & 247 & 16 & 7 & 120 &  0 & 16 & 0 & 0 &  3914 \\
10 & 12606 & 264 & 508 &  4 & 0 & 246 & 16 &  4 & 0 & 0 & 13648 \\
\br
\end{tabular}
\end{indented}
 \end{table}

%% file: symmetry-counting-with-tshift.tex
\begin{table}[!h]
\caption{
\label{tab:symmetry-counting-with-tshift}
The number of distinct codes $N(p,k)$ is tabulated
as a $p$-th row, $k$-th column  element,
where  $p|n$ counts the code-shift degeneracy and $k$ labels
symmetry class (see \tref{tab:symmetry-counting}) with $\sigma_{\mathrm{sym}}(k)$
in the square bracket.
Rank $n=9$ (upper) and 10 (lower).
This is obtained by dividing codes by orbit symmetry and by code-shift equivalence,
thus $\Sigma_{p|n}\Sigma_{k} p N(p,k)\sigma_{\mathrm{sym}}(k)=2^{2n}$.
The bottom row, $\Sigma_{p|n}  N(p,k)$ gives number of distinct codes in each symmetry class.
}
\begin{indented}
\item[]
\noindent
\begin{tabular}{crrrrrrrrrrr}
	\br
	$N(p,k)$ &   1[8] & 2[4] & 3[4]& 4[2]& 5[4]& 6[4]& 7[2] & 8[2] & 9[1] & 10[2] &\#codes$^{\rm a}$\\
	\mr
	9 & 3388 & 119 & 245 & 14 & 7 & 119 & 0 &14  & 0 & 0 & 262080\\
	3 &    0 & 1   & 2   & 1  & 0 & 1   & 0 & 1  & 0 & 0 &     60\\
	1 &    0 & 0   & 0   & 1  & 0 & 0   & 0 & 1  & 0 & 0 &     4\\
	\mr
	 3914& 3388 & 120  & 247   & 16 & 7    & 120  & 0 & 16 & 0 &  0  & $2^{18}$\\
	\mr
	\multicolumn{12}{c}{}\\
	\mr
	$N(p,k)$ & 1[8] & 2[4] & 3[4]& 4[2]& 5[4]& 6[4]& 7[2] & 8[2] & 9[1] & 10[2] &\#codes$^{\rm a}$\\
	\mr
	10 & 12594 & 258 & 495 & 0 & 0 & 240 & 15 & 0 & 0 & 0 & 1047540 \\
	5  &    12 &   6 &  12 & 3 & 0 &   6 &  0 & 3 & 0 & 0 & 1020\\
	2  &     0 &   0 &   1 & 0 & 0 &   0 &  1 & 0 & 0 & 0 & 12\\
	1  &     0 &   0 &   0 & 1 & 0 &   0 &  0 & 1 & 0 & 0 & 4\\
	\mr
	 13648 & 12606 & 264 & 508 & 4 & 0 & 246 & 16 & 4 & 0 & 0 & $2^{20}$\\
	\br
\end{tabular}
\item[] $^{\rm a}$ $\Sigma_{k}p N(p,k)\sigma_{\mathrm{sym}}(k)$.
\end{indented}
\end{table}

%% file: typical-pos-at-n9-Otype.tex
\begin{figure}[!h]
     \lineup
     \begin{tabular}{ll}
             \begin{minipage}{0.5\hsize}
                     \begin{center}
                         \includegraphics[keepaspectratio, scale=0.5]{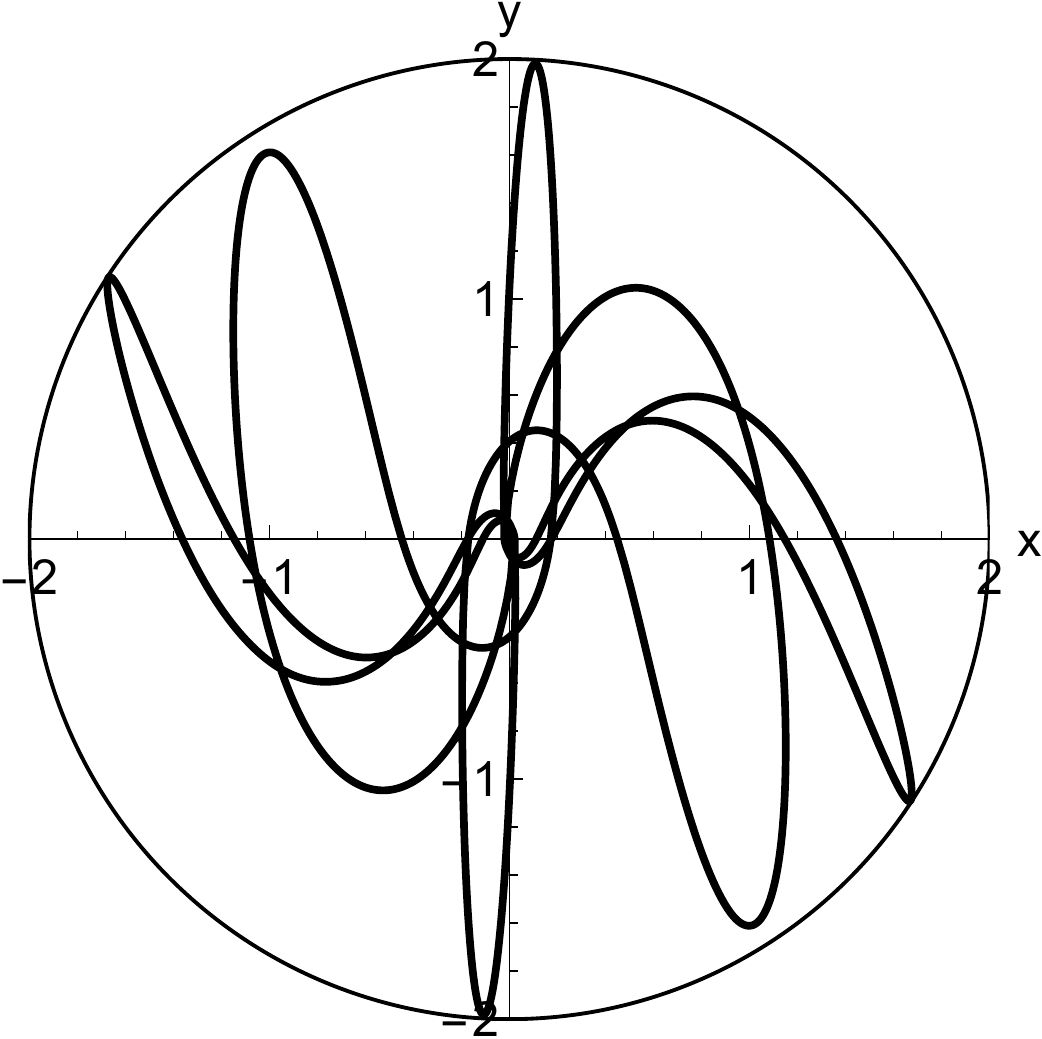}
                          \end{center}
             \end{minipage}
             \begin{minipage}{0.5\hsize}
                  \makeatletter
                    \def\@captype{table}
                    \makeatother
                      \scalebox{0.9}{
                       \begin{tabular}{c|ccccccccc}
                          \br
                            ($n$, Id)    & \multicolumn{9}{l}{ (9, 2285) }      \\
                            symmetry  & \multicolumn{9}{l}{ $A_X A_Y$ but $O$ }              \\
                            $X_0$        & \multicolumn{9}{D{.}{.}{29}}{ 1.6938906682974}  \\
                            $U_0$       & \multicolumn{9}{D{.}{.}{29}}{ -0.6845071038062}  \\
                            $\lambda_\mathrm{max}$  & \multicolumn{9}{D{.}{.}{29}}{ 14.73 }   \\

                            Action          & \multicolumn{9}{D{.}{.}{29}}{40.05 }   \\
                    \hline
                  &{\tiny 0}& {\tiny 1} & {\tiny 2} & {\tiny 3} &{\tiny 4} &{\tiny 5} &{\tiny 6} &{\tiny 7} & {\tiny 8}  \\
code &   $+$    &    $+$      &      $-$    &    $+$     &    $+$     &     $-$    &    $+$    &     $-$     &    $-$      \\
                   &{\tiny 9}&{\tiny 10}&{\tiny 11}& {\tiny 12}& {\tiny 13} &{\tiny 14} &{\tiny 15} &{\tiny 16} &{\tiny 17}   \\
                   &   $-$     &     $-$     &    $+$     &      $-$      &       $-$     &    $+$      &       $-$   &     $+$   &     $+$   \\
                   \br
                        \end{tabular}
                   }
             \end{minipage}
\end{tabular}
\caption{$O$ symmetric orbit at rank $n=9$.}
\label{fig:O-type}
\end{figure}

%% file: typical-pos-at-n10.tex
\begin{figure}[!h]
     \begin{tabular}{lc}
             \begin{minipage}{0.5\hsize}
                     \begin{center}
                         \includegraphics[keepaspectratio, scale=0.5]{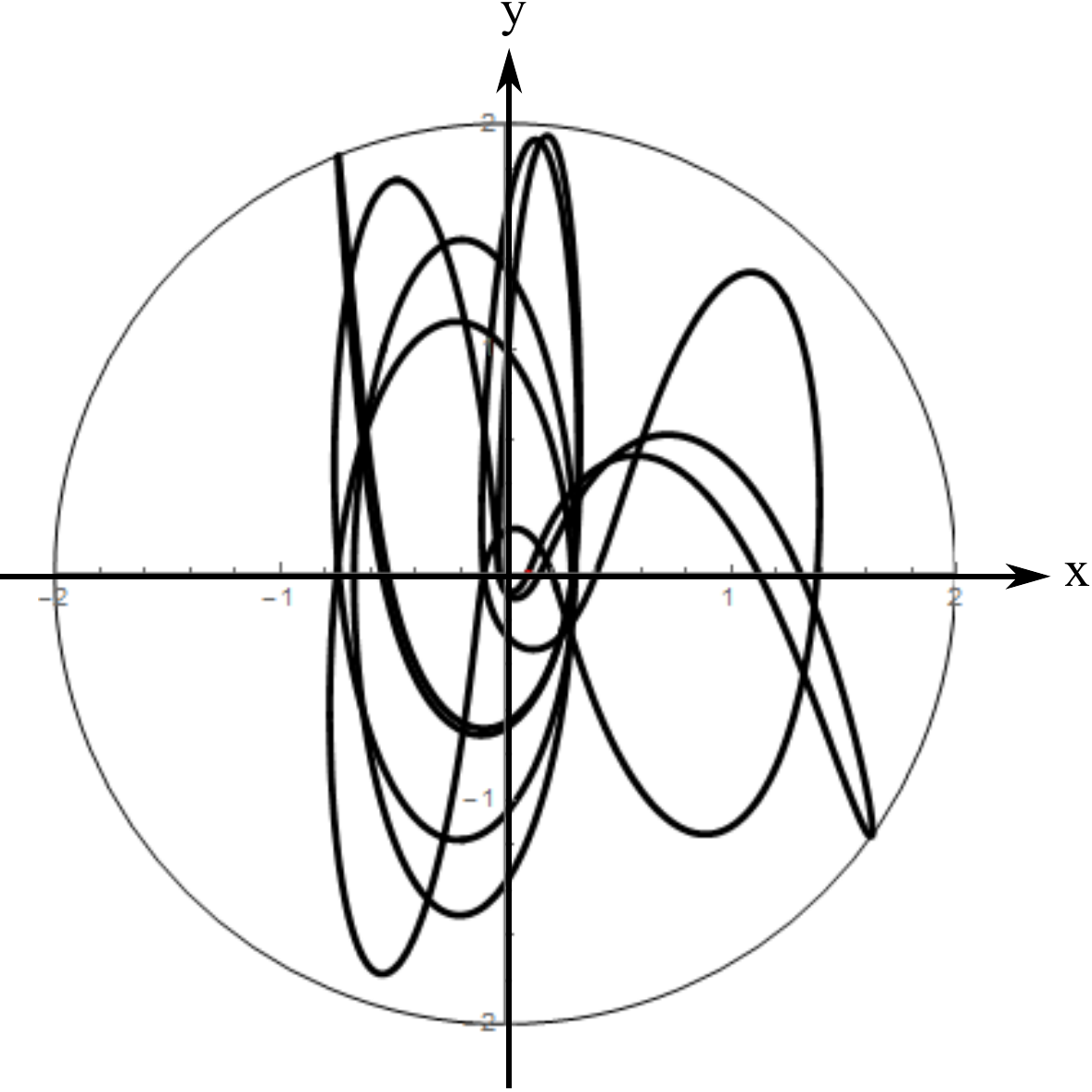}
                        \end{center}
             \end{minipage}
             \begin{minipage}{0.5\hsize}
                \begin{center}
                    \makeatletter
                    \def\@captype{table}
                    \makeatother
                           \scalebox{0.8}{
                                    \begin{tabular}{c|cccccccccc}
                                            ($n$, Id)     &  \multicolumn{10}{l}{ (10, 10000)          }    \\
                                            symmetry   &  \multicolumn{10}{l}{ $A_X, A_Y$ ~~(k=1)           }    \\
                                            $X_0$         & \multicolumn{10}{D{.}{.}{29}}{1.073770101110692 }   \\
                                            $U_0$         & \multicolumn{10}{D{.}{.}{29}}{0.806561621655999 }   \\
                       $\lambda_\mathrm{max}$  & \multicolumn{10}{D{.}{.}{29}}{14.31908020947697}   \\
                                            Action            & \multicolumn{10}{D{.}{.}{29}}{49.36092248823712}   \\
                                            \hline
                                            &{\tiny  0} & {\tiny 1} & {\tiny 2}& {\tiny 3} &{\tiny 4} &{\tiny 5} &{\tiny 6} &{\tiny 7} & {\tiny 8} & {\tiny 9} \\
                                            code     & $+$ & $+$ & $+$ & $+$ & $-$ & $+$ & $-$ & $-$ & $+$ & $-$ \\
                                            &{\tiny  10} & {\tiny 11} & {\tiny 12}& {\tiny 13} &{\tiny 14} &{\tiny 15} &{\tiny 16} &{\tiny 17} & {\tiny 18} & {\tiny 19} \\
                                                          & $+$ & $+$ & $+$ & $-$ & $-$ & $+$ & $-$ & $+$ & $-$ & $-$ \\
                                            \hline
                                        \end{tabular}
                                 }
                \end{center}
            \end{minipage}
\end{tabular}

     \begin{tabular}{lc}
             \begin{minipage}{0.5\hsize}
                     \begin{center}
                         \includegraphics[keepaspectratio, scale=0.5]{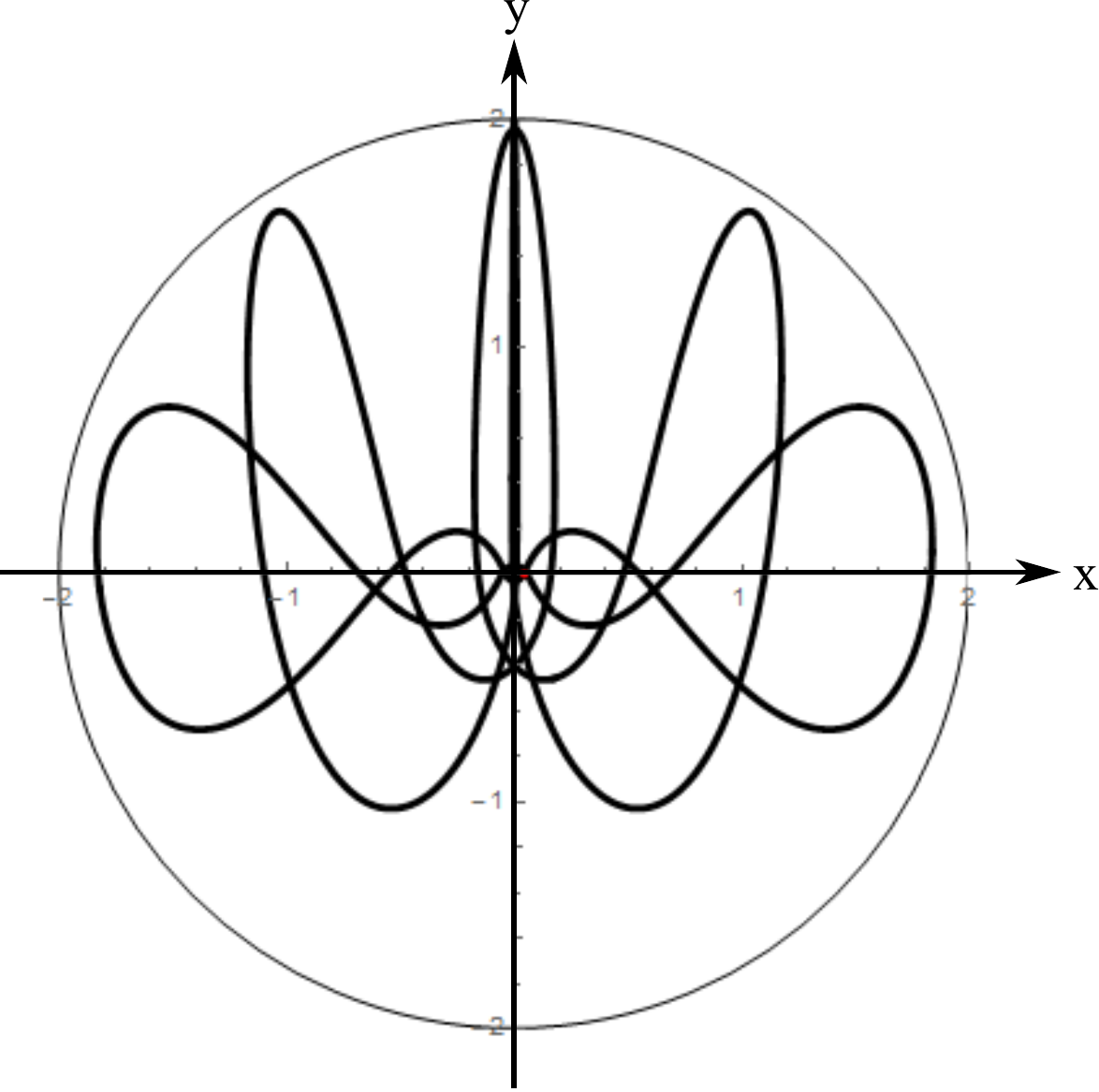}
                     \end{center}
             \end{minipage}
             \begin{minipage}{0.5\hsize}
                \begin{center}
                    \makeatletter
                    \def\@captype{table}
                    \makeatother
                           \scalebox{0.8}{
                                    \begin{tabular}{c|cccccccccc}
                                    ($n$, Id)     & \multicolumn{10}{l}{ (10, 5002) }             \\
                                    symmetry   & \multicolumn{10}{l}{ $S_X, A_Y$ ~~($k=2$)}              \\
                                    $X_0$         &  \multicolumn{10}{D{.}{.}{29}}{ 1.200325396087828 }   \\
                                    $U_0$         &  \multicolumn{10}{D{.}{.}{29}}{ 0.8860249568347829 }   \\
                                    $\lambda_\mathrm{max}$  &  \multicolumn{10}{D{.}{.}{29}}{ 17.46793360630122  }   \\
                                    Action          &  \multicolumn{10}{D{.}{.}{29}}{ 40.77425334607968 }    \\
                                    \hline
                                                 &{\tiny  0} & {\tiny 1} & {\tiny 2}& {\tiny 3} &{\tiny 4} &{\tiny 5} &{\tiny 6} &{\tiny 7} & {\tiny 8} & {\tiny 9} \\
                                    binary code     & $+$ & $+$ & $+$ & $+$ & $+$ & $-$ & $+$ & $+$ & $-$ & $+$ \\
                                                 &{\tiny  10} & {\tiny 11} & {\tiny 12}& {\tiny 13} &{\tiny 14} &{\tiny 15} &{\tiny 16} &{\tiny 17} & {\tiny 18} & {\tiny 19} \\
                                                  & $-$ & $-$ & $+$ & $-$ & $-$ & $-$ & $-$ & $-$ & $+$ & $-$ \\
                                    \hline
                                    \end{tabular}
                             }
                \end{center}
            \end{minipage}
\end{tabular}

     \begin{tabular}{lc}
             \begin{minipage}{0.5\hsize}
                     \begin{center}
                         \includegraphics[keepaspectratio, scale=0.5]{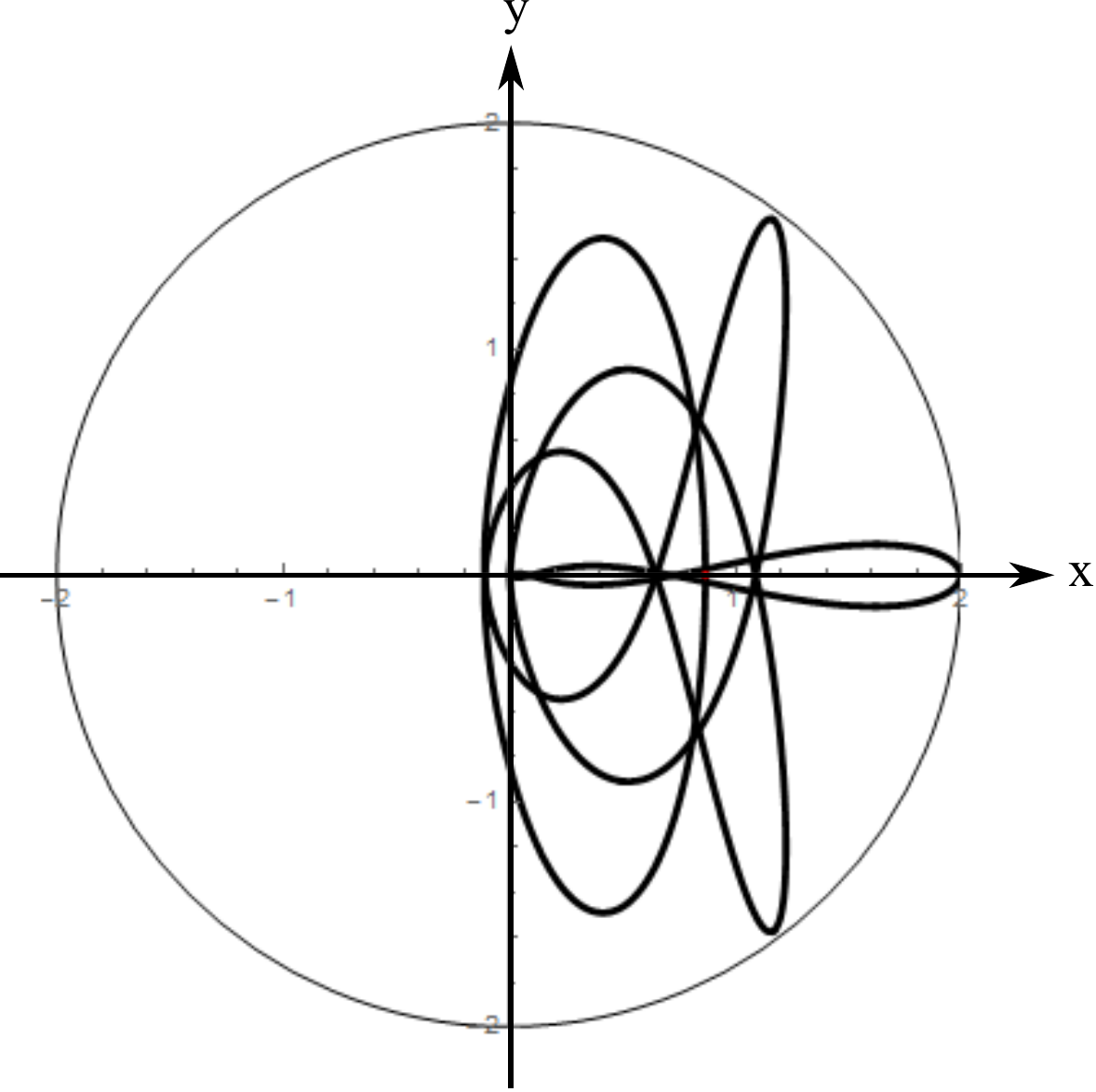}
                    \end{center}
             \end{minipage}
             \begin{minipage}{0.5\hsize}
                    \makeatletter
                    \def\@captype{table}
                    \makeatother
                      \scalebox{0.8}{
                        \begin{tabular}{c|cccccccccc}
                            ($n$, Id)     & \multicolumn{10}{l}{ (10, 100) }                \\
                            symmetry    & \multicolumn{10}{l}{ $A_X, S_Y$~~ ($k=3$)}              \\
                            $X_0$         &  \multicolumn{10}{D{.}{.}{29}}{ 0.8662601476594602 }  \\
                            $U_0$         &  \multicolumn{10}{D{.}{.}{29}}{ 0. }  \\
                            $\lambda_\mathrm{max}$  &  \multicolumn{10}{D{.}{.}{29}}{ 19.36441255533023 }   \\
                            Action          &  \multicolumn{10}{D{.}{.}{29}}{ 30.64876862775776 }   \\
                            \hline
                            &{\tiny  0} & {\tiny 1} & {\tiny 2}& {\tiny 3} &{\tiny 4} &{\tiny 5} &{\tiny 6} &{\tiny 7} & {\tiny 8} & {\tiny 9} \\
                            binary code     & $+$ & $-$ & $+$ & $+$ & $-$ & $+$ & $+$ & $+$ & $+$ & $+$ \\
                            &{\tiny  10} & {\tiny 11} & {\tiny 12}& {\tiny 13} &{\tiny 14} &{\tiny 15} &{\tiny 16} &{\tiny 17} & {\tiny 18} & {\tiny 19} \\
                                          & $+$ & $+$ & $+$ & $+$ & $+$ & $+$ & $-$ & $+$ & $+$ & $-$ \\
                            \hline
                        \end{tabular}
                        }
            \end{minipage}
\end{tabular}
\caption{Sample orbits at rank 10.
The first is asymmetric under both $X$ and $Y$; $k=1$ in the notation of \tref{tab:symmetry-counting}.
The third is $Y$ symmetric; class (c) because $NR$.
   }
\label{fig:three-sample-orbits}
\end{figure}

%% file: action_test.tex
\begin{figure}[!h]
	\begin{center}
	\includegraphics[width=8cm]{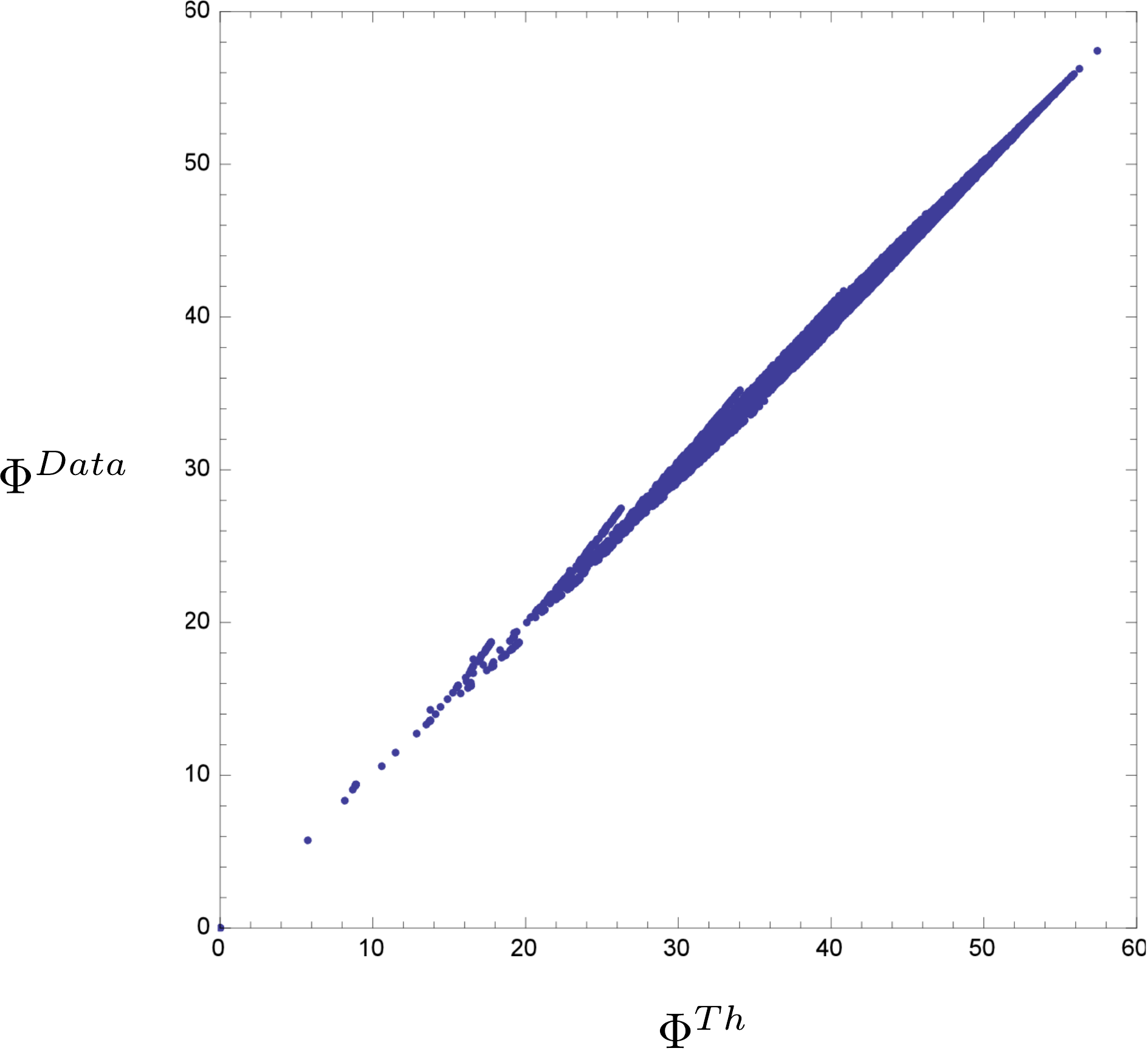}
	\end{center}
	\caption{The scatter plot of ($\Phi^{Th.}_{Id}$,
     $\Phi^{Data}_{Id}$). $Id$ labels each of  the distinct 13648
     POs in the $n=10$ family.
 	}
	\label{fig:action_test}
\end{figure}

%% file: separation-fig.tex
\begin{figure}[!h]
	\begin{center}
	\includegraphics[clip,width=13cm]{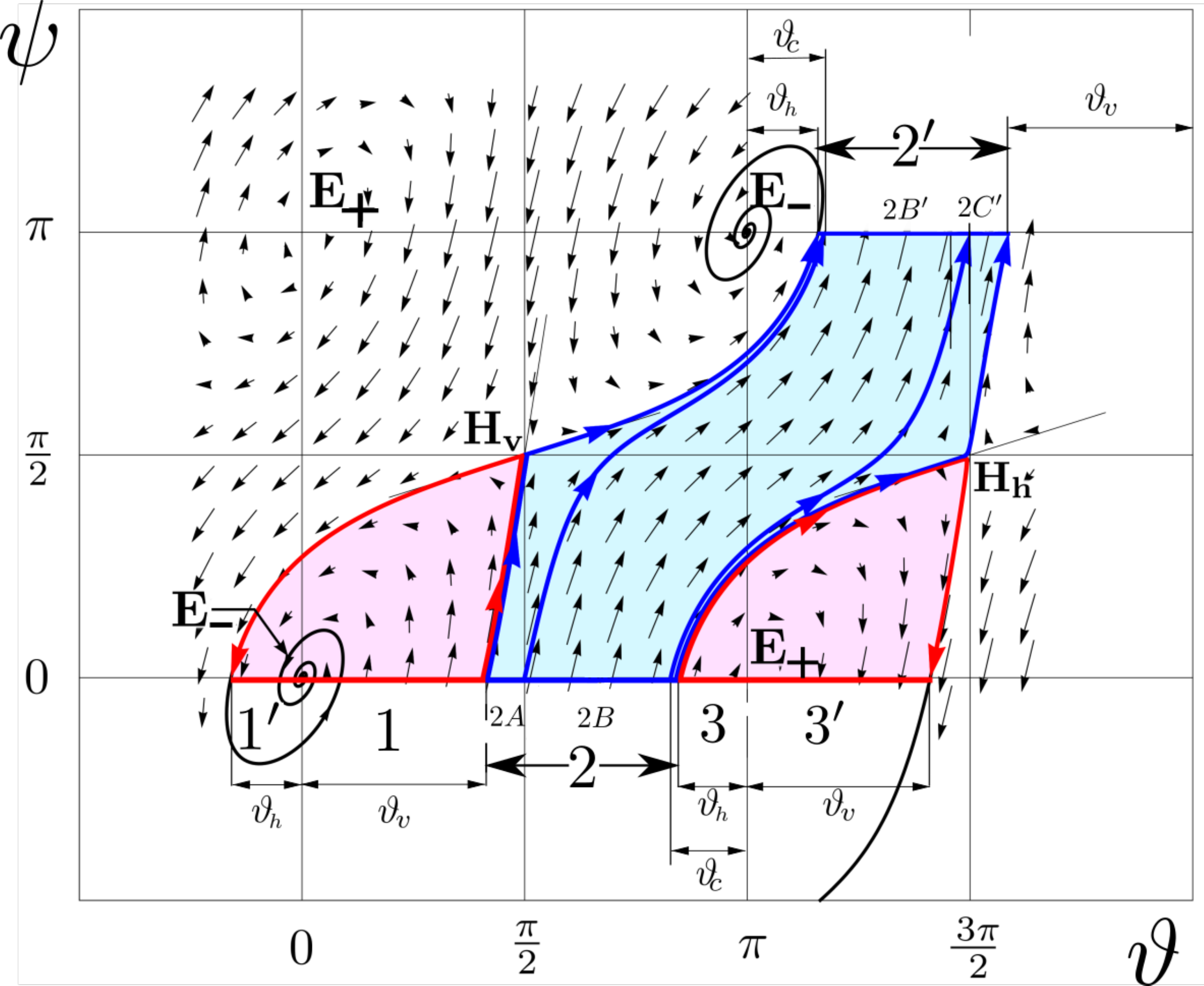}
	\end{center}
\vspace{-0.5cm}
	\caption{The flow given by the autonomous ODE (\ref{eq:autonomous-flow}
(The flow is the same with Fig.~1 in Gutzwiller \cite{gutzwiller77}
and equivalent to Fig.~2 in Devaney \cite{devaney78}.) The anisotropy is $\gamma=0.2$.
The initial fundamental domain is divided into three regions 1, 2, 3
due to the hyperbolic singularities
$H_v$ and $H_h$, The region 2 is further divided into
three sub-regions by the stream line with $\vartheta_0=\pi/2$
and that with $\vartheta_1=3\pi/2$. ($2C$ and $2A'$ are too
narrow to indicate by letters.)
}
	\label{fig:separation}
\end{figure}

%% file: variation-of-critical-angles.tex
\begin{figure}[!h]
	\centering
	\includegraphics[width=8cm]{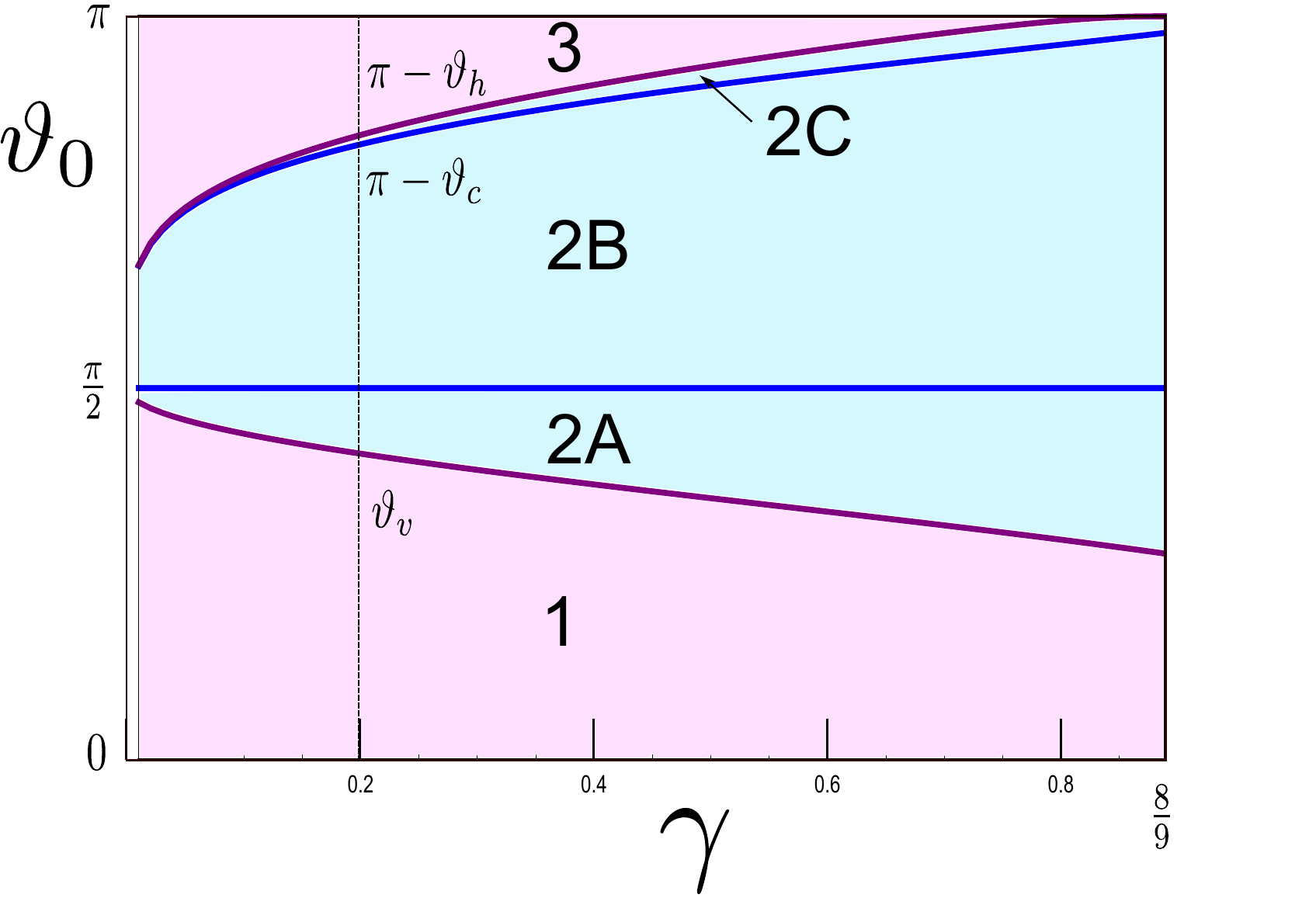}
	
	\caption{The variation of critical angles separating the fundamental
domain $\vartheta_0 \in [0,\pi], \phi=0$
into $1$, $2A$, $2B$, $2C$, $3$ for the anisotropy region $\gamma<8/9$.
At the canonical high anisotropy $\gamma=0.2$ corresponding to
\fref{fig:separation}, $\vartheta_h=0.319039058,
~\vartheta_c=0.344255689,~\vartheta_v=0.823746449$ in unit of $\pi/2$.
	}
\label{fig:variation-of-critical-angles}
\end{figure}

%% file: one_time_map_red_beak.tex
\begin{figure}[!h]
\centering	
\includegraphics[width=12cm]{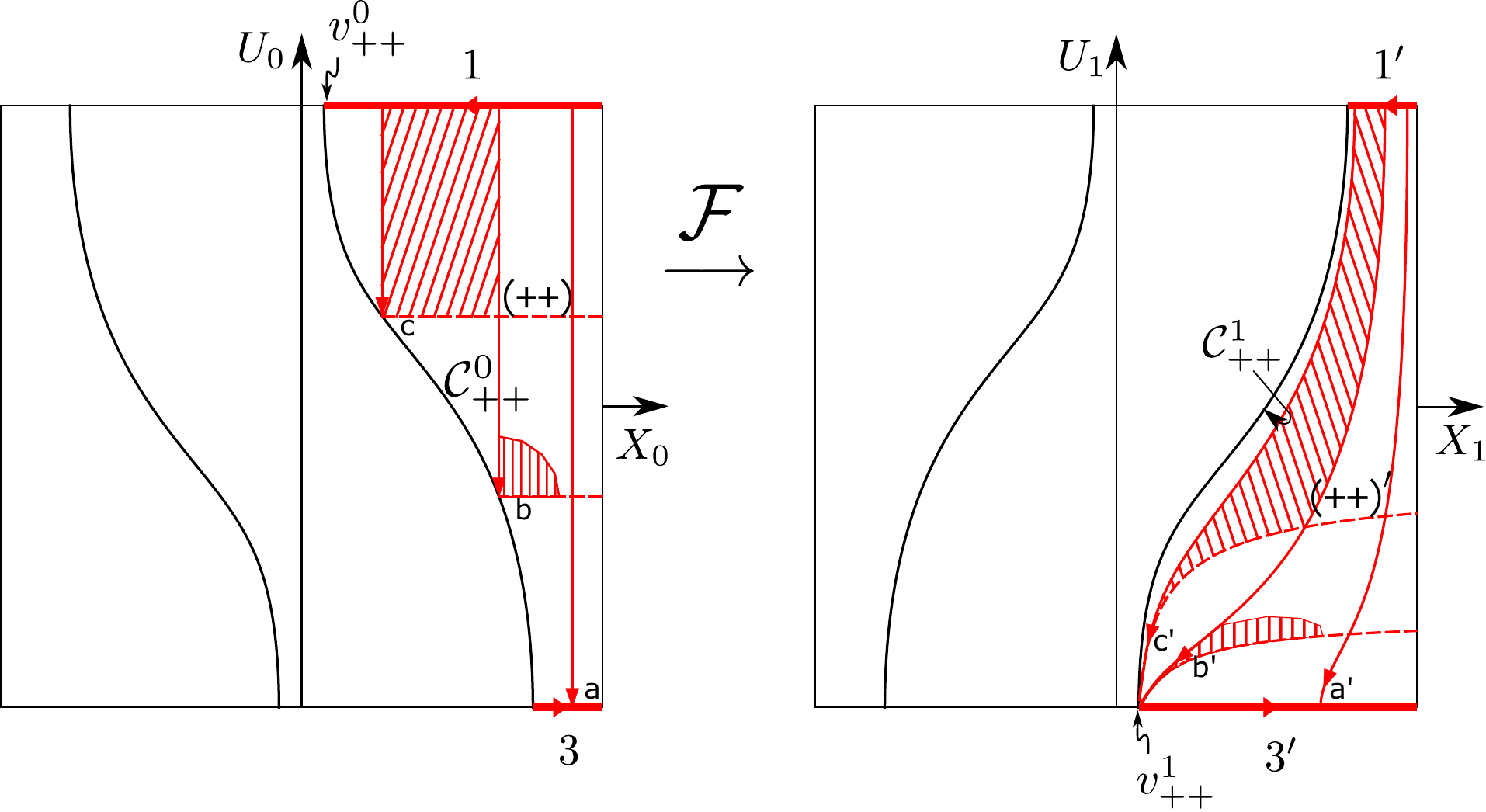}
\vspace{-0.5cm}
	\caption{ The action of ${\cal F}$ on $(+-)$.
            Figure is to scale ($\gamma=0.2$). Hatched regions
            are mapped preserving area and orientation.
            ${\cal C}^0_{++} \mapsto v^1_{++}$.}
	\label{fig:one_time_map_red_beak}
\end{figure}

%% file: one_time_map_blue_beak.tex
\begin{figure}[!h]
	\begin{center}
	\includegraphics[width=14cm]{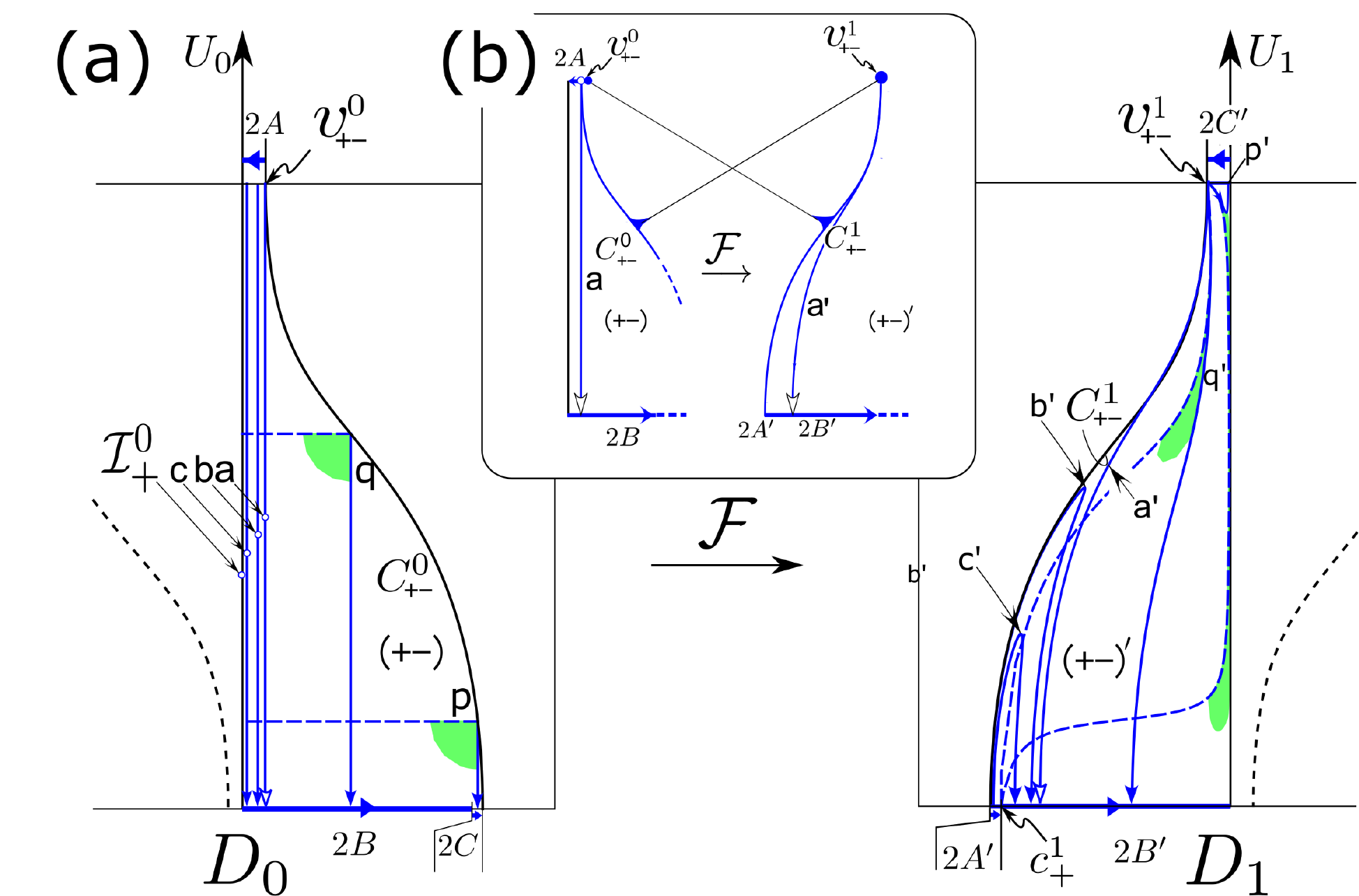}
	\end{center}
	\caption{
            ${\cal F}$ on $(+-)$.
            (a) The beaks ($p'$ and $q'$) show $C^0_{+-} \longmapsto v_{+-}^1$,
            while their bottoms reflect rotation (Compare a hook of $p'$ with a vertical curve of $q'$).
            Wings ($b',c'$) contracts a focusing point $c_{+}$, showing
            ${\cal I}_+ \longmapsto c_{+}$.
            (b) The vertical line segment $a$ is critical.
            Body of $a$ makes a curve $a'$, its top end $v_{+-}^0$ makes $C^1_{+-}$,
            and they together make a beak (enveloping wings),
            whose top $v_{+-}^1$ is then the image of the $C^0_{+-}$.
  	}
	\label{fig:one_time_map_blue_beak}
\end{figure}

%% file: congruence.tex
\begin{figure}[!h]
	\begin{center}
	\includegraphics[width=10cm]{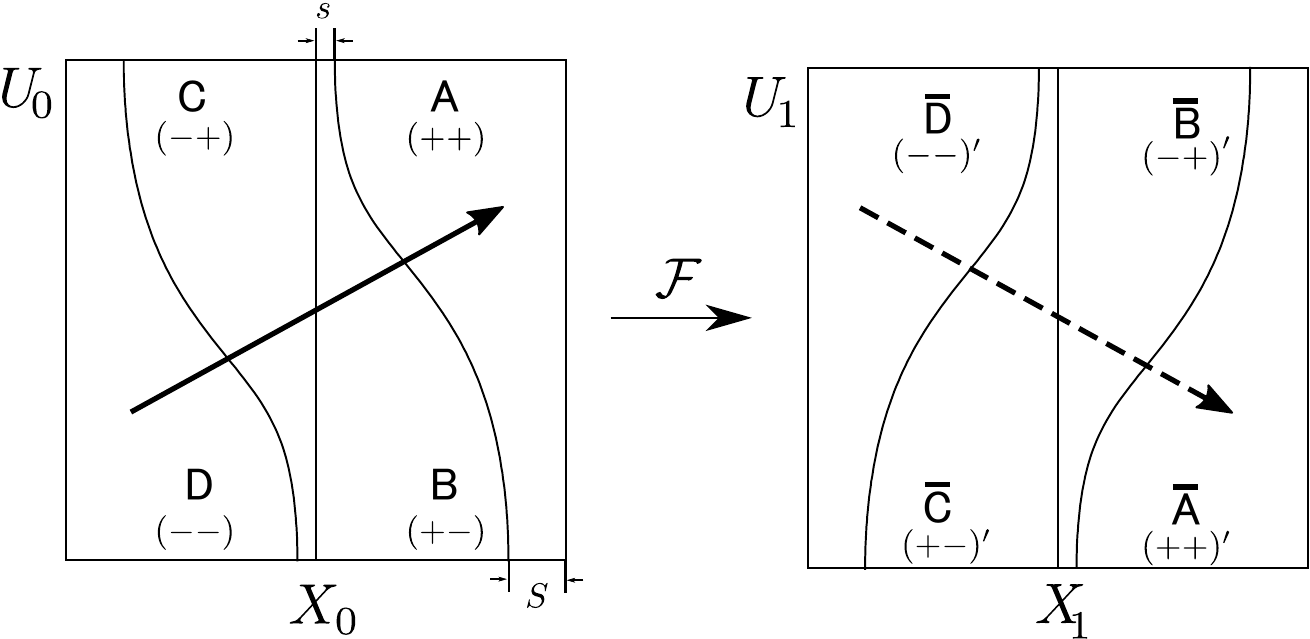}
	\end{center}
	\caption{Four regions in $D_0$;
$A\!(++),B\!(+-),C\!(-+),D\!(--)$ from right to left.
Similarly $\bar{A}\!(++)',\bar{B}\!(-+)',\bar{C}\!(+-)',\bar{D}(--)'$  $\in D_1$ also
from left to right.
(Note the swap between $(+-)$ and $(-+)$ by ${\cal F}$).
There are only two distinct shapes; $A$ with the shortest side
$S=2\cos^2 \vartheta_h$
and $B$ with $s=2\cos^2 \vartheta_v ~(S>s)$.
The direction of increasing DSS height is depicted by arrows, which are $U \ra -U$ mirror.
}
\label{fig:congruence}
\end{figure}

%% file: ribbon-longitudinal-splitting.tex
\begin{figure}[!h]
  \centering
  \includegraphics[width=80mm] {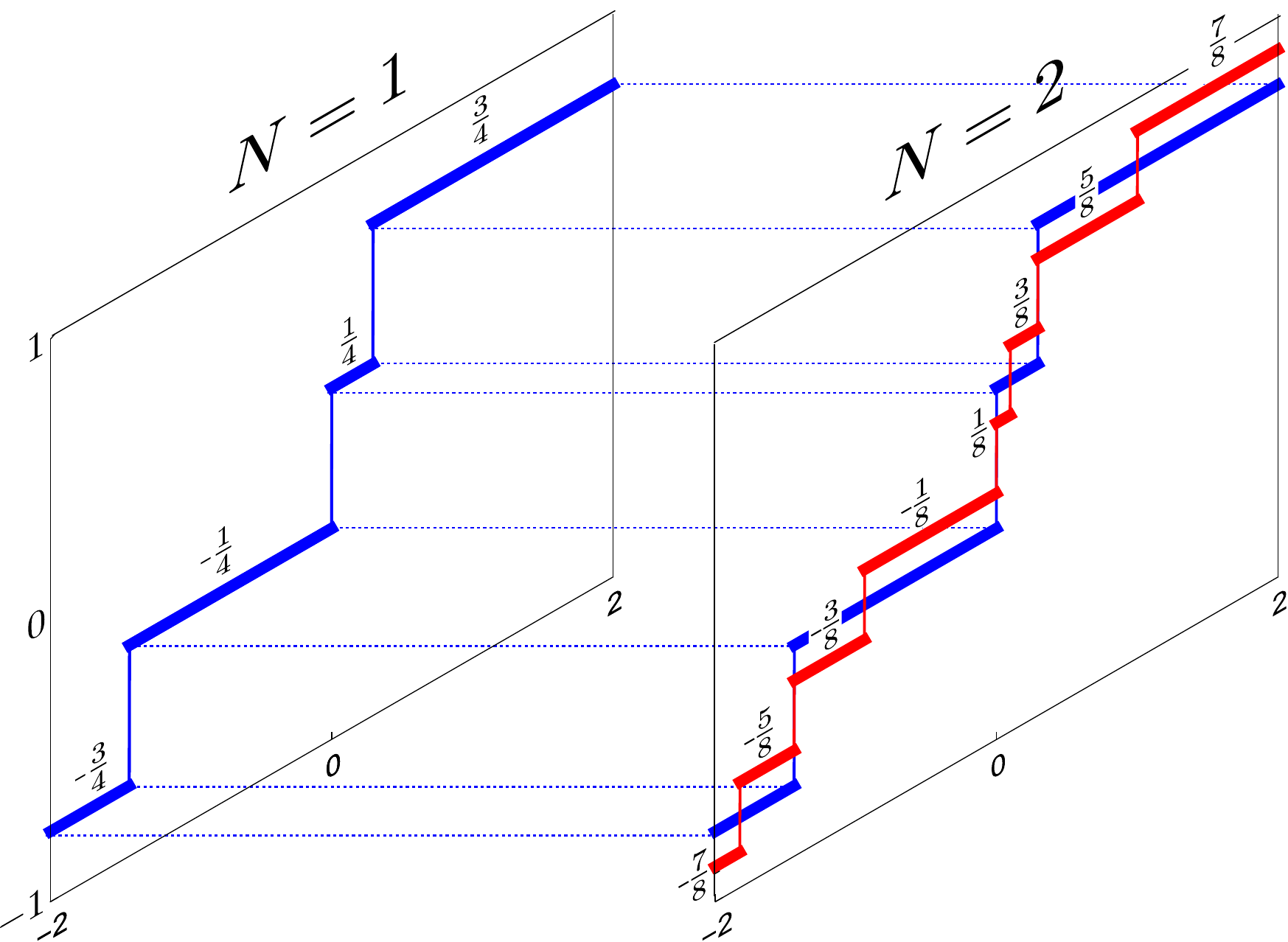}
\vspace{-0.5cm}
  \caption{
$N=1, 2$ DSS are compared by sections at $U_0=B/2$ ($\gamma=0.2$).
(See also $N=2\ra  3$ tiling-evolution in \fref{fig:crossing-of-future-past-ribbons}).
Next ribbons are created by {\it transverse} chopping+separation+elongation,
but it looks as a simple {\it longitudinal} splitting of every ribbon, if one does not care which
(half of) ribbon goes to which one of new ribbons.
   }
 \label{fig:ribbon-longitudinal-splitting}
\end{figure}

%% file: iteration-tree.tex
\begin{figure}[!h]
	\begin{center}
	\includegraphics[width=14cm]{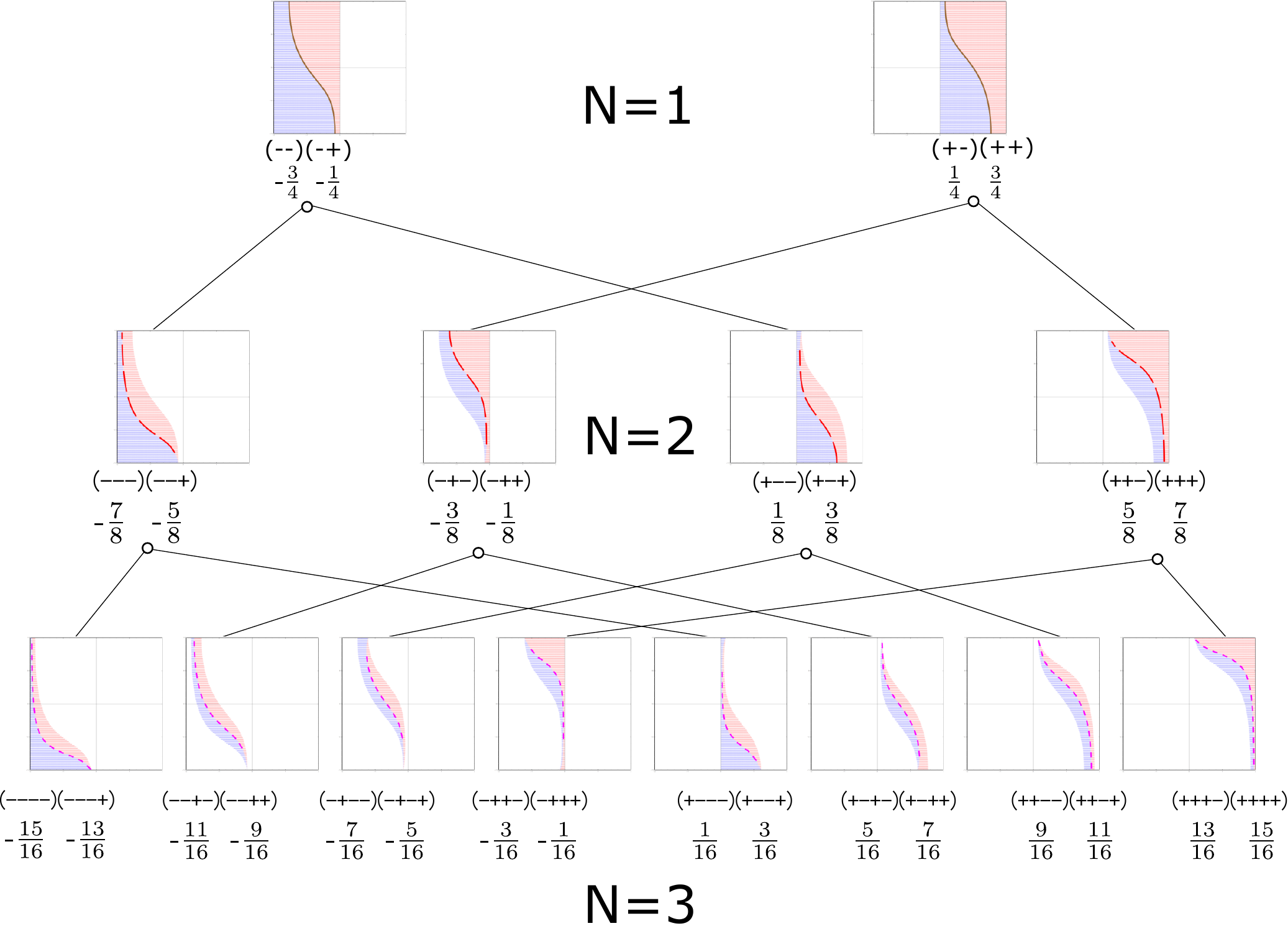}
	\end{center}
	\caption{From the level $N=1$ $2^2$ ribbons, higher level ribbons
    ($2^3$ at $N=2$, $2^4$ at $N=3$ are created
	by repeated application of $\mathcal{F}^{-1}$ and lines to track descendants
    are shown.
	It is extremely hard to track-back to the parent already
	through $\F^{-2}$ (twice  the {\it baker-map}).}
	\label{fig:iteration-tree}
\end{figure}

%% file: broucke_chi2_contours.tex
\begin{figure}[!h]
    \centering
  \includegraphics[width=15cm]{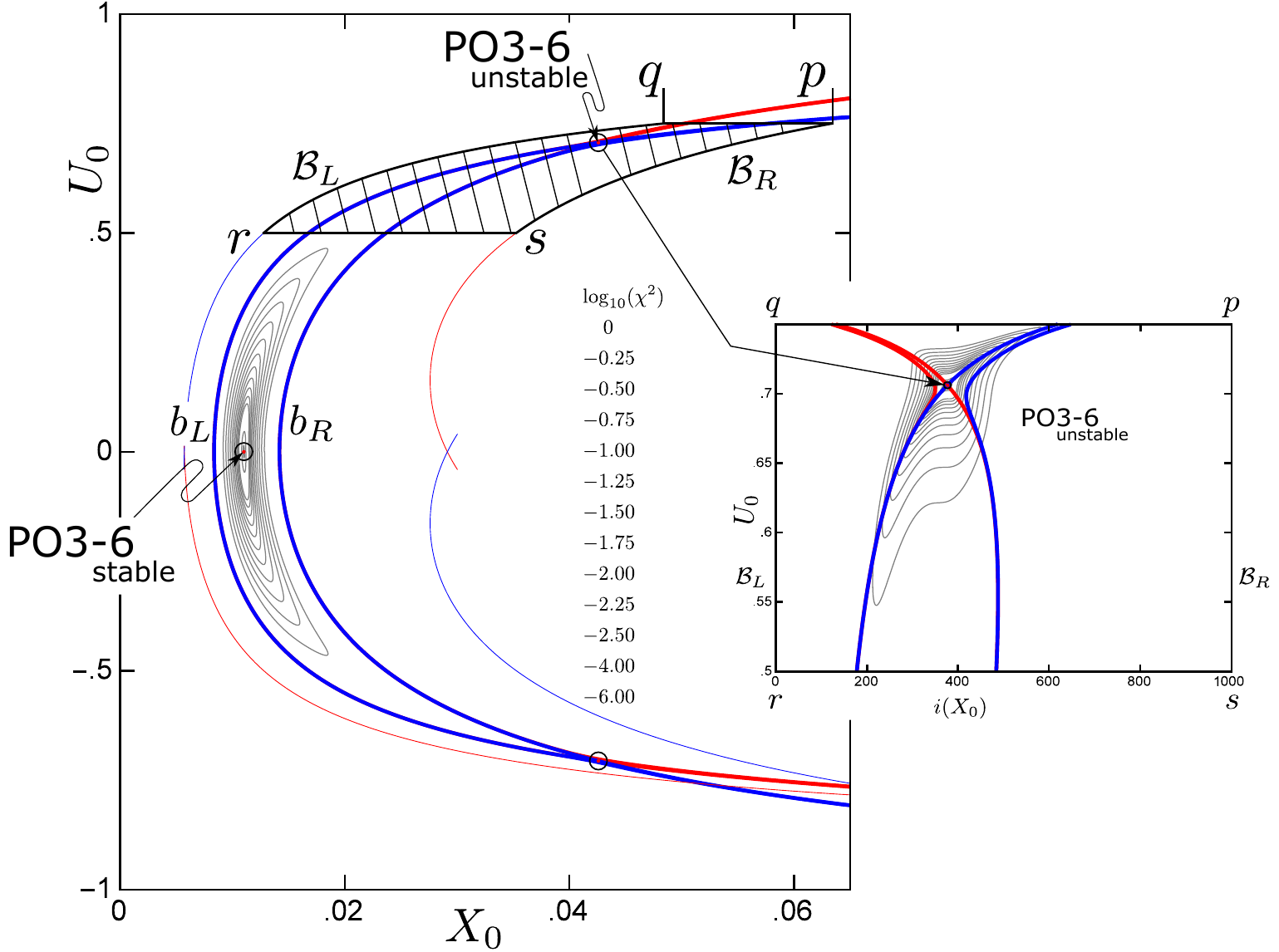}
    \caption{$\chi^2$ contours showing that no period $6$ PO other than PO3-6 (rank 3 id 6) $S$ and $U'$
    (\fref{fig:broucke_location_PO3-6}) exists inside Broucke's
    non-shrinking ribbon. $\gamma=0.6$.
    Union of future and past ribbons has asymptotic boundaries $b_L$ and $b_R$
    ($N=48$), which is well inside the $N=6$ union (${\cal B}_L$ to ${\cal B}_R$).
    Inside the latter, a fine mesh of initial points ($10^3 \times 10^3$) is set and $\chi^2$
    ((\ref{eq:chi2misfit}), $2n=6$)
    is calculated for each point.  This is a fail-safe procedure as well as a
    useful devise for magnifying near the narrow edge region.
    The equi-contours focus the $S$.
    Inset: The same $\chi^2$ contours in $pqrs$ (after a stretch in the $X$ direction).
    They focus the unstable $U'$ at the edge.
     }
    	\label{fig:broucke-chi2-contours}
\end{figure}